\documentclass[a4paper,onecolumn,11pt,unpublished]{quantumarticle}
\pdfoutput=1
\usepackage[utf8]{inputenc}
\usepackage[english]{babel}
\usepackage[T1]{fontenc}
\usepackage{amsmath,amssymb,amsthm,mathtools}
\usepackage{graphicx}
\usepackage{booktabs}
\usepackage{array}
\usepackage[numbers,sort&compress]{natbib}
\usepackage{hyperref}

\newtheorem{theorem}{Theorem}
\newtheorem{proposition}[theorem]{Proposition}
\newtheorem{lemma}[theorem]{Lemma}
\newtheorem{corollary}[theorem]{Corollary}
\newtheorem*{thmA}{Theorem A}
\newtheorem*{thmB}{Theorem B}
\newtheorem*{thmC}{Theorem C}
\theoremstyle{definition}
\newtheorem{definition}[theorem]{Definition}
\newtheorem{example}[theorem]{Example}
\theoremstyle{remark}
\newtheorem{remark}[theorem]{Remark}

\newcommand{\C}{\mathbb{C}}
\newcommand{\R}{\mathbb{R}}
\newcommand{\Z}{\mathbb{Z}}
\newcommand{\T}{\mathbb{T}}
\newcommand{\tr}{\operatorname{Tr}}
\newcommand{\rank}{\operatorname{rank}}
\newcommand{\conv}{\operatorname{conv}}
\newcommand{\diag}{\operatorname{diag}}

\newcommand{\ket}[1]{|#1\rangle}
\newcommand{\bra}[1]{\langle #1|}
\newcommand{\proj}[1]{|#1\rangle\langle #1|}
\newcommand{\tp}{\mathsf{T}}
\newcommand{\eps}{\varepsilon}
\newcommand{\SL}{\mathrm{SL}}
\newcommand{\GL}{\mathrm{GL}}
\newcommand{\Sym}{\mathrm{Sym}}
\newcommand{\Toep}{\mathcal{T}}
\newcommand{\Dec}{\mathcal{D}}
\newcommand{\Hank}{\mathcal{H}}
\newcommand{\Lat}{\mathcal{L}}
\newcommand{\Rg}{\mathcal{R}}

\begin{document}

\title{Bound entanglement as a trigonometric moment problem}

\author{Lilong Qian}
\email{ssqianlilong@gmail.com}
\affiliation{Independent Research}
\author{Teng Zhang}
\thanks{Corresponding author}
\email{zhangteng@gxu.edu.cn}
\affiliation{Guangxi University, Nanning, China}

\begin{abstract}
Entangled states that are positive under partial transposition (PPT) are the main obstruction to a simple description of entanglement. We study three basic questions about them within a class of multiqubit states, the mixtures, with real but possibly negative weights, of products of equatorial spin coherent states of several groups of qubits: which PPT entangled states belong to the class, how robust their entanglement is to white noise, and how small their rank can be. In this class each question becomes a question about a truncated trigonometric moment problem. Every such state is PPT across every bipartition, and it is fully separable exactly when an associated multi-level Toeplitz matrix is the moment matrix of a positive measure on the torus. We answer the three questions exactly in the following cases. For three qubits, every rank-four PPT entangled state with nonzero Lorentz invariant belongs to the class after invertible local operators, which proves a working hypothesis of Garberg, Irgens and Myrheim and yields a complete set of invariants for this seven-dimensional family; among the states with vanishing invariant, exactly a one-parameter family belongs to it. For every three-qubit state built from an unextendible product basis, the white-noise threshold for full separability is $\lambda^*=1-2/(1+\sqrt{1+s})$ with a local-unitary invariant $s\in(0,1]$, and $\lambda^*=3-2\sqrt2$ for the Shifts basis; a witness in closed form detects every entangled state of this family with eight local Pauli settings. For two groups of qubits we construct extremal states of minimal rank, and for three groups we determine the minimal rank of entanglement within the class in several configurations.
\end{abstract}

\maketitle

\section{Introduction}

Positivity under partial transposition (PPT) is the simplest necessary condition for separability \cite{Peres1996,Horodecki1996}, and states that are PPT but entangled cannot be distilled \cite{Horodecki1998}. Such bound entangled states are the main obstruction to a simple description of entanglement, and deciding separability is hard in general \cite{Gurvits2004NP,HorodeckiRMP2009}. Three basic questions about them are open even for a few qubits. Which bound entangled states are there? How robust is their entanglement to noise? How small can their rank be?

This paper studies these questions within a class of multiqubit states. Inside the class they read: which PPT entangled states belong to it, what are the exact noise thresholds of its states, and what is the smallest rank of an entangled state in it. Each of them becomes a question about a truncated trigonometric moment problem, and we answer them exactly in several cases. Consider $k$ groups of qubits, for instance $k$ atomic or spin ensembles, with $m_j$ qubits in group $j$, and the states that are mixtures of products of equatorial spin coherent states $\ket\theta^{\otimes m_j}$, $\ket\theta=(\ket0+e^{i\theta}\ket1)/\sqrt2$, with real weights that need not be positive. We call them Toeplitz states, because their matrix elements depend only on differences of Hamming weights and the coefficients form a multi-level Toeplitz matrix. Such a state is specified by the low-order Fourier moments of a phase quasi-distribution on the torus $\T^k$, in analogy with the Glauber--Sudarshan representation of optical fields \cite{Glauber1963,Sudarshan1963} and with the $P$-representation of spin states \cite{GiraudBraunBraun2008}. Two features make the class a natural laboratory for bound entanglement. Every Toeplitz state is PPT with respect to every bipartition, so all entanglement in the class is bound. And a Toeplitz state is fully separable exactly when its quasi-distribution can be replaced by a probability distribution, which is the truncated trigonometric moment problem in $k$ variables \cite{Rudin1963,Gabardo1998,BakonyiNaevdal2000,GeronimoWoerdeman2004,BakonyiWoerdeman2011}. Separability thus becomes a question about positive measures on a torus, with a dual description by nonnegative trigonometric polynomials, which yield entanglement witnesses with local equatorial measurements. For three qubits forming three groups, the Toeplitz states are exactly the states whose Pauli expansion contains no factor $Z$, and they are locally unitarily equivalent to the real states that are invariant under all partial transposes.

\emph{Which states?} For three qubits the PPT entangled states of the lowest possible rank, four, come in two types distinguished by a Lorentz invariant. Garberg, Irgens and Myrheim \cite{GIM2013} observed numerically that all their examples of the first type could be brought to a real form that is invariant under all partial transposes, adopted this as a working hypothesis, constructed states of this form and counted seven parameters, but a proof that their construction yields all such states was missing. Cheng and Chen \cite{ChengChen2026} gave an explicit form of the second type and a criterion for the states of the first type that arise from unextendible product bases (UPBs), and noted that a construction of all rank-four states was not known; explicit states of the first type that do not arise from UPBs were given by Ha and Kye \cite{HaKye2015} and by Cheng and Chen \cite[Example~16]{ChengChen2026}. We show that the Toeplitz class captures all states of the first type: after invertible local operators, each of them is a Toeplitz state. Among the states of the second type it captures exactly a one-parameter family.

\emph{How robust?} The UPB states are the standard examples of bound entanglement, yet the robustness to noise of the three-qubit UPB states was known only through the detection thresholds of particular witnesses. Wan and Guo \cite{WanGuo2026} proved that the projector witness of the Shifts state is optimal and asked whether another witness tolerates more white noise. Separability thresholds can be bounded numerically, for instance by symmetric extensions \cite{DPS2004}, Gilbert-type algorithms \cite{ShangGuhne2018} or adaptive polytopes \cite{Ohst2024}, but we are not aware of a previous value for the three-qubit UPB family. Exact thresholds of multipartite bound entangled states are available mainly for highly symmetric families such as GHZ-diagonal states \cite{Kay2011,Guhne2011,HanKye2017GHZ,HaHanKye2018}. Since the UPB states are Toeplitz states up to local unitaries, their thresholds become moment problems, which we solve exactly.

\emph{How small?} Bound entangled states of low rank are rare and rigid, and the smallest rank at which bound entanglement occurs is a basic invariant of a class of states. Within the Toeplitz class the question becomes: below which rank is every positive semidefinite Toeplitz matrix a moment matrix? By the duality between moment matrices and sums of squares, this is governed by the Hankel index of a product of projective lines, studied by Blekherman, Sinn and Velasco \cite{BSV2017}. We determine it in several configurations and construct extremal states that attain it.

\subsection{Main results}

\emph{The framework} (Theorems~\ref{thm:dict}, \ref{thm:cuts} and~\ref{thm:faces}). Toeplitz states are PPT across every bipartition, a one-line consequence of the absence of Pauli factors $Z$; they are fully separable exactly when the Toeplitz matrix is a moment matrix; and their entanglement can always be certified by local equatorial measurements. We determine which bipartitions can carry entanglement within the class, using the block Toeplitz separability theorem of Gurvits and Barnum \cite{Gurvits2001,Gurvits2003}, and show that across bipartitions splitting every group the minimal faces of the PPT cone at Toeplitz states are those of the Toeplitz cone. Table~\ref{tab:dictionary} summarises the correspondence. On this basis we prove the following three results, one for each question.

\begin{thmA}[Which states]
Let $\rho$ be a three-qubit PPT entangled state of rank four.
\begin{enumerate}
\item[(i)] If its Lorentz invariant is nonzero (type~I), then $\rho$ is equivalent under invertible local operators to a Toeplitz state. A cross-ratio, a normalised $4\times4$ Gram matrix and a sign, modulo an explicit finite group, form a complete set of invariants, and the moduli space has dimension seven.
\item[(ii)] If its Lorentz invariant vanishes (type~II), then $\rho$ is equivalent to a Toeplitz state if and only if the quadrilateral of unit vectors in its normal form, given in Proposition~\ref{prop:typeII}, is a rhombus.
\end{enumerate}
\end{thmA}

Part (i) consists of Theorems~\ref{thm:main3} and~\ref{thm:moduli} and Corollary~\ref{cor:invariants}, and part (ii) is Proposition~\ref{prop:typeIIT}. The proof of (i) is algebraic and independent of the moment problem. It proves the working hypothesis of \cite{GIM2013} and shows that the construction of \cite{GIM2013} yields all states of type~I; the sign distinguishes a state from its image under the exchange of two qubits, which is in general not locally equivalent to it.

\begin{thmB}[How robust]
For every three-qubit UPB state $\rho_{\rm UPB}$, the state $(1-\lambda)\rho_{\rm UPB}+\lambda I/8$ is fully separable if and only if
\[
\lambda\ge\lambda^*=1-\frac{2}{1+\sqrt{1+s}},
\]
where $s=|\sin2\theta_A\sin2\theta_B\sin2\theta_C|$ is computed from the angles of Bravyi's normal form \cite{Bravyi2004}. At $\lambda^*$ the state is a mixture of eight product states, and a witness given in closed form detects every entangled state of the family, is optimal in the sense of Lewenstein et al.\ \cite{LKCH2000} and is measured with the eight local Pauli settings $\{X,Z\}^{\otimes3}$. For the Shifts UPB, $\lambda^*=3-2\sqrt2$ and four settings suffice.
\end{thmB}

This is Theorem~\ref{thm:upb} and Corollary~\ref{cor:upb}, and it answers the question of Wan and Guo \cite{WanGuo2026}. In local Pauli settings a generic UPB needs one setting more than the standard projector witness \cite{Terhal2001}. Up to Pauli strings containing $Y$, our witness is the only one that detects the whole family, for the Shifts UPB and all UPBs close to it (Corollary~\ref{cor:unique}) and for all further UPBs we tested. As an illustration, in a sample of $159$ UPBs with angles drawn uniformly from $(0.02,\pi/2-0.02)$, the noise range detected by our witness exceeds that of the projector witness by a median factor of $6.7$. We give a measurement recipe for the Shifts state (Section~\ref{sec:recipe}) and compare with other rank-four states (Table~\ref{tab:thresholds}). UPB states are not the most robust: a GHZ-diagonal state of type~II exceeds all of them by a proof without computer, and Gabardo's state of type~I by a computer-assisted bound.

\begin{thmC}[How small]
\begin{enumerate}
\item[(i)] For two groups with $m_1,m_2\ge2$, every Toeplitz state of rank less than $2(m_1+m_2)-2$ is fully separable, and there is an integer Toeplitz matrix $T_\star$ of rank $2(m_1+m_2)-2$ that spans an extreme ray of the cone of positive semidefinite Toeplitz matrices and has no representing measure. The state $\rho_{T_\star}$ is an extremal PPT state whose range contains no product vector across any bipartition that splits both groups.
\item[(ii)] For three groups, the smallest rank of an entangled Toeplitz state is $4$ for $m=(1,1,1)$, $2a+2$ for $m=(a,1,1)$ with $a\ge2$, $6$ for $m=(2,2,1)$ and $m=(2,2,2)$, and $8$ for $m=(3,2,1)$, the last with computer assistance.
\end{enumerate}
\end{thmC}

These statements are Theorems~\ref{thm:twogroups} and~\ref{thm:rank}. The first statement of (i) is a result of Blekherman, Sinn and Velasco on toric surfaces \cite[Examples~6.16 and~6.18]{BSV2017}, transferred by Lemma~\ref{lem:cayley}, and the existence of a non-decomposable matrix of rank $2(m_1+m_2)-2$ can also be obtained from their upper bound for the Hankel index \cite[Theorem~4.1]{BSV2017}. New in (i) are the explicit integer extreme rays, their extremality in the PPT cones and the absence of product vectors in their ranges. In (ii), the value for $m=(1,1,1)$ follows from the separability of PPT states of rank at most three \cite{ChenDjokovic2013} and the UPB states of rank four; the other thresholds are new. As a consequence, positive semidefinite multi-level Toeplitz matrices below these ranks have Vandermonde decompositions with the minimal number of atoms.

Appendix~\ref{app:further} contains further examples for the three questions: separability after collective dephasing for extendible Toeplitz matrices and a $k$-qubit family of GHZ-diagonal states whose white-noise threshold $1-2^{1-k/2}$ tends to one, for the second question; a four-qubit state entangled across all balanced bipartitions and the Bell-nonlocal state of V\'ertesi and Brunner, for the first.

\subsection{Related work}

The separability of positive semidefinite block Toeplitz matrices goes back to Gurvits \cite{Gurvits2001} and Gurvits and Barnum \cite{Gurvits2003} and was developed by Farenick and coauthors \cite{Farenick2021,Farenick2022,Farenick2023}; there the Toeplitz structure concerns the blocks of a bipartite operator, whereas our Toeplitz states are multiqubit states whose matrix elements depend only on differences of Hamming weights. Faces of the two-qubit separable set were related to trigonometric moment curves by Kye \cite{Kye2013}. Multi-level Vandermonde decompositions of Toeplitz matrices under rank constraints appear in signal processing \cite{YangXieStoica2016,AnderssonCarlsson2017}. For a single group the connection between symmetric states, moment problems and Vandermonde decompositions was studied in \cite{BohnetWaldraff2017,QianChu2019,ChenChuQianShen2019,QianChenChu2020}, see also \cite{Eckert2002,TothGuhne2009,Augusiak2012}. Extremal PPT states and their faces were studied by Leinaas, Myrheim and coauthors \cite{LMO2007,LMS2010}, and rank-four PPT entangled states of two qutrits by Chen and {\DJ}okovi\'c \cite{ChenDjokovic2011}. For entanglement witnesses see \cite{GuhneToth2009}. Noise is a serious obstacle in experiments with bound entangled states; in an NMR experiment with a three-qubit bound entangled state of rank seven, for instance, the measured witness value was $-0.029\pm0.010$ against the ideal value $-0.11$ \cite{Kampermann2010}. Bell nonlocality from bound entanglement was established in \cite{Dur2001,VertesiBrunner2012,VertesiBrunner2014}, see also \cite{Moroder2014,Tendick2020}.

Section~\ref{sec:dict} introduces Toeplitz states and proves the correspondence with the moment problem. Sections~\ref{sec:three}, \ref{sec:noise} and~\ref{sec:structure} address the three questions in turn: which states belong to the class (rank-four states of three qubits), how robust their entanglement is (exact thresholds for UPB states), and how small their rank can be (extremal states and rank thresholds). Section~\ref{sec:discussion} lists open problems. Section~\ref{sec:three} uses from Section~\ref{sec:dict} only the description of Toeplitz states at the end of Section~\ref{sec:defs}; Section~\ref{sec:noise} uses Theorems~\ref{thm:dict} and~\ref{thm:cuts}(a) and, for the comparison with Section~\ref{sec:three}, Lemma~\ref{lem:pauli3}, Theorem~\ref{thm:moduli}(d) and Proposition~\ref{prop:typeIIT}; Section~\ref{sec:structure} uses all of Section~\ref{sec:dict}, and Theorem~\ref{thm:cuts}(b) in turn relies on Theorem~\ref{thm:twogroups}. The appendices follow the order of the sections: Appendix~\ref{app:proofs} contains the longer technical proofs of Sections~\ref{sec:three} and~\ref{sec:noise}, Appendix~\ref{app:rank} the proofs of Section~\ref{sec:structure} that use algebraic geometry, Appendix~\ref{app:further} further examples for the three questions, and Appendix~\ref{app:comp} the computations and the accompanying code.

\section{Toeplitz states and the moment problem}\label{sec:dict}

\subsection{Definitions}\label{sec:defs}

Let $m=(m_1,\dots,m_k)$ with $m_j\ge1$. We consider $n=m_1+\dots+m_k$ qubits divided into $k$ groups, group $j$ consisting of $m_j$ qubits. For a computational basis vector $\ket x$, $x\in\{0,1\}^n$, let $w(x)\in\Z^k$ be the vector of Hamming weights of $x$ restricted to the groups. Put
\[
\Lambda=\Lambda_m=\prod_{j=1}^k\{0,1,\dots,m_j\},\qquad D=\Lambda-\Lambda=\prod_{j=1}^k\{-m_j,\dots,m_j\}.
\]
A Hermitian matrix $T$ indexed by $\Lambda$ is a ($k$-level) Toeplitz matrix if $T_{\alpha,\beta}=t(\alpha-\beta)$ for a function $t:D\to\C$ with $t(-d)=\overline{t(d)}$. For $\theta\in\T^k=(\R/2\pi\Z)^k$ let $a(\theta)\in\C^\Lambda$ be the vector with entries $a(\theta)_\alpha=e^{i\langle\alpha,\theta\rangle}$. The matrix $a(\theta)a(\theta)^*$ is Toeplitz with $t(d)=e^{i\langle d,\theta\rangle}$. We write $\Toep^+_m$ for the cone of positive semidefinite (PSD) Toeplitz matrices and
\[
\Dec_m=\Bigl\{\sum_l p_l\,a(\theta_l)a(\theta_l)^*:\ p_l\ge0,\ \theta_l\in\T^k\Bigr\}\subseteq\Toep^+_m
\]
for the cone of decomposable matrices. A matrix in $\Toep^+_m$ belongs to $\Dec_m$ if and only if $t$ is the restriction to $D$ of the Fourier transform $d\mapsto\int e^{i\langle d,\theta\rangle}d\mu(\theta)$ of a positive measure $\mu$ on $\T^k$; this is the truncated trigonometric moment problem. The cone $\Dec_m$ is closed, since all its nonzero elements satisfy $t(0)>0$ and the atoms $a(\theta)a(\theta)^*$ form a compact set with $t(0)=1$.

\begin{definition}\label{def:toeplitz}
For a Toeplitz matrix $T$ indexed by $\Lambda_m$, let $\rho_T$ be the operator on $(\C^2)^{\otimes n}$ with matrix elements
\[
\bra x\rho_T\ket{x'}=t\bigl(w(x)-w(x')\bigr).
\]
If $T\ge0$ we call $\rho_T$ (normalised to unit trace when needed) a Toeplitz state.
\end{definition}

Let $\ket{E_\alpha}=\sum_{w(x)=\alpha}\ket x$, so that $\ket{E_\alpha}$ is an unnormalised product of Dicke states, and let $V:\C^\Lambda\to\bigotimes_j\Sym^{m_j}(\C^2)$ be the injective linear map with $Ve_\alpha=\ket{E_\alpha}$. Then
\begin{equation}\label{eq:lift}
\rho_T=VTV^*,\qquad \tr\rho_T=2^n\,t(0),\qquad Va(\theta)=2^{n/2}\bigotimes_{j=1}^k\ket{\theta_j}^{\otimes m_j},
\end{equation}
where $\ket\theta=(\ket0+e^{i\theta}\ket1)/\sqrt2$. In particular $\rho_T\ge0$ if and only if $T\ge0$, $\rank\rho_T=\rank T$, and $\rho_T$ is supported on the symmetric subspace of each group.

\begin{proposition}[Phase representation]\label{prop:Prep}
An operator $\rho$ on $(\C^2)^{\otimes n}$ is of the form $\rho_T$ for a Toeplitz matrix $T$ if and only if
\begin{equation}\label{eq:Prep}
\rho=\int_{\T^k}\bigotimes_{j=1}^k\proj{\theta_j}^{\otimes m_j}\,d\mu(\theta)
\end{equation}
for a finite real signed measure $\mu$. In this case $t(d)=2^{-n}\int e^{i\langle d,\theta\rangle}d\mu(\theta)$ for $d\in D$, and one may take $d\mu=2^nf_T(\theta)\,d\theta/(2\pi)^k$ with the real trigonometric polynomial $f_T(\theta)=\sum_{d\in D}t(d)e^{-i\langle d,\theta\rangle}$. The state $\rho_T$ admits a representation \eqref{eq:Prep} with a positive measure if and only if $T\in\Dec_m$.
\end{proposition}

\begin{proof}
By \eqref{eq:lift}, the right-hand side of \eqref{eq:Prep} equals $2^{-n}V\bigl(\int a(\theta)a(\theta)^*d\mu\bigr)V^*$, and $\int a(\theta)a(\theta)^*d\mu$ is the Toeplitz matrix with $t(d)=\int e^{i\langle d,\theta\rangle}d\mu$. Conversely $\int a(\theta)_\alpha\overline{a(\theta)_\beta}f_T(\theta)\,d\theta/(2\pi)^k=t(\alpha-\beta)$. For the last statement, $V$ is injective, and for a positive measure $\int a(\theta)a(\theta)^*d\mu$ lies in the closed convex cone generated by the atoms, which is $\Dec_m$; by Carath\'eodory's theorem a finite sum suffices.
\end{proof}

The coefficients $t(d)$ are, up to explicit factors, the moments of the collective ladder operators $J^{(j)}_\pm=\sum_{i\in\text{group }j}\sigma^{(i)}_\pm$, where $\sigma_+=\ket1\bra0$ and $\sigma_-=\ket0\bra1$. A direct count gives, for the normalised state $\rho=\rho_T/\tr\rho_T$,
\begin{equation}\label{eq:moments}
\Bigl\langle\prod_{j=1}^k J^{(j)}_{d_j}\Bigr\rangle_\rho=\frac{t(d)}{t(0)}\prod_{j=1}^k\frac{|d_j|!}{2^{|d_j|}}\binom{m_j}{|d_j|},\qquad J^{(j)}_{d_j}=\begin{cases}(J_-^{(j)})^{d_j},&d_j\ge0,\\(J_+^{(j)})^{-d_j},&d_j<0,\end{cases}
\end{equation}
and the joint distribution of the Hamming weights is the product of binomial distributions, as for the equatorial coherent states. Thus a Toeplitz state is completely specified by the phase moments \eqref{eq:moments}, and the question addressed in this paper is when these moments are moments of a genuine probability distribution of phases. The same structure appears when each group is regarded as a collective spin $J=m_j/2$ and $\ket{\theta}^{\otimes m_j}$ as a spin coherent state on the equator \cite{Arecchi1972}. For a single spin, states whose $P$-function can be chosen nonnegative are called classical \cite{GiraudBraunBraun2008}; Proposition~\ref{prop:Prep} is the analogue for products of equatorial coherent states, and Theorem~\ref{thm:dict}(b) below identifies it with full separability.

The Toeplitz description is compatible with merging groups. Let $\pi$ be a partition of $\{1,\dots,k\}$ into blocks $B_1,\dots,B_{k'}$ and let $\pi(\alpha)=(\sum_{j\in B_1}\alpha_j,\dots,\sum_{j\in B_{k'}}\alpha_j)$.

\begin{lemma}[Splitting and merging]\label{lem:merge}
Let $T'$ be a PSD Toeplitz matrix on the grid $\Lambda_{m'}$ with $m'_r=\sum_{j\in B_r}m_j$, and define $T$ on $\Lambda_m$ by $t(d)=t'(\pi(d))$. Then $T$ is a PSD Toeplitz matrix, $\rho_T=\rho_{T'}$ as operators on the same $n$ qubits, and $T\in\Dec_m$ if and only if $T'\in\Dec_{m'}$. In particular every Toeplitz state is a Toeplitz state for the finest partition, in which every qubit forms its own group.
\end{lemma}

\begin{proof}
Both matrix elements equal $t'(\pi(w(x))-\pi(w(x')))$. Positivity follows from $\rho_T=VTV^*$ with $V$ injective. The statement on decomposability follows from Theorem~\ref{thm:dict}(b) below, which identifies both conditions with full separability of the same operator.
\end{proof}

For the finest partition, $m=(1,\dots,1)$ and $\Lambda=\{0,1\}^n$. Writing $\sigma_\pm=(X\mp iY)/2$ one sees that the Toeplitz matrices on $\{0,1\}^n$ are exactly the Hermitian operators whose expansion in Pauli strings contains no factor $Z$. We use this description repeatedly.

\subsection{The dictionary}

Let $\mathcal E$ be the positive, but not completely positive, map on $2\times2$ matrices
\begin{equation}\label{eq:Phi}
\mathcal E(A)=\tfrac12\bigl(A+XA^{\tp}X\bigr)=\begin{pmatrix}\tfrac12(A_{00}+A_{11})&A_{01}\\A_{10}&\tfrac12(A_{00}+A_{11})\end{pmatrix}.
\end{equation}
$\mathcal E$ fixes $I,X,Y$ and annihilates $Z$, and it is self-adjoint with respect to the Hilbert--Schmidt inner product. A PSD $2\times2$ matrix with equal diagonal entries is a nonnegative combination of matrices $\proj\theta$: indeed $\begin{psmallmatrix}1&\bar z\\ z&1\end{psmallmatrix}$ with $|z|\le1$ is the average of $2\proj{\theta_1}$ and $2\proj{\theta_2}$ for $e^{i\theta_1}+e^{i\theta_2}=2z$. Hence $\mathcal E$ maps every state of a qubit to a mixture of equatorial states.

For a set $S$ of qubits write $\rho^{\Gamma_S}$ for the partial transpose on $S$ and $X_S$ for the product of the Pauli operators $X$ on the qubits in $S$.

\begin{theorem}\label{thm:dict}
Let $T\in\Toep^+_m$.
\begin{enumerate}
\item[(a)] For every set $S$ of qubits, $\rho_T^{\Gamma_S}=X_S\,\rho_T\,X_S$. In particular $\rho_T$ is PPT with respect to every bipartition, and every partial transpose has the same spectrum as $\rho_T$.
\item[(b)] $\rho_T$ is fully separable as a state of $n$ qubits if and only if $T\in\Dec_m$. In that case $\rho_T$ is a convex combination of the product states $\bigotimes_j\proj{\theta_j}^{\otimes m_j}$.
\item[(c)] If $T\notin\Dec_m$, there is an operator $W$ in the real span of Pauli strings with factors in $\{I,X,Y\}$ such that $\bra\psi W\ket\psi\ge0$ for every product vector $\psi$ of the $n$ qubits and $\tr(W\rho_T)<0$. In particular the entanglement of every Toeplitz state can be certified with local measurements in the equatorial plane of each qubit, using at most the $2^n$ settings $\{X,Y\}^{\otimes n}$.
\end{enumerate}
\end{theorem}

\begin{proof}
(a) Let $x,x'\in\{0,1\}^n$ and let $\bar x$ denote $x$ with the bits in $S$ flipped. Then $\bra x\rho_T^{\Gamma_S}\ket{x'}=t(w(x'')-w(x'''))$, where $x''$ agrees with $x$ outside $S$ and with $x'$ on $S$, and $x'''$ agrees with $x'$ outside $S$ and with $x$ on $S$. On the other hand $\bra xX_S\rho_TX_S\ket{x'}=t(w(\bar x)-w(\bar x'))$. Counting ones in each group, both differences equal $w(x|_{S^c})-w(x'|_{S^c})+w(x'|_S)-w(x|_S)$.

(b) If $T=\sum_lp_la(\theta_l)a(\theta_l)^*$ then $\rho_T=2^n\sum_lp_l\bigotimes_j\proj{\theta_{l,j}}^{\otimes m_j}$ by \eqref{eq:lift}. Conversely let $\rho_T=\sum_lp_l\proj{\psi_l}$ with product vectors $\psi_l=\bigotimes_{i=1}^nv_{l,i}$. Each $\psi_l$ lies in the range of $\rho_T$, hence in $\bigotimes_j\Sym^{m_j}(\C^2)$. If qubits $i,i'$ belong to the same group, the invariance of $\psi_l$ under their exchange gives $v_{l,i}\otimes v_{l,i'}=v_{l,i'}\otimes v_{l,i}$, so $v_{l,i}\parallel v_{l,i'}$. Hence $\psi_l=\bigotimes_ju_{l,j}^{\otimes m_j}$ with $u_{l,j}=(a_{l,j},b_{l,j})$.

Let $j$ be a group with $m_j\ge2$. All diagonal entries of $\rho_T$ equal $t(0)$. Fix the bits outside group $j$ to a string $r$ and let the bits in group $j$ have weight $w$. Then
\[
t(0)=\sum_lp_l\,c_l(r)\,|a_{l,j}|^{2(m_j-w)}|b_{l,j}|^{2w}\qquad(0\le w\le m_j),
\]
with $c_l(r)\ge0$ and $\sum_rc_l(r)>0$. Taking second differences in $w$ and summing over $r$ gives
\[
\sum_lp_l\Bigl(\sum_rc_l(r)\Bigr)|a_{l,j}|^{2(m_j-w-2)}|b_{l,j}|^{2w}\bigl(|a_{l,j}|^2-|b_{l,j}|^2\bigr)^2=0\qquad(0\le w\le m_j-2).
\]
All terms are nonnegative, and for $w=0$ and $w=m_j-2$ they force $|a_{l,j}|=|b_{l,j}|$ for every $l$. So each $u_{l,j}$ is a multiple of an equatorial vector.

Let $j$ be a group with $m_j=1$. The operator $\rho_T$ contains no Pauli factor $Z$ on this qubit, so it is fixed by $\mathcal E$ applied to this qubit. Applying $\mathcal E$ to the qubits of all groups with $m_j=1$ in the decomposition above, which is legitimate because $\mathcal E$ is positive and the terms are product operators, replaces each $\proj{u_{l,j}}$ by a mixture of equatorial states. Altogether $\rho_T$ is a nonnegative combination of the operators $\bigotimes_j\proj{\theta_j}^{\otimes m_j}=2^{-n}Va(\theta)a(\theta)^*V^*$, and the injectivity of $V$ gives $T\in\Dec_m$.

(c) By Lemma~\ref{lem:merge} we may assume that every qubit is its own group, so that $\Lambda=\{0,1\}^n$, $D=\{-1,0,1\}^n$ and $V$ is the identity. The cone $\Dec$ is closed and convex, so there is a real trigonometric polynomial $q(\theta)=\sum_{d\in D}q_de^{i\langle d,\theta\rangle}$, $q_{-d}=\overline{q_d}$, with $q\ge0$ on $\T^n$ and $\langle q,T\rangle:=\sum_dq_d\,t(d)<0$; indeed $\langle q,a(\theta)a(\theta)^*\rangle=q(\theta)$. Let $|d|$ be the number of nonzero entries of $d$ and put
\[
W=\sum_{d\in D}2^{|d|-n}q_d\bigotimes_{i=1}^n\Sigma_{d_i},\qquad \Sigma_0=I,\quad \Sigma_1=\sigma_-,\quad \Sigma_{-1}=\sigma_+ .
\]
Then $\bra\beta W\ket\alpha=2^{|\alpha-\beta|-n}q_{\alpha-\beta}$, so $\tr(W\rho_T)=\sum_\delta 2^{n-|\delta|}\cdot 2^{|\delta|-n}q_\delta t(\delta)=\langle q,T\rangle<0$, and for $\ket\theta=\bigotimes_i\ket{\theta_i}$ one obtains $\bra\theta W\ket\theta=2^{-n}q(\theta)\ge0$. For an arbitrary product state $\sigma=\bigotimes_i\sigma_i$, $\tr(W\sigma)=\tr(W\mathcal E^{\otimes n}(\sigma))$ because $W$ contains no $Z$, and $\mathcal E^{\otimes n}(\sigma)$ is a mixture of states $\proj\theta$. Hence $\tr(W\sigma)\ge0$. The operators $\Sigma_{\pm1}$ are combinations of $X$ and $Y$.
\end{proof}

Theorem~\ref{thm:dict} turns every PSD Toeplitz matrix without representing measure into a bound entangled state. Table~\ref{tab:dictionary} summarises the correspondence developed in this paper.

\begin{table}[t]
\centering
\small
\begin{tabular}{>{\raggedright\arraybackslash}p{0.34\textwidth}>{\raggedright\arraybackslash}p{0.36\textwidth}>{\raggedright\arraybackslash}p{0.2\textwidth}}
\toprule
quantum side & moment-problem side & reference\\
\midrule
Toeplitz state $\rho_T=VTV^*$ & multi-level Toeplitz matrix $T$, same rank & \eqref{eq:lift}\\
PPT across every bipartition & automatic & Theorem~\ref{thm:dict}(a)\\
fully separable & $T\in\Dec_m$: positive representing measure & Theorem~\ref{thm:dict}(b)\\
witness with equatorial settings & nonnegative $q$ with $\langle q,T\rangle<0$ & Theorem~\ref{thm:dict}(c)\\
bipartitions carrying entanglement & arrangement of the groups & Theorem~\ref{thm:cuts}\\
no semidefinite description & $\Dec_m$ not a spectrahedral shadow & Theorem~\ref{thm:classif}(b)\\
\midrule
\multicolumn{3}{l}{\emph{which states}}\\
three-qubit PPT entangled states of rank four, of type~I or of type~II with a rhombus, up to local operators & PSD Toeplitz matrices of rank four on $\{0,1\}^3$ without representing measure & Theorem~\ref{thm:main3}, Proposition~\ref{prop:typeIIT}\\
\midrule
\multicolumn{3}{l}{\emph{how robust}}\\
white-noise threshold & smallest $\lambda$ with $(1-\lambda)T+\lambda I\in\Dec_m$ & Section~\ref{sec:noise}\\
\midrule
\multicolumn{3}{l}{\emph{how small}}\\
extremal PPT states & extreme rays of $\Toep^+_m$ & Theorems~\ref{thm:faces}, \ref{thm:twogroups}\\
rank below which all states are separable & Hankel index, syzygies & Theorem~\ref{thm:rank}\\
\bottomrule
\end{tabular}
\caption{The correspondence between Toeplitz states and the truncated trigonometric moment problem.}
\label{tab:dictionary}
\end{table} By the Carath\'eodory--Toeplitz theorem, every PSD Toeplitz matrix is decomposable when there is a single group ($k=1$), which is also recovered in Theorem~\ref{thm:classif}; the relation between separability of symmetric states and Vandermonde decompositions was studied in \cite{QianChenChu2020,ChenChuQianShen2019}. For several groups Rudin \cite{Rudin1963} showed that positivity does not imply decomposability, and Gabardo \cite{Gabardo1998} gave explicit examples. Section~\ref{sec:structure} develops this systematically.

\subsection{Bipartitions}

Let $S\subset\{1,\dots,n\}$ be a nonempty proper set of qubits. We say that the bipartition $S|S^c$ splits group $j$ if group $j$ meets both $S$ and $S^c$.

\begin{lemma}[Product vectors]\label{lem:prodvec}
Let $\psi\in\bigotimes_j\Sym^{m_j}(\C^2)$ be a product vector with respect to $S|S^c$, and let $J$ be the set of groups split by $S|S^c$. Then
\[
\psi=\bigotimes_{j\in J}u_j^{\otimes m_j}\otimes\varphi\otimes\chi,
\]
where $\varphi$ is a vector on the qubits of the unsplit groups contained in $S$, $\chi$ a vector on the qubits of the unsplit groups contained in $S^c$, and both are symmetric within each group.
\end{lemma}

\begin{proof}
Write $\psi=\beta\otimes\gamma$ with $\beta$ on $S$ and $\gamma$ on $S^c$. Let $j\in J$, $i\in S$ and $i'\in S^c$ in group $j$, and expand $\beta=\sum_xe_x\otimes\beta_x$ and $\gamma=\sum_ye_y\otimes\gamma_y$ in the computational basis of qubits $i$ and $i'$. Invariance under the exchange of $i$ and $i'$ gives $\beta_x\otimes\gamma_y=\beta_y\otimes\gamma_x$ for $x,y\in\{0,1\}$. If $\beta_0,\beta_1$ were linearly independent, then $\beta_0\otimes\gamma_1=\beta_1\otimes\gamma_0$ would force $\gamma=0$. Hence $\beta=u\otimes\beta'$ with $u$ on qubit $i$, similarly $\gamma=v\otimes\gamma'$, and then $u\otimes v=v\otimes u$ gives $u\parallel v$. Since every qubit of group $j$ in $S$ can be paired with a qubit of group $j$ in $S^c$, all qubits of group $j$ factor out with a common vector $u_j$. Repeating this for every $j\in J$ gives the claim.
\end{proof}

The following lemma is due to Gurvits \cite{Gurvits2001} and Gurvits and Barnum \cite{Gurvits2003}. We include a short proof that also gives the number of terms.

\begin{lemma}[Block Toeplitz matrices]\label{lem:block}
Let $T=[B_{i-j}]_{i,j=0}^{N-1}$ be a PSD block Toeplitz matrix with blocks $B_d\in\C^{b\times b}$ and $r=\rank T$. Then $T=\sum_{l=1}^ru(\lambda_l)u(\lambda_l)^*\otimes P_l$ with $|\lambda_l|=1$, $u(\lambda)=(1,\lambda,\dots,\lambda^{N-1})$ and $P_l\ge0$ of rank one.
\end{lemma}

\begin{proof}
Write $T=YY^*$ with $Y=(Y_0;\dots;Y_{N-1})$, $Y_i\in\C^{b\times r}$. The Toeplitz property says that the stacked matrices $(Y_0;\dots;Y_{N-2})$ and $(Y_1;\dots;Y_{N-1})$ have the same Gram matrix of rows, so $(Y_1;\dots;Y_{N-1})=(Y_0;\dots;Y_{N-2})U$ for a unitary $U\in\mathrm U(r)$. Hence $Y_i=Y_0U^i$, and with $U=\sum_l\lambda_lv_lv_l^*$ we get $B_{i-j}=\sum_l\lambda_l^{i-j}(Y_0v_l)(Y_0v_l)^*$. Each $Y_0v_l$ is nonzero, since $Yv_l=u(\lambda_l)\otimes Y_0v_l$ and $Y$ has full column rank.
\end{proof}

\begin{theorem}[Bipartitions]\label{thm:cuts}
Let $S|S^c$ be a bipartition of the $n$ qubits.
\begin{enumerate}
\item[(a)] If there is a group $j_0$ such that all other groups are contained in $S$, or all other groups are contained in $S^c$, then $\rho_T$ is separable across $S|S^c$ for every $T\in\Toep^+_m$. For $k=1$ this hypothesis holds for every bipartition, since there are no other groups.
\item[(b)] Otherwise there is $T\in\Toep^+_m$ such that $\rho_T$ is entangled across $S|S^c$; by Theorem~\ref{thm:dict}(a) it is PPT across every bipartition. The proof of (b) uses the matrices of Theorem~\ref{thm:twogroups} in Section~\ref{sec:structure}, whose proof does not use (b).
\item[(c)] If each of $S$ and $S^c$ contains at most one unsplit group, and every unsplit group consists of a single qubit, then $\rho_T$ is separable across $S|S^c$ if and only if $T\in\Dec_m$. This applies in particular when every group is split.
\end{enumerate}
\end{theorem}

\begin{proof}
(a) Regard $T$ as a block Toeplitz matrix with respect to level $j_0$, with blocks indexed by the remaining levels. By Lemma~\ref{lem:block}, $T=\sum_lu(\lambda_l)u(\lambda_l)^*\otimes P_l$, and by \eqref{eq:lift} $\rho_T=\sum_l2^{m_{j_0}}\proj{\theta_l}^{\otimes m_{j_0}}\otimes V'P_lV'^*$, where $\lambda_l=e^{i\theta_l}$ and $V'$ is the lift for the remaining groups. The first factor is a product of single-qubit states and the second acts on one side of the bipartition.

(c) Let $\rho_T=\sum_lp_l\proj{\psi_l}$ with $\psi_l$ product across $S|S^c$. By Lemma~\ref{lem:prodvec} and the hypothesis, each $\psi_l$ is a product of single-qubit vectors that are equal within each group. The proof of Theorem~\ref{thm:dict}(b) now applies verbatim.

(b) Suppose (a) fails. We claim that the groups can be divided into two classes $C_1,C_2$ such that the qubits of each class meet both $S$ and $S^c$. If at least two groups are split, put one of them in $C_1$ and all other groups in $C_2$. If exactly one group $j$ is split, the failure of (a) for $j_0=j$ shows that the other groups lie on both sides; put $C_1=\{j\}$. If no group is split, the failure of (a) shows that each side contains at least two groups; put one group from each side into $C_1$. In each case the classes have $n_r=\sum_{j\in C_r}m_j\ge2$ qubits. By Theorem~\ref{thm:twogroups}(i) below there is $T'\in\Toep^+_{(n_1,n_2)}\setminus\Dec_{(n_1,n_2)}$. Let $T$ be obtained from $T'$ by Lemma~\ref{lem:merge}, so that $\rho_T=\rho_{T'}$. For the two-group description the bipartition splits both groups, so by (c) $\rho_{T'}$ is separable across $S|S^c$ only if $T'\in\Dec$, which is not the case.
\end{proof}

For two groups with $m_1,m_2\ge2$, Theorem~\ref{thm:cuts} says that a Toeplitz state is separable across every bipartition that leaves a group intact, and that across the bipartitions splitting both groups separability is equivalent to decomposability. For example, for $m=(2,2)$ five of the seven bipartitions of four qubits always give separable states, namely the four bipartitions of type $1|3$ and the bipartition between the groups, while the two bipartitions $A_1B_1|A_2B_2$ and $A_1B_2|A_2B_1$ are entangled exactly when $T\notin\Dec_{(2,2)}$. Patterns of bipartite separability of multiqubit mixed states were classified by D\"ur and Cirac \cite{DurCirac2000}; Theorem~\ref{thm:cuts} determines which bipartitions can carry entanglement in the Toeplitz class; which joint patterns occur is discussed in Section~\ref{sec:discussion}.

\subsection{Faces}

For a bipartition $S|S^c$ let $\mathcal P_S=\{\rho\ge0:\rho^{\Gamma_S}\ge0\}$. The minimal face of $\mathcal P_S$ containing $\rho$ spans the space of Hermitian $H$ with $\Rg(H)\subseteq\Rg(\rho)$ and $\Rg(H^{\Gamma_S})\subseteq\Rg(\rho^{\Gamma_S})$, since the minimal faces of the two PSD cones both contain $\rho$ in their relative interior. Similarly the minimal face of $\Toep^+_m$ containing $T$ spans the space of Hermitian Toeplitz $H$ with $\Rg(H)\subseteq\Rg(T)$.

\begin{theorem}[Faces]\label{thm:faces}
Let $S|S^c$ split every group and let $T\in\Toep^+_m$. A Hermitian operator $H$ satisfies $\Rg(H)\subseteq\Rg(\rho_T)$ and $\Rg(H^{\Gamma_S})\subseteq\Rg(\rho_T^{\Gamma_S})$ if and only if $H=\rho_h$ for a Hermitian Toeplitz matrix $h$ with $\Rg(h)\subseteq\Rg(T)$. Consequently the minimal face of $\mathcal P_S$ containing $\rho_T$ is the image of the minimal face of $\Toep^+_m$ containing $T$, and if $T$ spans an extreme ray of $\Toep^+_m$ then $\rho_T$ spans an extreme ray of $\mathcal P_S$ and of the cone of states that are PPT with respect to all bipartitions.
\end{theorem}

\begin{proof}
If $H=\rho_h$ with $\Rg(h)\subseteq\Rg(T)$, then $\Rg(H)\subseteq V\Rg(T)=\Rg(\rho_T)$ and $\Rg(H^{\Gamma_S})=X_S\Rg(H)\subseteq\Rg(\rho_T^{\Gamma_S})$ by Theorem~\ref{thm:dict}(a). Conversely, since $\Rg(H)\subseteq\bigotimes_j\Sym^{m_j}$, the entry $\bra xH\ket{x'}$ is invariant under permutations of the bits of $x$ within each group and under permutations of the bits of $x'$ within each group. Since $\Rg(\rho_T^{\Gamma_S})=X_S\Rg(\rho_T)$, the operator $H_S=X_SH^{\Gamma_S}X_S$ also has range in $\bigotimes_j\Sym^{m_j}$. In terms of $H$, $\bra rH_S\ket{r'}=\bra{\tilde r}H\ket{\tilde r'}$, where $\tilde r_i=r_i$, $\tilde r'_i=r'_i$ for $i\notin S$ and $\tilde r_i=1-r'_i$, $\tilde r'_i=1-r_i$ for $i\in S$. Fix a group $j$ and attach to it the $2m_j$ binary labels consisting of the bits of $x$ in group $j$ and the complemented bits $1-x'_i$ of $x'$ in group $j$. The symmetries of $H$ permute the first $m_j$ labels and the last $m_j$ labels, and the row symmetry of $H_S$ exchanges a label of the first kind at a qubit $i\notin S$ with a label of the second kind at a qubit $i'\in S$. Because the group is split, these permutations generate the full symmetric group on the $2m_j$ labels. Hence $\bra xH\ket{x'}$ depends only on the number of ones among the labels, which is $w_j(x)-w_j(x')+m_j$. Thus $\bra xH\ket{x'}=h(w(x)-w(x'))$ and $H=\rho_h=VhV^*$, and $\Rg(h)\subseteq\Rg(T)$ because $V$ is injective. The statements on faces follow from the description of the spans of minimal faces given before the theorem; an extreme ray of $\mathcal P_S$ that lies in the smaller cone is an extreme ray of the smaller cone.
\end{proof}

Theorem~\ref{thm:faces} concerns minimal faces only, and only bipartitions that split every group; for such bipartitions it reduces the facial structure of the PPT cone at Toeplitz states to that of the Toeplitz cone.

\section{Rank-four bound entanglement of three qubits}\label{sec:three}

This section addresses the first question of the introduction for three qubits: which bound entangled states belong to the Toeplitz class? The answer is that, up to invertible local operators, every PPT entangled state of rank four with nonzero Lorentz invariant does, and a one-parameter subfamily of the states with vanishing Lorentz invariant (Proposition~\ref{prop:typeIIT}). The argument is algebraic and does not use the moment problem.

In this section each of three qubits forms its own group, $m=(1,1,1)$. Let $\Toep_3$ denote the real space of Hermitian operators on $(\C^2)^{\otimes3}$ whose Pauli expansion contains no factor $Z$, which by Section~\ref{sec:dict} is the space of Toeplitz matrices on $\{0,1\}^3$, and let $\Hank_3$ denote the real symmetric operators that are invariant under each of the partial transposes $\Gamma_1,\Gamma_2,\Gamma_3$.

\begin{lemma}\label{lem:pauli3}
$\Hank_3$ is the real span of the Pauli strings with factors in $\{I,X,Z\}$. With $U_{\rm R}=e^{-i\pi X/4}$ one has $U_{\rm R}^{\otimes3}\Hank_3U_{\rm R}^{*\otimes3}=\Toep_3$.
\end{lemma}

\begin{proof}
An operator is invariant under $\Gamma_j$ if and only if no Pauli string in its expansion has $Y$ at position $j$, and a Hermitian operator with real Pauli coefficients and an even number of factors $Y$ in each string is real. Since $U_{\rm R}XU_{\rm R}^*=X$ and $U_{\rm R}ZU_{\rm R}^*=-Y$, the second statement follows.
\end{proof}

\begin{lemma}\label{lem:zfree}
Let $\rho$ be a Hermitian operator on three qubits and $V\in\GL(2,\C)$ act on qubit $j$. The operator $V\rho V^{*}$ has no Pauli string with $Z$ at position $j$ if and only if $\tr_j[(H\otimes I)\rho]=0$ for $H=V^{*}ZV$. The matrices of the form $V^{*}ZV$ are exactly the indefinite Hermitian $2\times2$ matrices.
\end{lemma}

\begin{proof}
The coefficients of the Pauli strings of $\sigma$ with $Z$ at position $j$ are the coefficients of the operator $\tr_j[(Z\otimes I)\sigma]$ on the remaining qubits, and $\tr_j[(Z\otimes I)V\rho V^{*}]=\tr_j[(V^{*}ZV\otimes I)\rho]$. The second statement is Sylvester's law of inertia.
\end{proof}

In this section a three-qubit state is called PPT if it is PPT with respect to each of the bipartitions $1\,|\,23$, $2\,|\,13$ and $3\,|\,12$, and entangled if it is not fully separable. A PPT state of rank four is separable across each of these bipartitions by the theorem of Kraus, Cirac, Karnas and Lewenstein on $2\times N$ systems \cite{KCKL2000}, so for the states considered here entanglement always refers to full separability. The lowest rank at which three-qubit PPT states can be entangled is four \cite{ChenDjokovic2013}. Garberg, Irgens and Myrheim \cite{GIM2013} studied these states numerically and divided them into two types according to a quadratic Lorentz invariant, so called because $\SL(2,\C)$ is the double cover of the proper orthochronous Lorentz group and acts on the Bloch vectors of each qubit by Lorentz transformations. They observed that their numerical examples of the first type could be transformed to real states symmetric under all partial transposes and adopted this as a working hypothesis \cite[Sec.~4.3]{GIM2013}. On this basis they constructed states of the first type from real vectors \cite[Sec.~4.3.1, Eqs.~(46)--(50)]{GIM2013}, counted seven parameters \cite[Sec.~4.3.2]{GIM2013}, and used an invariant ratio of determinants of the four points on the first qubit \cite[App.~C]{GIM2013}, which is equivalent to their cross-ratio. Cheng and Chen \cite{ChengChen2026} gave an explicit form of the second type and a criterion for the states of the first type that come from UPBs, and noted that a construction of all rank-four states was not known. Ha and Kye \cite{HaKye2015} constructed explicit rank-four PPT entangled states, including states that do not arise from UPBs, by moving along simplicial faces of the separable set, and Cheng and Chen gave states of type~I that do not arise from UPBs \cite[Example~16]{ChengChen2026}.

The main result of this section, Theorem~\ref{thm:main3}, proves the working hypothesis of \cite{GIM2013}. Consequently the construction of \cite{GIM2013}, which we restate in coordinate-free form in Theorem~\ref{thm:normal}(a), yields all states of the first type (Theorem~\ref{thm:normal}(b)). Theorem~\ref{thm:moduli} adds a Gram-matrix invariant and a sign which, together with the cross-ratio, form a complete set of invariants.

\begin{theorem}\label{thm:main3}
Let $\rho$ be a three-qubit PPT entangled state of rank four with nonzero Lorentz invariant. Then there are $V_1,V_2,V_3\in\SL(2,\C)$ such that $(V_1\otimes V_2\otimes V_3)\rho(V_1\otimes V_2\otimes V_3)^{*}$ is a positive semidefinite Toeplitz matrix of rank four on $\{0,1\}^3$. Equivalently, $\rho$ is $\SL(2,\C)^{3}$-equivalent to a real state that is invariant under each of the three partial transposes.
\end{theorem}

The proof rests on a determinant identity for the partial transpose of two-qubit operators (Lemma~\ref{lem:det}) and on the decomposition of Kraus, Cirac, Karnas and Lewenstein \cite{KCKL2000} in the form given by Chen and {\DJ}okovi\'c \cite{ChenDjokovic2013}. The invariants of Theorem~\ref{thm:moduli} are a cross-ratio and a normalised $4\times 4$ Gram matrix, seven parameters in total as counted in \cite{GIM2013}, together with a sign that distinguishes a state from its image under the exchange of qubits $2$ and $3$.

\subsection{Known structure}

We use the following results of Chen and {\DJ}okovi\'c, which build on \cite{KCKL2000}.

\begin{proposition}[\cite{ChenDjokovic2013}, Propositions 25 and 26]\label{prop:CD}
Let $\rho$ be a three-qubit PPT entangled state of rank four. Then $\rank\rho^{\Gamma_S}=4$ for every $S$, every two-qubit reduction of $\rho$ has rank four, and the range of $\rho$ contains no product vector $x\otimes y\otimes z$. Moreover
\begin{equation}\label{eq:kraus}
\rho=\sum_{i=1}^{4}X_i\otimes U_i,\qquad X_i=\proj{x_i},\quad U_i=\proj{u_i},
\end{equation}
where the $x_i\in\C^2$ are pairwise linearly independent, the $u_i\in\C^2\otimes\C^2$ are linearly independent and entangled, and $x_i\otimes u_i$, $i=1,\dots,4$, are the only vectors of the form $x\otimes u$ in the range of $\rho$. The kernel of $\rho$ contains exactly four vectors of the form $x\otimes u$, namely $x_i^{\perp}\otimes u_i'$, where $(u_i')$ is the basis reciprocal to $(u_i)$.
\end{proposition}

The statement on the vectors $x\otimes u$ is not formulated explicitly in \cite{ChenDjokovic2013}, but it follows from \eqref{eq:kraus}: the range of $\rho$ is spanned by the $x_i\otimes u_i$, and if $x\otimes u=\sum_ic_ix_i\otimes u_i$ and $y\perp x$, contraction with $y$ gives $\sum_ic_i\langle y,x_i\rangle u_i=0$, so $c_i=0$ unless $x_i$ is proportional to $x$, by the linear independence of the $u_i$ and the pairwise independence of the $x_i$. The same statements hold for the bipartitions $2\,|\,13$ and $3\,|\,12$, and for each partial transpose of $\rho$, which is again a PPT entangled state of rank four.

Let
\[
\eps=\begin{pmatrix}0&1\\-1&0\end{pmatrix},\qquad F=\eps\otimes\eps,\qquad E=\eps\otimes\eps\otimes\eps .
\]
For $u\in\C^2\otimes\C^2$ written as a $2\times2$ matrix, $u^{\tp}Fu=2\det u$, so $u$ is a product vector if and only if $u^{\tp}Fu=0$. The Lorentz invariant \cite{GIM2013,ChengChen2026} is
\[
Q(\rho)=-\tr\bigl(\rho^{\tp}E\,\rho\,E\bigr).
\]
It is invariant under $\rho\mapsto(V_1\otimes V_2\otimes V_3)\rho(V_1\otimes V_2\otimes V_3)^{*}$ with $V_j\in\SL(2,\C)$, because $V^{\tp}\eps V=\eps$ for $V\in\SL(2,\C)$. A three-qubit PPT entangled state of rank four is said to be of type~I if $Q\ne0$ and of type~II if $Q=0$. For a state of unit trace, $Q=-I_\rho$ for the invariant $I_\rho=\tr(\rho^{\tp}\eps^{\otimes3}\rho\,\eps^{\otimes3})$ of \cite[Def.~12]{ChengChen2026}; for instance $Q=3/16$ for the Shifts state, for which \cite[Lemma~15]{ChengChen2026} gives $I_\rho=-3/16$. Only the distinction between $Q=0$ and $Q\ne0$, which does not depend on the normalisation, is used in the proofs.

\begin{lemma}\label{lem:Q}
For $\rho$ as in \eqref{eq:kraus} let $G$ be the complex symmetric matrix $G_{ij}=u_i^{\tp}Fu_j$. Then
\[
Q(\rho)=\sum_{i\ne j}|x_i^{\tp}\eps x_j|^{2}\,|G_{ij}|^{2}.
\]
Consequently $Q(\rho)=0$ if and only if $G$ is diagonal.
\end{lemma}

\begin{proof}
We have $\tr(\rho^{\tp}E\rho E)=\sum_{i,j}\tr(X_i^{\tp}\eps X_j\eps)\tr(U_i^{\tp}FU_jF)$, and
\begin{align*}
\tr(X_i^{\tp}\eps X_j\eps)&=(x_i^{\tp}\eps x_j)(x_j^{*}\eps\bar x_i)=-|x_i^{\tp}\eps x_j|^2,\\
\tr(U_i^{\tp}FU_jF)&=(u_i^{\tp}Fu_j)(u_j^{*}F\bar u_i)=|G_{ij}|^2 .
\end{align*}
The last statement holds because the $x_i$ are pairwise independent.
\end{proof}

\subsection{A determinant identity}

For a Hermitian operator $M$ on two qubits let $M^{\Gamma}$ be its partial transpose on the first qubit and $\widetilde M=F\,\overline M\,F$.

\begin{lemma}\label{lem:det}
For every Hermitian operator $M$ on $\C^2\otimes\C^2$,
\begin{equation}\label{eq:det}
16\det M^{\Gamma}=8\det M-2\tr\bigl[(M\widetilde M)^2\bigr]+\bigl(\tr M\widetilde M\bigr)^2 .
\end{equation}
\end{lemma}

The identity can be verified by direct expansion; a conceptual proof is given in Appendix~\ref{app:proofs}.

\subsection{The key lemma}

\begin{lemma}\label{lem:key}
Let $\rho$ be a three-qubit PPT entangled state of rank four with decomposition \eqref{eq:kraus}. If $X_1,\dots,X_4$ are linearly independent, then $G$ is diagonal.
\end{lemma}

The proof is given in Appendix~\ref{app:proofs}. It applies Lemma~\ref{lem:det} to the operators $M(a)=\sum_ia_iU_i$, $a\in\R^4$, whose partial transposes are diagonal in the basis provided by Proposition~\ref{prop:CD} for $\rho^{\Gamma_2}$. This gives an identity between a quartic form built from $G$ and a product of four real linear forms in $a$, and comparing the real roots of both sides on the coordinate planes forces $G_{ij}=0$ for $i\ne j$.

\subsection{Proof of Theorem~\ref{thm:main3}}

Let $\rho$ be of type I. By Lemma~\ref{lem:Q} the matrix $G$ is not diagonal, so by Lemma~\ref{lem:key} the matrices $X_1,\dots,X_4$ are linearly dependent. Hence there is a nonzero Hermitian $H_1$ with $\bra{x_i}H_1\ket{x_i}=0$ for all $i$. If $H_1$ were semidefinite, all $x_i$ would lie in its kernel, which is impossible because the $x_i$ are pairwise independent. So $H_1$ is indefinite and $\tr_1[(H_1\otimes I)\rho]=\sum_i\bra{x_i}H_1\ket{x_i}U_i=0$. By Lemma~\ref{lem:zfree} there is $V_1$ such that $V_1\rho V_1^{*}$ has no $Z$ on qubit~$1$. The invariant $Q$ is symmetric in the three qubits. For $j=2,3$, Proposition~\ref{prop:CD} applied to the bipartition $j\,|\,\text{rest}$ gives a decomposition $\rho=\sum_iX_i^{(j)}\otimes U_i^{(j)}$ with $X_i^{(j)}$ acting on qubit $j$, and Lemmas~\ref{lem:Q} and~\ref{lem:key}, applied with qubit $j$ in the role of qubit $1$ and with $\rho^{\Gamma_k}$ for a qubit $k\ne j$ in the role of $\rho^{\Gamma_2}$, show that the $X_i^{(j)}$ are linearly dependent; the hypotheses of these lemmas are symmetric under relabelling the qubits, because Proposition~\ref{prop:CD} holds for every bipartition and every partial transpose of $\rho$. The argument above then gives $V_2$ and $V_3$. Operators on different qubits do not interfere, so $(V_1\otimes V_2\otimes V_3)\rho(V_1\otimes V_2\otimes V_3)^{*}$ has no $Z$ at any position and belongs to $\Toep_3$ by Lemma~\ref{lem:pauli3}. Rescaling gives $V_j\in\SL(2,\C)$. The second formulation follows from $U_{\rm R}^{\otimes3}\Hank_3 U_{\rm R}^{*\otimes3}=\Toep_3$. \qed

\subsection{Construction and normal form}

Four pairwise independent vectors $x_i\in\R^2$ satisfy a linear relation
\begin{equation}\label{eq:c}
\sum_{i=1}^4c_i\,x_ix_i^{\tp}=0 ,
\end{equation}
unique up to scale, because any three of the matrices $x_ix_i^{\tp}$ are linearly independent in the three-dimensional space of real symmetric $2\times2$ matrices. All $c_i$ are nonzero and their signs alternate along the cyclic order of the points $[x_i]$ on the real projective line.

\begin{theorem}[part (a): \cite{GIM2013}, Sec.~4.3.1]\label{thm:normal}
\begin{enumerate}
\item[(a)] Let $x_1,\dots,x_4\in\R^2$ be pairwise independent with relation \eqref{eq:c}, let $\gamma\ne0$, and let $u_1,\dots,u_4\in\R^4$ be linearly independent with
\begin{equation}\label{eq:diagG}
u_i^{\tp}Fu_i=\gamma\,c_i\qquad(i=1,\dots,4).
\end{equation}
Then $\rho=\sum_ix_ix_i^{\tp}\otimes u_iu_i^{\tp}$ belongs to $\Hank_3$, has rank four, is PPT and is entangled. It is of type~I if and only if $G=(u_i^{\tp}Fu_j)$ is not diagonal.
\item[(b)] Every three-qubit PPT entangled state of rank four and type~I is $\SL(2,\C)^3$-equivalent, up to a positive factor, to a state of the form in (a).
\end{enumerate}
\end{theorem}

\begin{proof}
For real $u\in\R^2\otimes\R^2$ one checks that
\begin{equation}\label{eq:ptreal}
(uu^{\tp})^{\Gamma}-uu^{\tp}=\tfrac12\,(u^{\tp}Fu)\,K_0,\qquad K_0=\ket{01}\bra{10}+\ket{10}\bra{01}-\ket{00}\bra{11}-\ket{11}\bra{00},
\end{equation}
and the same holds for the partial transpose on the other factor.

(a) $\rho$ is real and invariant under $\Gamma_1$. By \eqref{eq:ptreal}, \eqref{eq:diagG} and \eqref{eq:c}, $\rho^{\Gamma_2}-\rho=\tfrac\gamma2\sum_ic_ix_ix_i^{\tp}\otimes K_0=0$, and likewise for $\Gamma_3$. So $\rho\in\Hank_3$ and $\rho$ is PPT. The vectors $x_i\otimes u_i$ are linearly independent, so the rank is four. If $\sum_i\mu_ix_i\otimes u_i=y\otimes w$, contracting with a vector orthogonal to $y$ shows that $\mu_i\ne0$ only if $x_i$ is proportional to $y$, which happens for at most one $i$. Hence the only vectors $y\otimes w$ in the range are the $x_i\otimes u_i$, and none of the $u_i$ is a product vector because $u_i^{\tp}Fu_i=\gamma c_i\ne0$. So the range contains no product vector $x\otimes y\otimes z$ and $\rho$ is entangled. The statement on the type is Lemma~\ref{lem:Q}.

(b) By Theorem~\ref{thm:main3} we may assume $\rho\in\Hank_3$. Let \eqref{eq:kraus} be its decomposition. Since $\rho=\rho^{\Gamma_1}=\sum_i\overline{X_i}\otimes U_i$, the vectors $\bar x_i\otimes u_i$ lie in the range, and the uniqueness in Proposition~\ref{prop:CD} gives $\bar x_i\propto x_i$. So we may take $x_i$ real. Since $\rho=\bar\rho=\sum_iX_i\otimes\overline{U_i}$, uniqueness gives $\bar u_i\propto u_i$, so we may take $u_i$ real. Now $0=\rho^{\Gamma_2}-\rho=\tfrac12\sum_i(u_i^{\tp}Fu_i)\,x_ix_i^{\tp}\otimes K_0$ and the uniqueness of \eqref{eq:c} give \eqref{eq:diagG} with some $\gamma$, and $\gamma\ne0$ because the $u_i$ are entangled.
\end{proof}

Part (a) is a coordinate-free form of the construction of \cite[Sec.~4.3.1]{GIM2013}, where the $x_i$ are fixed to a standard form and the conditions \eqref{eq:diagG} appear as Eqs.~(46)--(50). Part (b), which rests on Theorem~\ref{thm:main3}, is new: together they answer the completeness question of \cite{ChengChen2026} for the first type. Every such state is obtained, up to local filtering, by choosing four points on a line and four real vectors on prescribed level sets of the quadratic form $u\mapsto u^{\tp}Fu=2\det u$ on $\R^{2\times2}$.

\begin{proposition}\label{prop:typeII}
A three-qubit PPT entangled state of rank four is of type~II if and only if it is $\SL(2,\C)^3$-equivalent, up to a positive factor, to
\begin{equation}\label{eq:typeII}
\rho=\sum_{i=1}^4(I+n_i\cdot\vec\sigma)\otimes\proj{\Phi_i},
\end{equation}
where $\Phi_1,\dots,\Phi_4$ is the Bell basis and $n_1,\dots,n_4\in\R^3$ are pairwise distinct unit vectors with $n_1+n_2+n_3+n_4=0$.
\end{proposition}

\begin{proof}
Let $\rho$ be of type II, so $G$ is diagonal, and scale the $u_i$ so that $G=I$. Let $\Phi$ be the matrix of a basis of maximally entangled vectors with $\Phi^{\tp}F\Phi=I$. Then $O=\tilde U\Phi^{-1}$ satisfies $O^{\tp}FO=F$. After replacing $u_1$ by $-u_1$ if necessary, $O$ lies in the identity component of the complex orthogonal group of $F$, which is $\{A\otimes B:A,B\in\SL(2,\C)\}$. Applying $A^{-1}\otimes B^{-1}$ to qubits $2,3$ we get $\rho=\sum_iX_i\otimes B_i$ with $B_i=\proj{\Phi_i}$. Now $B_i^{\Gamma}=\tfrac12I-B_{\pi(i)}$ for a permutation $\pi$, so $\rho^{\Gamma_2}=\sum_k(\tfrac12S_0-X_{\pi^{-1}(k)})\otimes B_k$ with $S_0=\sum_iX_i$, which has rank two. Since $\rho^{\Gamma_2}\ge0$ has rank four, each block $\tfrac12S_0-X_i$ is positive semidefinite of rank one. Filtering qubit $1$ so that $S_0=2I$, each $X_i$ is a projector $\tfrac12(I+n_i\cdot\vec\sigma)$, and $\sum_in_i=0$. The $n_i$ are pairwise distinct because the $x_i$ are pairwise linearly independent (Proposition~\ref{prop:CD}). Conversely \eqref{eq:typeII} is PPT, has $G$ diagonal, and for pairwise distinct $n_i$ its range contains only the vectors $x_i\otimes\Phi_i$ of the form $x\otimes u$, so it is entangled.
\end{proof}

Garberg, Irgens and Myrheim constructed a family of type~II states described by one complex parameter \cite[Sec.~4.4]{GIM2013}, and an explicit form of all type~II states with one complex parameter was obtained in \cite{ChengChen2026}; \eqref{eq:typeII} is a geometric version: a closed quadrilateral with unit sides in $\R^3$, up to rotation. In the proof of Proposition~\ref{prop:typeII}, $B_i^{\Gamma}=\tfrac12I-B_{\pi(i)}$, where $\pi$ exchanges $\Phi^+$ with $\Psi^-$ and $\Phi^-$ with $\Psi^+$; this holds for the partial transpose on either qubit, because the Bell projectors are real symmetric. The next result determines which type~II states belong to the Toeplitz class.

\begin{proposition}\label{prop:typeIIT}
For a three-qubit PPT entangled state $\rho$ of rank four and type~II the following are equivalent.
\begin{enumerate}
\item[(a)] $\rho$ is $\SL(2,\C)^3$-equivalent to a Toeplitz state.
\item[(b)] Up to a positive factor, $\rho$ is $\SL(2,\C)^3$-equivalent to a state of the form in Theorem~\ref{thm:normal}(a) with $G$ diagonal.
\item[(c)] In a representation \eqref{eq:typeII} of $\rho$, the vectors $n_1,\dots,n_4$ are $n,-n,m,-m$ with $m\ne\pm n$; that is, the quadrilateral is a rhombus.
\end{enumerate}
\end{proposition}

\begin{proof}
(a)$\Leftrightarrow$(b). By Lemma~\ref{lem:pauli3}, (a) means that $\rho$ is equivalent to a state in $\Hank_3$. The proof of Theorem~\ref{thm:normal}(b) after its first sentence does not use the type, so every PPT entangled state of rank four in $\Hank_3$ has the form in Theorem~\ref{thm:normal}(a), with $G$ diagonal by Lemma~\ref{lem:Q}. Conversely the states in Theorem~\ref{thm:normal}(a) belong to $\Hank_3$.

(b)$\Rightarrow$(c). An invertible local operator maps the vectors of the form $x\otimes u$ in the range of a state to those of its image, so by the uniqueness in Proposition~\ref{prop:CD} the first factors $I+n_i\cdot\vec\sigma$ in \eqref{eq:typeII} are, up to order and positive factors, the images $Vx_ix_i^{\tp}V^*$ of the first factors of a state as in Theorem~\ref{thm:normal}(a). The $x_ix_i^{\tp}$ lie in the three-dimensional space of real symmetric matrices, so the $I+n_i\cdot\vec\sigma$ are linearly dependent, and there is a nonzero Hermitian $H=h_0I+h\cdot\vec\sigma$ with $\tr\bigl(H(I+n_i\cdot\vec\sigma)\bigr)=2(h_0+h\cdot n_i)=0$ for all $i$. Summing over $i$ and using $\sum_in_i=0$ gives $h_0=0$, so the $n_i$ lie in the plane orthogonal to $h\ne0$. Four pairwise distinct unit vectors in a plane with $n_1+n_2=-(n_3+n_4)=w$ consist of two antipodal pairs: if $w=0$ this is clear, and otherwise $|w|<2$ and the unordered pair of unit vectors in the plane with sum $w$ is unique, so $\{n_3,n_4\}=\{-n_1,-n_2\}$.

(c)$\Rightarrow$(a). A unitary on qubit~$1$ moves $n$ and $m$ into the $XZ$ plane, so that the matrices $I+n_i\cdot\vec\sigma$ are real; the Bell vectors are real as well. The local unitaries $V_{\rm H}\otimes V_{\rm H}$ and $V_{\rm P}\otimes V_{\rm P}$ on qubits $2$ and $3$, with the Hadamard matrix $V_{\rm H}$ and the phase gate $V_{\rm P}=\diag(1,i)$, permute the Bell projectors: the first exchanges those of $\Phi^-$ and $\Psi^+$, the second those of $\Phi^+$ and $\Phi^-$. They act transitively on the three partitions of the Bell basis into two pairs, so after relabelling we may assume $n_{\pi(k)}=-n_k$ for all $k$. With $X_i=I+n_i\cdot\vec\sigma$ and $\sum_iX_i=4I$, the computation in the proof of Proposition~\ref{prop:typeII} gives $\rho^{\Gamma_2}=\sum_k(2I-X_{\pi(k)})\otimes B_k=\sum_k(I-n_{\pi(k)}\cdot\vec\sigma)\otimes B_k=\rho$, and likewise $\rho^{\Gamma_3}=\rho$. Since $\rho$ is real and its first factors are real symmetric, also $\rho^{\Gamma_1}=\rho$, so $\rho\in\Hank_3$, and Lemma~\ref{lem:pauli3} gives (a).
\end{proof}

Up to rotations, the vectors in (c) depend on one real parameter, the angle $\varphi$ between $n$ and $m$, whereas the quadrilaterals in \eqref{eq:typeII} depend on two. The angle is an invariant up to a finite ambiguity: the points $[x_i]$ are determined by $\rho$ (Proposition~\ref{prop:CD}), invertible operators on qubit~$1$ act on them by M\"obius transformations, and for the order $n,-n,m,-m$ their cross-ratio is $-\tan^2(\varphi/2)$, as one sees by placing $n$ at the north pole. So these states form a genuinely one-parameter family up to equivalence. The state of Proposition~\ref{prop:mabk} with $k=3$ is a Toeplitz state of rank four and type~II, so it belongs to this family. Together with Theorem~\ref{thm:main3}, Proposition~\ref{prop:typeIIT} determines which three-qubit PPT entangled states of rank four belong to the Toeplitz class up to invertible local operators: all states of type~I, and the states of type~II whose quadrilateral is a rhombus.

\subsection{Moduli}

Let $\rho$ be as in Theorem~\ref{thm:normal}(a) with $G$ not diagonal, and fix a labelling of its decomposition into the terms $x_ix_i^{\tp}\otimes u_iu_i^{\tp}$, $i=1,\dots,4$. Let $\mathrm{cr}$ be the cross-ratio of the points $[x_1],\dots,[x_4]$, let $\tilde U=[u_1,\dots,u_4]$ be the real $4\times4$ matrix whose columns are the $u_i$ in the basis $\ket{00},\ket{01},\ket{10},\ket{11}$, let $o=\operatorname{sign}\det\tilde U$, and let
\[
\widehat G_{ij}=\frac{G_{ij}}{\sqrt{|G_{ii}G_{jj}|}} .
\]
On such triples we let the finite group of triples $(\pi,\nu,\kappa)$, with $\pi$ a permutation of $\{1,2,3,4\}$, $\nu\in\{\pm1\}^4$ and $\kappa=\pm1$, act by
\[
(\mathrm{cr},\widehat G,o)\mapsto\Bigl(\pi\cdot\mathrm{cr},\ \bigl(\kappa\,\nu_i\nu_j\widehat G_{\pi(i)\pi(j)}\bigr)_{i,j},\ \operatorname{sgn}(\pi)\,\nu_1\nu_2\nu_3\nu_4\,o\Bigr),
\]
where $\pi\cdot\mathrm{cr}$ is the cross-ratio of $[x_{\pi(1)}],\dots,[x_{\pi(4)}]$, given by the corresponding anharmonic map. The three kinds of generators correspond to relabelling the terms, to changing the signs of the $u_i$, and to an operator with negative determinant on qubit $2$.

\begin{theorem}\label{thm:moduli}
\begin{enumerate}
\item[(a)] Two PPT entangled states $\rho,\rho'\in\Hank_3$ of rank four and type I are $\SL(2,\C)^3$-equivalent up to a positive factor if and only if $\rho'=(W_1\otimes W_2\otimes W_3)\rho(W_1\otimes W_2\otimes W_3)^{\tp}$ up to a positive factor with $W_j\in\GL(2,\R)$.
\item[(b)] The orbit of $(\mathrm{cr},\widehat G,o)$ under the finite group does not depend on the choice of the labelled decomposition and is invariant under $\GL(2,\R)^3$ and positive factors. Conversely, two such states whose triples lie in the same orbit are $\SL(2,\C)^3$-equivalent up to a positive factor. The exchange of qubits $2$ and $3$ fixes $\mathrm{cr}$ and $\widehat G$ and changes $o$ into $-o$.
\item[(c)] The admissible pairs $(\mathrm{cr},\widehat G)$ are: $\mathrm{cr}\in\R\setminus\{0,1\}$, and $\widehat G$ real symmetric of signature $(2,2)$, not diagonal, with diagonal entries $\pm1$ that alternate in sign along the cyclic order of the points $[x_1],\dots,[x_4]$ on the real projective line, in the labelling fixed above; both values of $o$ occur for each admissible pair. In particular the moduli space has dimension seven.
\item[(d)] The number of product vectors $x\otimes y\otimes z$ in the kernel of $\rho$ equals the number of zeros on the diagonal of $\widehat G^{-1}$.
\item[(e)] If the six off-diagonal entries of $\widehat G$ are nonzero and have pairwise distinct absolute values, then $\rho$ is not $\SL(2,\C)^3$-equivalent, up to a positive factor, to its image under the exchange of qubits $2$ and $3$.
\end{enumerate}
\end{theorem}

\begin{proof}
(a) Let $\rho'=(V_1\otimes V_2\otimes V_3)\rho(V_1\otimes V_2\otimes V_3)^{*}$. The $X_i$ span the real symmetric matrices, so the Hermitian matrices $H$ with $\tr_1[(H\otimes I)\rho]=0$ are the multiples of $Y$, and the same holds for $\rho'$. Since $\tr_1[(Y\otimes I)\rho']=0$ is equivalent to $\tr_1[(V_1^{*}YV_1\otimes I)\rho]=0$, we get $V_1^{*}YV_1=cY$ with real $c\ne0$. The matrices $V$ with $V^*YV=cY$ for some real $c\ne0$ are exactly the elements of $\C^{\times}\GL(2,\R)$. Indeed $W^{\tp}YW=(\det W)Y$ for real $W$; conversely, after rescaling $V$ and composing it with a real matrix of determinant $-1$ if necessary, we may assume $V^*YV=Y$ and $\det V=1$, and then $V^*\eps V=\eps=V^{\tp}\eps V$ because $\eps=iY$ and $\det V=1$, which forces $\overline V=V$. Hence $V_1\in\C^{\times}\GL(2,\R)$. The same applies to qubits $2$ and $3$.

(b) A real operator on qubit $1$ acts projectively on the $[x_i]$ and preserves $\mathrm{cr}$. Operators $A\otimes B$ with $A,B\in\GL(2,\R)$ on qubits $2,3$ replace $\tilde U$ by $(A\otimes B)\tilde U$ and $G$ by $(\det A\det B)\,G$; since $\det(A\otimes B)=(\det A)^2(\det B)^2>0$ they preserve $o$. The decomposition \eqref{eq:kraus} is unique up to the order of the terms and rescalings $x_i\mapsto\mu_ix_i$, $u_i\mapsto\nu_i|\mu_i|^{-1}u_i$ with $\mu_i\ne0$ and $\nu_i=\pm1$. These act on $G$ by a diagonal congruence and on $o$ by $\prod_i\nu_i$, and a permutation of the terms permutes the columns of $\tilde U$. This gives the invariance. Conversely, let $\rho,\rho'$ have triples in the same orbit. After relabelling and sign changes of the $u_i'$, and after an operator with negative determinant on qubit $2$ if the orbit element involves $\widehat G\mapsto-\widehat G$ (it replaces $G'$ by a negative multiple and preserves $o$), we may assume that the triples are equal. A real projective map on qubit $1$ and rescalings $x_i'\mapsto\mu_ix_i'$, compensated by $u_i'\mapsto|\mu_i|^{-1}u_i'$, bring $x_i'$ to $\pm x_i$ without changing $o$; then $G'_{ii}=\gamma'c_i$ and $G_{ii}=\gamma c_i$ with the same $c_i$. Since $\widehat G'=\widehat G$, the diagonal signs agree, so $\gamma'/\gamma>0$, $G'=(\gamma'/\gamma)G$, and after a positive factor $G'=G$. Then $\tilde U'=O\tilde U$ with $O^{\tp}FO=F$ and $\det O=\det\tilde U'/\det\tilde U>0$. In the identification $\R^4=\R^{2\times2}$ the form $u^{\tp}Fu$ is twice the determinant. Its real orthogonal group consists of the maps $m\mapsto AmB^{\tp}$ with $\det A\det B=1$, which have determinant $1$ on $\R^4$, and of their compositions with the transposition $m\mapsto m^{\tp}$, which has determinant $-1$ and is the exchange of qubits $2$ and $3$. Hence $O=A\otimes B$, and $\rho'$ is obtained from $\rho$ by a real local operator. The exchange of qubits $2$ and $3$ replaces each $u_i$ by its transpose, which preserves $G$ because $F$ is invariant under the exchange, and changes the sign of $\det\tilde U$.

(c) $G=\tilde U^{\tp}F\tilde U$ has the signature $(2,2)$ of $F$, and $G_{ii}=\gamma c_i$ alternates in sign. Conversely, given admissible data choose $x_i$ with cross-ratio $\mathrm{cr}$, put $\Delta=\diag(\sqrt{|c_1|},\dots,\sqrt{|c_4|})$ and define $G=\Delta\widehat G\Delta$. By Sylvester's law of inertia every real symmetric matrix of signature $(2,2)$ can be written as $G=\tilde U^{\tp}F\tilde U$ with real invertible $\tilde U$, and replacing $\tilde U$ by the composition with the transposition changes $o$; Theorem~\ref{thm:normal}(a) applies. The matrix $\widehat G$ has six free entries.

(d) By Proposition~\ref{prop:CD} the vectors of the form $x\otimes u$ in the kernel are $x_i^{\perp}\otimes u_i'$, where $u_i'$ is the $i$th column of $\tilde U^{-\tp}$. Since $F^{-1}=F$, the Gram matrix of the $u_i'$ is $\tilde U^{-1}F\tilde U^{-\tp}=G^{-1}$, so $u_i'$ is a product vector if and only if $(G^{-1})_{ii}=0$.

(e) Suppose that $\rho$ is equivalent to its image under the exchange. By (b) the triples $(\mathrm{cr},\widehat G,o)$ and $(\mathrm{cr},\widehat G,-o)$ lie in the same orbit, so there are a permutation $\pi$, signs $\nu_i$ and $\kappa=\pm1$ with $\kappa\nu_i\nu_j\widehat G_{\pi(i)\pi(j)}=\widehat G_{ij}$ for all $i,j$ and $\operatorname{sgn}(\pi)\prod_i\nu_i=-1$. Comparing absolute values of the off-diagonal entries shows that $\pi$ maps every pair $\{i,j\}$ to itself, so $\pi$ is the identity. The diagonal entries give $\kappa=1$, and the nonzero off-diagonal entries give $\nu_i\nu_j=1$ for all $i\ne j$. Hence all $\nu_i$ are equal and $\prod_i\nu_i=1$, a contradiction.
\end{proof}

The triple is defined for states in $\Hank_3$. By Theorem~\ref{thm:main3} it yields invariants of every state of type~I.

\begin{corollary}\label{cor:invariants}
Let $\rho$ be a three-qubit PPT entangled state of rank four and type~I. For $j=1,2,3$ let $H_j$ span the kernel of the linear map $H\mapsto\tr_j[(H\otimes I)\rho]$ on Hermitian $2\times2$ matrices, write $H_j=V_j^*ZV_j$ with $V_j\in\SL(2,\C)$ up to a positive factor, and put $\rho_{\mathcal H}=U_{\rm R}^{*\otimes3}(V_1\otimes V_2\otimes V_3)\rho(V_1\otimes V_2\otimes V_3)^*U_{\rm R}^{\otimes3}$ with $U_{\rm R}$ as in Lemma~\ref{lem:pauli3}. Then $\rho_{\mathcal H}\in\Hank_3$, and the orbit of the triple $(\mathrm{cr},\widehat G,o)$ of $\rho_{\mathcal H}$ under the finite group does not depend on the choices made. Two such states are $\SL(2,\C)^3$-equivalent up to a positive factor if and only if these orbits coincide.
\end{corollary}

\begin{proof}
The kernel is one-dimensional: by the proof of Theorem~\ref{thm:main3} it contains an indefinite $H_j$, and it is at most one-dimensional because any three of the projectors $X_i^{(j)}$ are linearly independent. By Lemma~\ref{lem:zfree} and Lemma~\ref{lem:pauli3}, $\rho_{\mathcal H}\in\Hank_3$. Different choices of $V_j$, and $\SL(2,\C)^3$-equivalent states, lead to states in $\Hank_3$ that are $\SL(2,\C)^3$-equivalent, hence related by real local operators by Theorem~\ref{thm:moduli}(a), and Theorem~\ref{thm:moduli}(b) applies.
\end{proof}

\begin{remark}\label{rem:jinv}
A second, elementary criterion for a state to be inequivalent to its image under the exchange of qubits $2$ and $3$ uses the other two qubits. For a state of type~I, the decompositions of Proposition~\ref{prop:CD} across $2\,|\,13$ and $3\,|\,12$ give four points on the projective lines of qubits $2$ and $3$. Their cross-ratios are defined up to the anharmonic action, so their $j$-invariants $j_2$ and $j_3$ are $\SL(2,\C)^3$-invariants, and the exchange of qubits $2$ and $3$ interchanges them. Hence a state with $j_2\ne j_3$ is not equivalent to its image under the exchange.
\end{remark}

Numerically, for $200$ random states of the form in Theorem~\ref{thm:normal}(a), the triple recovered from the state agrees with the one used in the construction, it is unchanged by random real local operators, and the exchange of qubits $2$ and $3$ changes $o$. In all $200$ cases $j_2\ne j_3$, with smallest relative difference $3.9\times10^{-4}$, and the exchange interchanges $j_2$ and $j_3$ (Appendix~\ref{app:comp}).

The invariant in (d) was used by Singh and Banik \cite{SinghBanik2026} to separate filtering orbits; in particular they showed that the three-qubit Bell-nonlocal state of V\'ertesi and Brunner does not arise from a UPB by invertible local operators. A state obtained from a UPB by invertible local operators has four product vectors in its kernel, so it lies on the locus $\diag(\widehat G^{-1})=0$. These are four equations on a seven-dimensional space, and Bravyi's family of UPBs has three parameters; for states on this locus whose four kernel product vectors are in general position, a criterion for coming from a UPB is given in \cite[Thm.~17]{ChengChen2026}.

\begin{remark}\label{rem:gabardo}
Gabardo's example \cite{Gabardo1998} of a positive semidefinite Toeplitz matrix on $\{0,1\}^3$ without representing measure,
\[
t(0)=t(\pm(1,-1,-1))=t(\pm(1,-1,1))=t(\pm(1,1,-1))=4,\qquad t(d)=-1\ \text{otherwise},
\]
has rank four. As a three-qubit state it is of type~I, with $Q=15/512$ at unit trace, its points $[x_i]$ are $[\pm1:1]$ and $[e^{\pm2\pi i/3}:1]$, which are in harmonic position, and no diagonal entry of $\widehat G^{-1}$ vanishes; these statements are verified in exact arithmetic (Appendix~\ref{app:comp}). It therefore lies outside the UPB locus. Its white-noise threshold, computed by column generation, is $\lambda^*\approx0.1933$, larger than the maximum $3-2\sqrt2$ over all UPB states (Section~\ref{sec:noise}). Explicit rank-four three-qubit PPT entangled states outside the UPB family were constructed in the quantum-information literature by Ha and Kye \cite{HaKye2015} and by Cheng and Chen \cite[Example~16]{ChengChen2026}; Gabardo's matrix gives an independent example from the literature on the moment problem.
\end{remark}

\section{Exact noise thresholds for UPB states}\label{sec:noise}

For a PSD Toeplitz matrix $T$ with $t(0)=1$ consider the family $T_\lambda=(1-\lambda)T+\lambda I$. It corresponds to $(1-\lambda)\rho_T/\tr\rho_T+\lambda\sigma_0$, where $\sigma_0=\int\bigotimes_j\proj{\theta_j}^{\otimes m_j}d\theta/(2\pi)^k$ is the state with uniformly distributed phases; when every group consists of a single qubit, $\sigma_0=I/2^n$ is the maximally mixed state. We call $\lambda^*(T)=\min\{\lambda:T_\lambda\in\Dec\}$ the noise threshold of $T$; for single-qubit groups it is the white-noise threshold for full separability. This section addresses the second question of the introduction: how robust is bound entanglement to noise? We determine $\lambda^*$ exactly for all three-qubit UPB states, using the moment problem and its dual description by nonnegative trigonometric polynomials. Semidefinite bounds on $\lambda^*(T)$ from extendibility and collective dephasing, and a family of $k$-qubit states with $\lambda^*\to1$, are given in Appendix~\ref{app:further}.

\subsection{Unextendible product bases}\label{sec:upb}

UPBs and the bound entangled states they define were introduced in \cite{Bennett1999,DiVincenzo2003}. By Bravyi's classification \cite{Bravyi2004} every three-qubit UPB is, up to local unitaries and a relabelling of its four members, of the form
\[
v_1=\ket{0,0,0},\quad v_2=\ket{1,B,C},\quad v_3=\ket{A,1,C^{\perp}},\quad v_4=\ket{A^{\perp},B^{\perp},1},
\]
with $\ket{A}=\cos\theta_A\ket0+\sin\theta_A\ket1$, $\ket{A^{\perp}}=-\sin\theta_A\ket0+\cos\theta_A\ket1$, and similarly for $B,C$. Applying $Z$ to a qubit changes the sign of the corresponding angle, so we may assume $\theta_A,\theta_B,\theta_C\in(0,\pi/2)$. Let $\Pi=\sum_i\proj{v_i}$ and $\rho_{\rm UPB}=(I-\Pi)/4$. We put $(a,b,c)=(2\theta_A,2\theta_B,2\theta_C)\in(0,\pi)^3$ and
\begin{equation}\label{eq:phi}
\varphi_1=(0,0,0),\quad\varphi_2=(\pi,b,c),\quad\varphi_3=(a,\pi,c+\pi),\quad\varphi_4=(a+\pi,b+\pi,\pi).
\end{equation}

\begin{proposition}\label{prop:upbT}
Let $U_{\rm B}=\diag(1,i)\,e^{-i\pi X/4}$. Then $U_{\rm B}^{\otimes3}\rho_{\rm UPB}\,U_{\rm B}^{*\otimes3}=T/8$, where $T\in\Toep_3$ is given by
\[
T=2I-\frac14\sum_{i=1}^4a(\varphi_i)a(\varphi_i)^{*},\qquad t(d)=2\delta_{d,0}-\frac14\sum_{i=1}^4e^{i\langle d,\varphi_i\rangle}.
\]
\end{proposition}

\begin{proof}
$U_{\rm B}(\cos\vartheta\ket0+\sin\vartheta\ket1)=\frac{e^{-i\vartheta}}{\sqrt2}\begin{psmallmatrix}1\\ e^{2i\vartheta}\end{psmallmatrix}$, so $U_{\rm B}^{\otimes3}v_i$ is a unimodular multiple of $a(\varphi_i)/\sqrt8$.
\end{proof}

Any two of the points $\varphi_i$ differ by $\pi$ in at least one coordinate, which expresses the orthogonality of the $v_i$. All one-body moments $t(\pm e_j)$ vanish.

\subsection{The exact threshold}

Every UPB state is PPT and entangled \cite{Bennett1999}, and Proposition~\ref{prop:upbT} shows that it is a Toeplitz state of rank four; it has four product vectors in its kernel, so by Theorem~\ref{thm:moduli}(d) it lies on the locus $\diag(\widehat G^{-1})=0$. We now determine their robustness to white noise exactly. For $0\le\lambda\le1$ let $T_\lambda=(1-\lambda)T+\lambda I$, which corresponds to $(1-\lambda)\rho_{\rm UPB}+\lambda I/8$.

\begin{theorem}\label{thm:upb}
Let $s=\sin a\,\sin b\,\sin c=\sin2\theta_A\sin2\theta_B\sin2\theta_C\in(0,1]$ and
\[
\lambda^*=1-\frac{2}{1+\sqrt{1+s}} .
\]
For $0\le\lambda\le1$ the state $(1-\lambda)\rho_{\rm UPB}+\lambda I/8$ is fully separable if and only if $\lambda\ge\lambda^*$. Equivalently, $T_\lambda\in\Dec$ if and only if $1-\lambda\le\tau^*$, where $\tau^*=2/(1+\sqrt{1+s})$ is the positive root of $s\tau^2+4\tau-4=0$.
\end{theorem}

The two formulations are equivalent by Theorem~\ref{thm:dict}(b). For a UPB given in an arbitrary local basis, the local vectors of party $j$ form exactly two orthonormal bases $\{e_j,e_j^\perp\}$ and $\{f_j,f_j^\perp\}$, and
\[
s=\prod_{j=1}^32\,\bigl|\langle e_j|f_j\rangle\langle f_j|e_j^\perp\rangle\bigr| ,
\]
which is invariant under local unitaries, since each factor is unchanged by a unitary acting on both bases and by phases of the basis vectors, and which reduces to $s$ as defined above for Bravyi's form, where for instance $2|\langle0|A\rangle\langle A|1\rangle|=2|\cos\theta_A\sin\theta_A|$; so the threshold can be computed without transforming to that form. The proof occupies the rest of this subsection. We write $\tau=1-\lambda$.

\subsubsection*{Symmetry}
Consider the maps of $\T^3$
\[
g_{12}(\theta)=(\theta_1+\pi,\ b-\theta_2,\ c-\theta_3),\qquad g_{13}(\theta)=(a-\theta_1,\ \theta_2+\pi,\ c+\pi-\theta_3),
\]
and $g_{14}=g_{12}g_{13}$. They are commuting involutions, and together with the identity they form a Klein four-group $\mathcal K$ that permutes $\varphi_1,\dots,\varphi_4$ simply transitively. A shift $\theta_j\mapsto\theta_j+\pi$ is implemented on $a(\theta)a(\theta)^{*}$ by conjugation with $Z$ and a reflection $\theta_j\mapsto k-\theta_j$ by conjugation with $\diag(1,e^{ik})X$. Hence $\mathcal K$ acts on $\Toep_3$ by local unitaries that fix $T$ and $I$, and if $T_\lambda$ has a representing measure it has a $\mathcal K$-invariant one.

\subsubsection*{Reduction}
Write $\theta=(a/2,b/2,c/2)+(x,y,z)$ and define $\mathbf m=(\mu_1,\dots,\mu_5):\T^3\to\R^5$ by
\begin{equation}\label{eq:M}
\begin{gathered}
\mu_1=\sin x\sin y,\qquad \mu_2=\cos x\sin z,\qquad \mu_3=\cos y\cos z,\\
\mu_4=\cos x\sin y\cos z,\qquad \mu_5=\sin x\cos y\sin z .
\end{gathered}
\end{equation}
Here $\mu_1$, $\mu_2$ and $\mu_3$ collect the $\mathcal K$-invariant combinations of the two-body moments of the qubit pairs $(1,2)$, $(1,3)$ and $(2,3)$, and $\mu_4$, $\mu_5$ those of the three-body moments. A computation gives, for the orbit sums $O_d(\theta)=\sum_{g\in \mathcal K}e^{i\langle d,g\theta\rangle}$ with $d\in\{-1,0,1\}^3$ and $\omega_d=e^{i\langle d,(a,b,c)\rangle/2}$,
\begin{align*}
&O_{\pm e_j}=0, && O_{(1,d_2,0)}=-4d_2\,\omega_d\,\mu_1, && O_{(1,0,d_3)}=4id_3\,\omega_d\,\mu_2,\\
&O_{(0,1,d_3)}=4\,\omega_d\,\mu_3, && O_{(1,d_2,d_3)}=\omega_d\,(4id_2\,\mu_4-4d_3\,\mu_5),
\end{align*}
and $O_{-d}=\overline{O_d}$. The point $\varphi_1$ corresponds to $(x,y,z)=-(a/2,b/2,c/2)$, where $\mathbf m$ takes the value
\[
\mathbf m_0=(A_0,B_0,C_0,-e_1,e_2),\qquad
\begin{aligned}
&A_0=\sin\tfrac a2\sin\tfrac b2,\quad B_0=-\cos\tfrac a2\sin\tfrac c2,\quad C_0=\cos\tfrac b2\cos\tfrac c2,\\
&e_1=\cos\tfrac a2\sin\tfrac b2\cos\tfrac c2,\quad e_2=\sin\tfrac a2\cos\tfrac b2\sin\tfrac c2 .
\end{aligned}
\]
Note that $A_0B_0C_0=-s/8$, $e_1e_2=s/8$ and $|\mathbf m_0|=1$. For a $\mathcal K$-invariant probability measure $\mu$ the moment equations $\int e^{i\langle d,\theta\rangle}d\mu=\tau\,t(d)=-\tfrac\tau4O_d(\varphi_1)$ for $d\ne0$ are equivalent to $\int\mathbf m\,d\mu=-\tau \mathbf m_0$. Therefore
\begin{equation}\label{eq:convV}
T_\lambda\ \text{is decomposable}\iff-\tau \mathbf m_0\in\conv\mathbf m(\T^3).
\end{equation}

\subsubsection*{The image of \texorpdfstring{$\mathbf m$}{m}}

\begin{lemma}\label{lem:V}
$\mathbf m(\T^3)=\mathcal M:=\{m\in\R^5:\ |m|=1,\ \mu_4\mu_5=\mu_1\mu_2\mu_3\}$.
\end{lemma}

The proof is elementary and given in Appendix~\ref{app:proofs}.

\subsubsection*{Decomposability at \texorpdfstring{$\tau^{*}$}{tau*}}
Let $\tau=\tau^{*}$ and
\[
(A,B,C)=-\tau(A_0,B_0,C_0),\qquad\delta=\tau(e_1+e_2),
\]
and let $u,v>0$ be defined by $v-u=\delta$ and $uv=\tau^3s/8$. Put
\begin{equation}\label{eq:Ppm}
P_-=(A,B,C,-u,-v),\qquad P_+=(A,B,C,v,u).
\end{equation}
Then the products of the last two coordinates of $P_-$ and of $P_+$ both equal $uv=ABC$, and
\[
|P_-|^2=|P_+|^2=\tau^2(A_0^2+B_0^2+C_0^2)+\delta^2+2uv=\tau^2\bigl(1+\tfrac s4\bigr)+\tau^3\tfrac s4 ,
\]
which equals $1$ because $s\tau^2=4(1-\tau)$. So $P_\pm\in\mathcal M$. With
\[
p_-=\frac12\Bigl(1-\frac{\tau(e_1-e_2)}{u+v}\Bigr),\qquad p_+=1-p_-,
\]
one checks $p_-P_-+p_+P_+=-\tau \mathbf m_0$, and $0<p_-<1$ because $(u+v)^2-\tau^2(e_1-e_2)^2=\tau^2s(1+\tau)/2>0$. By Lemma~\ref{lem:V} and \eqref{eq:convV}, $T_{\lambda^{*}}$ is decomposable, with a representing measure supported on two $\mathcal K$-orbits, that is on eight points. Since $I=\int a(\theta)a(\theta)^{*}\,d\theta/(2\pi)^3$ is decomposable, so is $T_\lambda$ for all $\lambda\ge\lambda^{*}$.

\subsubsection*{The upper bound}

\begin{lemma}\label{lem:concave}
The function $\Omega\mapsto\sqrt{\det\Omega}$ is concave on the set of positive semidefinite real symmetric $3\times3$ matrices with unit diagonal.
\end{lemma}

The proof is given in Appendix~\ref{app:proofs}.

Let
\begin{equation}\label{eq:ell}
\ell=(A+BC,\ B+AC,\ C+AB,\ \delta,\ -\delta).
\end{equation}
For $m\in\mathcal M$,
\[
(\mu_4-\mu_5)^2=1-\mu_1^2-\mu_2^2-\mu_3^2-2\mu_1\mu_2\mu_3=\det\Omega(\mu),
\]
where
\begin{equation}\label{eq:Omega}
\Omega(\mu)=\begin{pmatrix}1&-\mu_1&-\mu_2\\-\mu_1&1&-\mu_3\\-\mu_2&-\mu_3&1\end{pmatrix},
\end{equation}
and $\Omega(\mu)\ge0$ because its principal minors are nonnegative. Hence
\begin{equation}\label{eq:Psi}
\ell\cdot m\ \le\ \Psi(\mu_1,\mu_2,\mu_3):=(A+BC)\mu_1+(B+AC)\mu_2+(C+AB)\mu_3+\delta\sqrt{\det\Omega(\mu)} .
\end{equation}
By Lemma~\ref{lem:concave}, $\Psi$ is concave on the convex set where $\Omega\ge0$. At $(\mu_1,\mu_2,\mu_3)=(A,B,C)$ we have $\det N=(v-u)^2=\delta^2>0$, so this is an interior point, and $\partial\Psi/\partial \mu_1=(A+BC)-\delta(A+BC)/\delta=0$, and likewise for $\mu_2,\mu_3$. Therefore
\[
\max_{m\in\mathcal M}\ell\cdot m=\Psi(A,B,C)=A^2+B^2+C^2+3ABC+\delta^2=1+ABC .
\]
On the other hand a direct computation gives $\ell\cdot \mathbf m_0=-\tau^{*}\bigl(1+\tfrac s4+\tfrac38\tau^{*}s\bigr)$. If $-\tau \mathbf m_0\in\conv\mathcal M$ then $-\tau\,\ell\cdot \mathbf m_0\le1+ABC$, that is
\[
\tau\le\frac{1+(\tau^{*})^3s/8}{\tau^{*}\bigl(1+\tfrac s4+\tfrac38\tau^{*}s\bigr)}=\tau^{*},
\]
where the last equality follows from $s(\tau^{*})^2=4(1-\tau^{*})$. Together with \eqref{eq:convV} this proves Theorem~\ref{thm:upb}. \qed

\subsubsection*{Consequences}

\begin{corollary}\label{cor:upb}
\begin{enumerate}
\item[(a)] Among all three-qubit UPB states the Shifts state $\theta_A=\theta_B=\theta_C=\pi/4$ tolerates the largest amount of white noise, $\lambda^{*}=3-2\sqrt2$.
\item[(b)] $I-\gamma\Pi$ is fully separable if and only if $\gamma\le4/(3+\sqrt{1+s})$. For $4/(3+\sqrt{1+s})<\gamma<1$ it is a PPT entangled state of full rank that is separable across every bipartition.
\item[(c)] At the threshold the state, which has rank eight as every state of the family with $\lambda>0$, is a convex combination of eight product states.
\item[(d)] There is a witness $W$ that detects every entangled state of the family $(1-\lambda)\rho_{\rm UPB}+\lambda I/8$, so that it is optimal with respect to white noise: $\bra\psi W\ket\psi\ge0$ for all product vectors $\psi$, and $\tr\bigl(W[(1-\lambda)\rho_{\rm UPB}+\lambda I/8]\bigr)<0$ if and only if $\lambda<\lambda^*$. In the frame of Bravyi's form,
\begin{multline}\label{eq:Wclosed}
W=\tfrac18\bigl[(1+ABC)\,I\otimes I\otimes I-\ell_1R_A^\perp\otimes R_B^\perp\otimes I-\ell_2R_A\otimes I\otimes R_C^\perp-\ell_3I\otimes R_B\otimes R_C\\-\ell_4R_A\otimes R_B^\perp\otimes R_C-\ell_5R_A^\perp\otimes R_B\otimes R_C^\perp\bigr],
\end{multline}
where $R_j=\cos\theta_j\,Z+\sin\theta_j\,X$ and $R_j^\perp=\cos\theta_j\,X-\sin\theta_j\,Z$ for $j=A,B,C$, and $A,B,C$ and $\ell=(\ell_1,\dots,\ell_5)$ are given in the proof of Theorem~\ref{thm:upb}, in particular \eqref{eq:ell}. Its Pauli expansion contains only the identity, two-body and three-body strings with factors in $\{I,X,Z\}$, so $W$ can be measured with the eight local Pauli settings $\{X,Z\}^{\otimes3}$; the coefficient of $XXX$ is $\tau^*(e_1^2-e_2^2)/8$, which vanishes if and only if $\tan\theta_B=\tan\theta_A\tan\theta_C$. For $\theta_A=\theta_B=\theta_C=\pi/4$ one can take
\begin{multline*}
W_{\rm S}=(4\sqrt2-2)\,III+IXX+IXZ+IZX+IZZ-XIX+XIZ+XXI-XZI\\-ZIX+ZIZ-ZXI+ZZI+\sqrt2\,(ZZZ+XZX-XXZ-ZXX),
\end{multline*}
which can be measured with the four settings $XXZ$, $XZX$, $ZXX$ and $ZZZ$.
\item[(e)] The witness $W$ of (d) is optimal in the sense of Lewenstein et al.\ \cite{LKCH2000}, as is every witness that vanishes on a fully separable state of full rank.
\end{enumerate}
\end{corollary}

\begin{proof}
(a) is clear from Theorem~\ref{thm:upb}. For (b), $(1-\lambda)\rho_{\rm UPB}+\lambda I/8$ is proportional to $I-\gamma\Pi$ with $\gamma=2(1-\lambda)/(2-\lambda)$, and the state has full rank for $\lambda>0$. It is PPT by Theorem~\ref{thm:dict}(a) and separable across every bipartition by Theorem~\ref{thm:cuts}(a), because every bipartition of three qubits has a single qubit on one side. (c) was shown in the proof of Theorem~\ref{thm:upb}. For (d), the functional $\ell$ defines the real trigonometric polynomial $q(\theta)=1+ABC-\ell\cdot\mathbf m(\theta-(a,b,c)/2)$, which by the formulas for $O_d$ has degree at most one in each variable and no terms of degree one in total, and which is nonnegative on $\T^3$ by the upper bound. Let $W_q$ be the operator of Theorem~\ref{thm:dict}(c) and $W=U_{\rm B}^{*\otimes3}W_qU_{\rm B}^{\otimes3}$. Then $\bra\psi W\ket\psi\ge0$ for product vectors, and $\tr\bigl(W[(1-\lambda)\rho_{\rm UPB}+\lambda I/8]\bigr)=\tr(W_qT_\lambda)/8=\langle q,T_\lambda\rangle/8=\bigl(1+ABC+(1-\lambda)\,\ell\cdot \mathbf m_0\bigr)/8$. This is affine in $\lambda$, vanishes at $\lambda^*$ by the computation of $\ell\cdot \mathbf m_0$ above, and is negative for $\lambda<\lambda^*$ because $\ell\cdot \mathbf m_0<0$. Since $U_{\rm B}XU_{\rm B}^*=Y$ and $U_{\rm B}ZU_{\rm B}^*=X$, the strings with factors in $\{I,X,Y\}$ in $W_q$ correspond to strings with factors in $\{I,Z,X\}$ in $W$. For the closed form, note that $W_q$ is the operator without Pauli factors $Z$ with $a(\theta)^*W_qa(\theta)=q(\theta)$, and that for a single qubit $a(\theta)^*Ia(\theta)=2$, $a(\theta)^*Xa(\theta)=2\cos\theta$ and $a(\theta)^*Ya(\theta)=2\sin\theta$. Hence a product of the functions $1$, $\cos(\theta_j-\alpha)$ and $\sin(\theta_j-\alpha)$ of the three variables corresponds to the tensor product of the operators $I/2$, $(\cos\alpha\,X+\sin\alpha\,Y)/2$ and $(\cos\alpha\,Y-\sin\alpha\,X)/2$, and conjugation with $U_{\rm B}^*$ maps these to $I/2$, $R_j/2$ and $R_j^\perp/2$ for $\alpha=\theta_j$. Inserting \eqref{eq:M} into $q$ gives the closed form, and the coefficient of $XXX$ is $-(\ell_4\sin\theta_A\cos\theta_B\sin\theta_C+\ell_5\cos\theta_A\sin\theta_B\cos\theta_C)/8=\delta(e_1-e_2)/8$, where $\delta=\tau^*(e_1+e_2)$ and $e_1-e_2=\cos\theta_A\cos\theta_B\cos\theta_C(\tan\theta_B-\tan\theta_A\tan\theta_C)$. The vanishing of the one-body moments $O_{\pm e_j}$ removes the one-body strings, and every two-body string is a marginal of a three-body setting. The closed form of $W_{\rm S}$ is obtained by evaluating these expressions at $a=b=c=\pi/2$; one checks directly that $\tr(W_{\rm S}\rho_{\rm UPB})=3\sqrt2-5<0$ and $\tr W_{\rm S}/8=4\sqrt2-2$, so that $\tr\bigl(W_{\rm S}[(1-\lambda)\rho_{\rm UPB}+\lambda I/8]\bigr)$ vanishes exactly at $\lambda=3-2\sqrt2$, and the four settings contain every two-body string of $W_{\rm S}$ as a marginal. For (e), recall that a witness $W''$ is finer than a witness $W'$ if it detects every state detected by $W'$, and that $W'$ is optimal if every witness finer than $W'$ is a positive multiple of $W'$ \cite{LKCH2000}. Let $W'$ be a witness and $\sigma=\sum_lp_l\proj{\psi_l}$ a fully separable state of full rank with $\tr(W'\sigma)=0$, $p_l>0$ and product vectors $\psi_l$. Since $\sigma$ has full rank, the $\psi_l$ span $\C^8$, and since $\bra\psi W'\ket\psi\ge0$ for product vectors, $\bra{\psi_l}W'\ket{\psi_l}=0$ for all $l$. Let $W''$ be finer than $W'$. A nonzero operator that is nonnegative on product vectors has positive trace, because its average over product states is a positive multiple of its trace and product states span all operators; so the cone $K=\{\rho\ge0:\tr(W''\rho)\ge0\}$ contains the identity in its interior. $W'$ is nonnegative on $K$, and by conic duality $W'=\mu W''+P$ with $\mu\ge0$ and $P\ge0$, where $\mu>0$ because $W'$ is not positive semidefinite. Then $\bra{\psi_l}P\ket{\psi_l}=0$ for all $l$, so $P=0$ and $W''$ is a positive multiple of $W'$. This argument is the one of \cite{LKCH2000}, written for product vectors of three qubits. For $W$ we take $\sigma=\rho_{\lambda^*}$, which has rank eight by (c) and satisfies $\tr(W\rho_{\lambda^*})=0$ by (d).
\end{proof}

Up to the Pauli strings that contain $Y$, the witness of Corollary~\ref{cor:upb}(d) is the only one that detects every entangled state of the family, for the Shifts UPB, for all UPBs close to it, and for all further UPBs we have tested.

\begin{proposition}\label{prop:unique}
Let $W'$ be a witness, written in the frame of Bravyi's form, with $\tr\bigl(W'[(1-\lambda)\rho_{\rm UPB}+\lambda I/8]\bigr)<0$ for all $\lambda<\lambda^*$, and let $\overline{W'}=2^{-3}\sum_SW'^{\Gamma_S}$ be its average over the partial transposes with respect to all sets $S$ of qubits. Then $\overline{W'}$ is a witness with the same property, and its Pauli expansion is that of $W'$ without the strings that contain $Y$. Moreover $\overline{W'}=U_{\rm B}^{*\otimes3}W_{q'}U_{\rm B}^{\otimes3}$ for a nonnegative trigonometric polynomial $q'$ of degree at most one in each variable that vanishes, together with its gradient, at the eight points $\theta_1,\dots,\theta_8$ of the representing measure of $T_{\lambda^*}$ constructed in the proof of Theorem~\ref{thm:upb}. If these $32$ linear conditions determine $q'$ up to a factor, then $\overline{W'}$ is a positive multiple of $W$, and $W'$ contains the string $XXX$ unless $\tan\theta_B=\tan\theta_A\tan\theta_C$.
\end{proposition}

\begin{proof}
For a product vector $\psi$, $\bra\psi W'^{\Gamma_S}\ket\psi=\bra{\psi_S}W'\ket{\psi_S}\ge0$, where $\psi_S$ is the product vector obtained by complex conjugation of the factors in $S$, so $\overline{W'}$ is a witness. Partial transposition of a qubit changes the sign of $Y$ and fixes $I$, $X$ and $Z$, which gives the statement on Pauli expansions. Let $\rho_\lambda=(1-\lambda)\rho_{\rm UPB}+\lambda I/8$. By Proposition~\ref{prop:upbT}, $U_{\rm B}^{\otimes3}\rho_\lambda U_{\rm B}^{*\otimes3}$ contains no Pauli factor $Z$, and since $U_{\rm B}^*ZU_{\rm B}=Y$, $\rho_\lambda$ contains no factor $Y$, so $\rho_\lambda^{\Gamma_S}=\rho_\lambda$ and $\tr(\overline{W'}\rho_\lambda)=\tr(W'\rho_\lambda)$. The operator $U_{\rm B}^{\otimes3}\overline{W'}U_{\rm B}^{*\otimes3}$ contains no factor $Z$, so it equals $W_{q'}$ with $q'(\theta)=a(\theta)^*W_{q'}a(\theta)\ge0$, and $\tr(\overline{W'}\rho_\lambda)=\langle q',T_\lambda\rangle/8$. By continuity $\langle q',T_{\lambda^*}\rangle\le0$, and since $T_{\lambda^*}=\sum_lp_la(\theta_l)a(\theta_l)^*$ with $p_l>0$, also $\langle q',T_{\lambda^*}\rangle=\sum_lp_lq'(\theta_l)\ge0$. Hence $q'(\theta_l)=0$, and since $q'\ge0$, its gradient vanishes at $\theta_l$. The polynomial $q$ of Corollary~\ref{cor:upb}(d) satisfies the same conditions. If they determine $q'$ up to a factor, then $q'=cq$, with $c>0$ because $\langle q',T_0\rangle$ and $\langle q,T_0\rangle$ are negative, so $\overline{W'}=cW$. The last statement follows from the coefficient of $XXX$ in Corollary~\ref{cor:upb}(d).
\end{proof}

\begin{corollary}\label{cor:unique}
For the Shifts UPB, and for every UPB whose angles are sufficiently close to $(\pi/4,\pi/4,\pi/4)$, the conditions of Proposition~\ref{prop:unique} determine $q'$ up to a factor. Hence every witness that detects all entangled states of the family coincides with a positive multiple of $W$ up to Pauli strings that contain $Y$.
\end{corollary}

\begin{proof}
For the Shifts UPB the eight points $\theta_l$ have coordinates with $(\cos\theta_{l,j},\sin\theta_{l,j})\in\{(\pm\beta,\pm\alpha),(\pm\alpha,\pm\beta)\}$, where $\alpha=\sqrt{\sqrt2-1/2}$ and $\beta=1-1/\sqrt2$ satisfy $\alpha^2+\beta^2=1$, so all conditions have coefficients in the field $\mathbb Q(\alpha)$. An exact computation in this field shows that the $32\times27$ matrix of the conditions has rank $26$, that $q$ and its gradient vanish at the eight points, and that the Hessian of $q$ is positive definite there (Appendix~\ref{app:comp}). The coefficients of $q$ depend continuously on the angles. Since the zeros of $q$ are nondegenerate minima, for angles close to $\pi/4$ the zeros of $q$ lie close to those for the Shifts UPB, with at most one near each; the eight distinct points of the representing measure in the proof of Theorem~\ref{thm:upb} are zeros of $q$, so they depend continuously on the angles. The rank of the matrix of conditions is therefore at least $26$ near the Shifts UPB, and at most $26$ because $q$ satisfies the conditions.
\end{proof}

For nine further angle triples, six of them random, the $32$ conditions at the explicit points of the proof of Theorem~\ref{thm:upb} also have a one-dimensional solution space, with smallest nonzero singular value at least $0.098$ (Appendix~\ref{app:comp}).

\begin{figure}[t]
\centering
\includegraphics[width=0.62\textwidth]{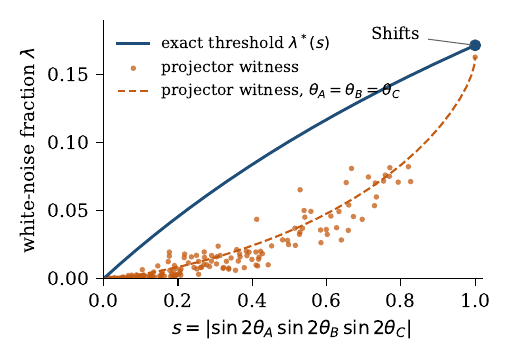}
\caption{White-noise thresholds for three-qubit UPB states as a function of $s=|\sin2\theta_A\sin2\theta_B\sin2\theta_C|$. Solid line: the exact threshold $\lambda^*(s)$ of Theorem~\ref{thm:upb}, below which $(1-\lambda)\rho_{\rm UPB}+\lambda I/8$ is entangled. Dots: the detection threshold $2\varepsilon$ of the projector witness $\Pi-\varepsilon I$, where $\varepsilon$ is the minimal overlap of $\Pi$ with product states, for the Shifts UPB and $159$ UPBs with angles drawn independently and uniformly from $(0.02,\pi/2-0.02)$; dashed line: the same for the symmetric family $\theta_A=\theta_B=\theta_C$. Since $\varepsilon$ is computed by local minimisation, the computed value is an upper bound on $\varepsilon$, and the dots are upper bounds on the detection thresholds of the projector witnesses. The large dot marks the Shifts UPB.}
\label{fig:upb}
\end{figure}

The witness $\Pi-\varepsilon I$ built from the UPB projector, with $\varepsilon$ the minimal overlap of $\Pi$ with product states, goes back to Terhal \cite{Terhal2001}. It is optimal in the sense of Lewenstein et al.\ \cite{LKCH2000} for the Shifts state by the work of Wan and Guo \cite{WanGuo2026}, who asked whether other witnesses tolerate more noise. For the Shifts UPB the minimal overlap is $\varepsilon=1-3\sqrt6/8$ \cite{Branciard2010}, so the projector witness detects $(1-\lambda)\rho_{\rm UPB}+\lambda I/8$ only for $\lambda<2\varepsilon=2-3\sqrt6/4\approx0.1629$, whereas the witness $W_{\rm S}$ of Corollary~\ref{cor:upb}(d) detects it for $\lambda<3-2\sqrt2\approx0.1716$, which by Theorem~\ref{thm:upb} is the largest possible range. Thus both witnesses are optimal in the sense of \cite{LKCH2000}, but only $W$ detects every entangled state of the family. Each $v_i$ has a factor $\ket0$ or $\ket1$, on which $X$ has expectation zero, so $\Pi$ contains no string $XXX$ and can be measured with at most seven of the settings $\{X,Z\}^{\otimes3}$. The witness $W$ needs the setting $XXX$ in addition unless $\tan\theta_B=\tan\theta_A\tan\theta_C$, and by Proposition~\ref{prop:unique} and Corollary~\ref{cor:unique} this cannot be avoided by another witness that detects the whole family, for the UPBs near the Shifts UPB and for those tested numerically; for the Shifts UPB both witnesses use the same four settings. These counts refer to local Pauli settings. If arbitrary local bases are allowed, $\Pi$ can be measured with four product-basis settings, one for each $v_i$, and $W$ with five settings in the bases $\{R_j,R_j^\perp\}$: those of its two three-body terms and those of its three two-body terms, which are not marginals of the former and are pairwise incompatible. We do not claim that these numbers are minimal. For generic UPBs the gain in noise tolerance is large. As an illustration, in the sample of $159$ UPBs of Figure~\ref{fig:upb} other than Shifts, with angles drawn independently and uniformly from $(0.02,\pi/2-0.02)$, the ratio of $\lambda^*$ to the computed detection threshold of the projector witness has median $6.7$, with bootstrap $95\%$ interval $[5.0,8.6]$ for the median, quartiles $3.5$ and $18$, and minimum $1.57$. These ratios are lower bounds for the true ratios, because the detection thresholds are computed by local minimisation over product vectors and are therefore upper bounds.

\subsection{Measurement recipe for the Shifts state}\label{sec:recipe}

In the standard form of the Shifts UPB, $\ket{0,1,+}$, $\ket{1,+,0}$, $\ket{+,0,1}$, $\ket{-,-,-}$ \cite{Bennett1999}, which is obtained from Bravyi's form with $\theta_A=\theta_B=\theta_C=\pi/4$ by the local unitary $I\otimes X\otimes \mathrm{H}$ with the Hadamard matrix $\mathrm H$, the witness $W_{\rm S}$ of Corollary~\ref{cor:upb}(d) reads
\begin{multline*}
W_{\rm S}'=(4\sqrt2-2)\,III+IXX+IXZ-IZX-IZZ+XIX-XIZ+XXI+XZI\\+ZIX-ZIZ-ZXI-ZZI-\sqrt2\,(XXX+XZZ+ZXZ+ZZX).
\end{multline*}
Each of the four settings $XXX$, $XZZ$, $ZXZ$ and $ZZX$ yields one three-body correlator and three two-body marginals, and these twelve marginals are exactly the two-body strings of $W_{\rm S}'$. In the Shifts state each of the sixteen correlators equals $\pm\frac14$: $-\frac14$ for $IXX$, $IXZ$, $XIX$, $XXI$, $XZI$, $ZIX$, and $+\frac14$ for $IZX$, $IZZ$, $XIZ$, $ZIZ$, $ZXI$, $ZZI$ and the four three-body strings. Every term therefore contributes negatively, and $\tr(W_{\rm S}'\rho_{\rm Shifts})=3\sqrt2-5$. As a witness inequality in correlator form, which assumes trusted qubit measurements, $B=(4\sqrt2-2)I-W_{\rm S}'$ satisfies $\langle B\rangle\le4\sqrt2-2\approx3.657$ for every fully separable state, while $(1-\lambda)\rho_{\rm Shifts}+\lambda I/8$ gives $\langle B\rangle=(1-\lambda)(3+\sqrt2)$. Entanglement is detected for every visibility $1-\lambda>2\sqrt2-2\approx0.828$, and by Theorem~\ref{thm:upb} not below it. The projector witness uses the same four settings and requires visibility above $\tfrac34\sqrt6-1\approx0.837$.

At the threshold $\lambda=3-2\sqrt2$, the state is the uniform mixture of the eight product states in Table~\ref{tab:shifts}, whose Bloch vectors lie in the $XZ$ planes. These are the zeros of $W_{\rm S}'$ used in the proof of Corollary~\ref{cor:upb}(e), and they give an explicit separable model of the state at the threshold.

\begin{table}[t]
\centering
\small
\begin{tabular}{cccc}
\toprule
$l$ & qubit 1 & qubit 2 & qubit 3\\
\midrule
1 & $(\alpha,-\beta)$ & $(-\alpha,\beta)$ & $(-\beta,\alpha)$\\
2 & $(-\alpha,\beta)$ & $(-\beta,\alpha)$ & $(\alpha,-\beta)$\\
3 & $(-\beta,\alpha)$ & $(\alpha,-\beta)$ & $(-\alpha,\beta)$\\
4 & $(\beta,-\alpha)$ & $(\beta,-\alpha)$ & $(\beta,-\alpha)$\\
5 & $(\alpha,\beta)$ & $(-\beta,-\alpha)$ & $(-\alpha,-\beta)$\\
6 & $(-\alpha,-\beta)$ & $(\alpha,\beta)$ & $(-\beta,-\alpha)$\\
7 & $(\beta,\alpha)$ & $(\beta,\alpha)$ & $(\beta,\alpha)$\\
8 & $(-\beta,-\alpha)$ & $(-\alpha,-\beta)$ & $(\alpha,\beta)$\\
\bottomrule
\end{tabular}
\caption{The eight product states whose uniform mixture is $(1-\lambda^*)\rho_{\rm Shifts}+\lambda^*I/8$, $\lambda^*=3-2\sqrt2$, in the standard frame of the Shifts UPB. Each entry gives the Bloch vector $(x,z)$ of a qubit, with $y=0$, $\alpha=\sqrt{\sqrt2-1/2}\approx0.956$ and $\beta=1-1/\sqrt2\approx0.293$.}
\label{tab:shifts}
\end{table}

\subsection{Comparison with other rank-four states}

Table~\ref{tab:thresholds} collects white-noise thresholds of PPT entangled states that appear in this paper. Within rank four, the UPB states are not the most robust. By Theorem~\ref{thm:upb}, every UPB state has $\lambda^*\le3-2\sqrt2\approx0.1716$. Gabardo's state of Remark~\ref{rem:gabardo}, of type~I, has $\lambda^*=0.1932774$ to seven digits by a computer-assisted proof (Appendix~\ref{app:comp}), and a local search in the seven-dimensional family of Theorem~\ref{thm:normal} starting from it finds a type~I state with $\lambda^*\approx0.2635$; the search had not converged when stopped. This state has rank four, is PPT, has no product vector in its range and has $Q\approx0.0305$ at unit trace (Appendix~\ref{app:comp}). The state of Proposition~\ref{prop:mabk} with $k=3$, $\rho=\bigl[I+\tfrac12(XXX+XYY-YXY-YYX)\bigr]/8$, is a GHZ-diagonal PPT entangled state of rank four and type~II, a Toeplitz state whose quadrilateral is a rhombus by Proposition~\ref{prop:typeIIT}, with exact threshold $1-1/\sqrt2\approx0.293$, so it exceeds all UPB states by a proof without computer; this value also follows from the separability criterion for three-qubit GHZ-diagonal states of G\"uhne \cite{Guhne2011}, which Han and Kye showed to be necessary and sufficient \cite{HanKye2017GHZ}. The value $0.2635$ is therefore only a lower bound for the largest threshold among states of type~I, and which rank-four state tolerates the most white noise is open.

\begin{table}[t]
\centering
\footnotesize
\begin{tabular}{lcll}
\toprule
state & rank, type & $\lambda^*$ & source\\
\midrule
\multicolumn{4}{l}{\emph{three qubits, noise $I/8$}}\\
UPB states & 4, I & $1-2/(1+\sqrt{1+s})$, exact & Theorem~\ref{thm:upb}\\
Shifts UPB & 4, I & $3-2\sqrt2\approx0.1716$, exact & Corollary~\ref{cor:upb}\\
Shifts UPB, projector witness & 4, I & detection only, $0.1629$ & \cite{Branciard2010,WanGuo2026}\\
V\'ertesi--Brunner & 4, I & $0.0917$, computer-assisted & Appendix~\ref{sec:bell}\\
Gabardo & 4, I & $0.1933$, computer-assisted & Remark~\ref{rem:gabardo}\\
best found by local search & 4, I & $0.2635$, computer-assisted & this section\\
GHZ-diagonal, $k=3$ & 4, II & $1-1/\sqrt2\approx0.2929$, exact & \cite{Guhne2011,HanKye2017GHZ}, Prop.~\ref{prop:mabk}\\
\midrule
\multicolumn{4}{l}{\emph{four qubits, noise $I/16$}}\\
$\rho_{T_\star}$, $m=(2,2)$ & 6 & $0.1420$, numerical & Appendix~\ref{app:dephasing}\\
tetracode state & 7 & $0.1367$, numerical & Example~\ref{ex:tetracode}\\
\bottomrule
\end{tabular}
\caption{White-noise thresholds $\lambda^*$ for full separability, with states normalised to unit trace. For the projector witness the value is the largest $\lambda$ up to which it detects entanglement, not a separability threshold. Numerical values are computed by column generation for the moment problem. The primal solution is an explicit decomposition, whose moment residual, below $10^{-14}$, is absorbed by a change of $\lambda$ of the same size, since a neighbourhood of $\sigma_0$ consists of decomposable matrices; weak duality gives the lower bound $\lambda_{\rm LP}+\min q$, where $q$ is the dual trigonometric polynomial. For the three-qubit states the minimum of $q$ is bounded from below on every cell of an adaptively refined grid by a Taylor estimate with explicit remainder and a margin for rounding errors, which gives $\min q\ge-2\times10^{-8}$; these values are therefore correct in all digits shown, by a computer-assisted proof. For the four-qubit states the minimum of $q$ is located by a grid search refined by local minimisation, which gives a numerical certificate to the digits shown (Appendix~\ref{app:comp}). The value $0.2635$ is a lower bound for the largest threshold of type~I states.}
\label{tab:thresholds}
\end{table}

\section{Extremal bound entanglement and rank thresholds}\label{sec:structure}

This section addresses the third question of the introduction: how small can the rank of a bound entangled Toeplitz state be? We construct non-decomposable Toeplitz matrices from lattices, obtain extremal states of minimal rank for two groups, classify the configurations in which bound entanglement occurs at all, and determine the smallest rank in several configurations. We use the Hankel index of Blekherman, Sinn and Velasco \cite{BSV2017}: for a real projective variety $X$ that is not contained in a hyperplane, it is the smallest rank larger than one of an extreme ray of the cone of positive semidefinite linear functionals on quadratic forms restricted to $X$, the dual of the cone of sums of squares. By Lemma~\ref{lem:cayley}, for the embedding $X$ of $(\mathbb P^1)^k$ by $\mathcal O(m_1,\dots,m_k)$ it is the smallest rank larger than one of an extreme ray of $\Toep^+_m$.

\subsection{A lattice construction}

Let $\Lat\subset\Z^k$ be a sublattice of finite index $N$ and $\mathcal G=\Z^k/\Lat$. For $d\in\Z^k$ write $[d]\in \mathcal G$ for its class. The annihilator $\Lat^\perp=\{\theta\in\T^k:e^{i\langle\gamma,\theta\rangle}=1\ \forall\gamma\in\Lat\}$ is a group of order $N$, and $\theta\mapsto\chi_\theta$, $\chi_\theta([d])=e^{i\langle d,\theta\rangle}$, identifies it with the character group $\widehat{\mathcal G}$.

\begin{theorem}[Lattice construction]\label{thm:lattice}
Suppose that (L1) $\Lat$ is generated by $\Lat\cap D$, and (L2) every class of $\mathcal G$ meets $D$. Let $n_\Lambda$ be the number of classes met by $\Lambda$, assume $n_\Lambda<N$, and for $n_\Lambda\le\mu<N$ let $T_\mu$ be the Toeplitz matrix on $\Lambda$ with
\[
t_\mu(d)=\mu\,\mathbf 1_\Lat(d)-1 .
\]
Then $T_\mu\in\Toep^+_m\setminus\Dec_m$, and $\rank T_\mu=n_\Lambda$ for $\mu>n_\Lambda$, $\rank T_{n_\Lambda}=n_\Lambda-1$.
\end{theorem}

\begin{proof}
For $x\in\C^\Lambda$ and a class $g$ put $y_g=\sum_{\alpha\in\Lambda,[\alpha]=g}x_\alpha$. Then $x^*T_\mu x=\mu\sum_g|y_g|^2-|\sum_gy_g|^2$, and by the Cauchy--Schwarz inequality this is nonnegative for $\mu\ge n_\Lambda$. On the $n_\Lambda$-dimensional space of the vectors $y$ the form is $\mu I-J$, which has rank $n_\Lambda$ for $\mu>n_\Lambda$ and $n_\Lambda-1$ for $\mu=n_\Lambda$, and $T_\mu$ has the same rank.

If $\alpha-\alpha'\in\Lat$, rows $\alpha$ and $\alpha'$ of $T_\mu$ coincide, so $e_\alpha-e_{\alpha'}\in\ker T_\mu$. Suppose $T_\mu=\sum_lp_la(\theta_l)a(\theta_l)^*$ with $p_l>0$. Then $a(\theta_l)\in\Rg(T_\mu)$ is orthogonal to these vectors, so $e^{i\langle\gamma,\theta_l\rangle}=1$ for $\gamma\in\Lat\cap D$, and by (L1) $\theta_l\in\Lat^\perp$. Let $\nu(\chi)\ge0$ be the total weight of the $\theta_l$ with $\chi_{\theta_l}=\chi$. The function $f:\mathcal G\to\R$, $f(0)=\mu-1$ and $f(g)=-1$ for $g\ne0$, satisfies $t_\mu(d)=f([d])$, and by (L2) $f(g)=\sum_\chi\nu(\chi)\chi(g)$ for all $g\in \mathcal G$. Fourier inversion gives $\nu(\chi)=N^{-1}\sum_gf(g)\overline{\chi(g)}=N^{-1}(\mu-N\delta_{\chi,1})$, so $\nu(1)=(\mu-N)/N<0$, a contradiction.
\end{proof}

For $\mu=N$ the same computation shows that $T_N$ is decomposable,
\[
T_N=\sum_{\theta\in\Lat^\perp\setminus\{0\}}a(\theta)a(\theta)^*.
\] Thus along the segment $\mu\in[n_\Lambda,N]$ the states $\rho_{T_\mu}$ interpolate between a low-rank bound entangled state and a separable one.

\subsection{Two groups}

\begin{theorem}[Minimal-rank extremal states]\label{thm:twogroups}
Let $k=2$, $m_1,m_2\ge2$, $M=m_1+m_2$, and $\Lat=\Z(m_1,m_2)+\Z(2,-2)$, of index $N=2M$. Then (L1) and (L2) hold and $\Lambda$ meets $n_\Lambda=N-1$ classes. The matrix $T_\star=T_{N-1}$, explicitly
\[
t_\star(a,b)=\begin{cases}2M-2,&a+b=0\text{ and }a\text{ even, or }|a+b|=M\text{ and }a\equiv m_1\ (\mathrm{mod}\ 2),\\-1,&\text{otherwise},\end{cases}
\]
has the following properties.
\begin{enumerate}
\item[(i)] $T_\star\in\Toep^+_m\setminus\Dec_m$ and $\rank T_\star=2M-2=2(m_1+m_2)-2$.
\item[(ii)] $T_\star$ spans an extreme ray of $\Toep^+_m$.
\item[(iii)] For every bipartition $S|S^c$ that splits both groups, $\rho_{T_\star}$ spans an extreme ray of $\mathcal P_S$ and of the cone of states that are PPT with respect to all bipartitions, and its range contains no vector that is a product across $S|S^c$.
\end{enumerate}
For $m=(2,2)$, $T_\star$ is the matrix with $t(0,0)=t(\pm2,\pm2)=6$ and $t(d)=-1$ otherwise, which is the case $p=q=2$ of Gabardo's Theorem~6.1 \cite{Gabardo1998}.
\end{theorem}

\begin{proof}
The map $(a,b)\mapsto(a+b,\,a\bmod2)$ from $\Z^2$ onto $\Z\times\Z_2$ has kernel $\Z(2,-2)$, so $\mathcal G\cong(\Z\times\Z_2)/\langle(M,\mu_1)\rangle$ with $\mu_1=m_1\bmod2$; we write classes as $(s,\varepsilon)$ with $s=a+b$. In $\Lambda$, for $1\le s\le M-1$ the antidiagonal $a+b=s$ contains at least two consecutive values of $a$, so both parities occur; $s=0$ gives only $(0,0)$, and $s=M$ gives only $(m_1,m_2)$, which lies in the class of $(0,0)$. Hence $\Lambda$ meets all classes except $c_\star=(0,1)$, and $n_\Lambda=N-1$. The class $c_\star$ meets $D$ at $(1,-1)$, so (L2) holds, and (L1) holds because $(m_1,m_2)$ and $(2,-2)$ lie in $D$. The description of $\Lat\cap D$ in the statement follows from $\Lat=\{x(m_1,m_2)+y(2,-2)\}$ and $|a+b|\le M$ on $D$. Part (i) is Theorem~\ref{thm:lattice}.

(ii) Let $H$ be a Hermitian Toeplitz matrix with $\ker H\supseteq\ker T_\star$, with symbol $h$. Since $e_\alpha-e_{\alpha'}\in\ker H$ for $\alpha-\alpha'\in\Lat$, we have $h(\alpha-\beta)=h(\alpha'-\beta)$ for all $\beta\in\Lambda$. We show that $h(d)=h(d')$ whenever $d,d'\in D$ and $d-d'\in\Lat$. First, if $d$ and $d+(2,-2)$ lie in $D$, then $\alpha=(\max(0,d_1),\max(2,d_2))$, $\alpha'=\alpha+(2,-2)$ and $\beta=\alpha-d$ lie in $\Lambda$, so $h(d+(2,-2))=h(d)$. Since $D\cap\{a+b=s\}$ is an interval in $a$, this connects all points of $D$ in the same class with the same value of $s$. Second, if $d\in-\Lambda$, then $\alpha=0$, $\alpha'=(m_1,m_2)$ and $\beta=-d$ give $h(d+(m_1,m_2))=h(d)$. For $-M+1\le s\le-1$ the antidiagonal $a+b=s$ in $-\Lambda$ contains both parities, and the points $(0,0)$ and $(-m_1,-m_2)$ lie in $-\Lambda$; this connects the two values of $s$ that occur in each class. Hence $h(d)=\bar h([d])$ for a function $\bar h$ on $\mathcal G$.

Let $x\in\C^\Lambda$ satisfy $\sum_{[\alpha]=g}x_\alpha=1$ for each of the $N-1$ classes $g\ne c_\star$. Then $x^*T_\star x=(N-1)^2-(N-1)^2=0$, so $x\in\ker T_\star\subseteq\ker H$, and $(Hx)_\beta=\sum_{g\ne c_\star}\bar h([\beta]-g)=0$ for every $\beta\in\Lambda$. Since $[\beta]$ ranges over $\mathcal G\setminus\{c_\star\}$, the convolution $\bar h*\mathbf 1_{\mathcal G\setminus\{c_\star\}}$ equals $\varsigma\,\delta_{c_\star}$ for some $\varsigma$. With $\hat f(\chi)=\sum_gf(g)\overline{\chi(g)}$ this gives $\hat{\bar h}(\chi)\bigl(N\delta_{\chi,1}-\overline{\chi(c_\star)}\bigr)=\varsigma\overline{\chi(c_\star)}$, so $\hat{\bar h}(\chi)=-\varsigma$ for $\chi\ne1$ and $\hat{\bar h}(1)=\varsigma/(N-1)$. Thus $H$ is determined by $\varsigma$, the space of such $H$ is one-dimensional, and the minimal face of $\Toep^+_m$ containing $T_\star$ is the ray through $T_\star$.

(iii) Extremality follows from (ii) and Theorem~\ref{thm:faces}. By Lemma~\ref{lem:prodvec}, a vector in $\Rg(\rho_{T_\star})$ that is a product across $S|S^c$ has the form $u_1^{\otimes m_1}\otimes u_2^{\otimes m_2}$ with $u_j=(p_j,q_j)$, that is $Vx$ with $x_\alpha=\prod_jp_j^{m_j-\alpha_j}q_j^{\alpha_j}$, and $x\in\Rg(T_\star)$. Suppose first $p_1p_2\ne0$, and normalise $p_j=1$, $z_j=q_j$. Orthogonality to $e_\alpha-e_{\alpha'}$ gives $z_1^2=z_2^2$ and $z_1^{m_1}z_2^{m_2}=1$, hence $z_1^M=\pm1$, $|z_1|=|z_2|=1$ and $z=e^{i\theta}$ with $\theta\in\Lat^\perp$. Orthogonality to the vector $x$ used in (ii) gives $\sum_{g\ne c_\star}\chi_\theta(g)=0$, but this sum equals $N-1$ for $\chi_\theta=1$ and $-\chi_\theta(c_\star)\ne0$ otherwise. If $p_1=0$, then $x$ is supported on $\alpha_1=m_1$; orthogonality to $e_{(m_1-2,\beta)}-e_{(m_1,\beta-2)}$ forces $x_{(m_1,\beta)}=0$ for $\beta\le m_2-2$, hence $p_2=0$ and $x$ is a multiple of $e_{(m_1,m_2)}$, which is not orthogonal to $e_{(0,0)}-e_{(m_1,m_2)}$. The case $p_2=0$ is symmetric.
\end{proof}

Since the sum of $e^{i\langle d,\theta\rangle}$ over $\theta\in\Lat^\perp$ is $N\mathbf 1_\Lat(d)$, the matrices of Theorem~\ref{thm:lattice} are $T_\mu=\frac\mu N\sum_{\theta\in\Lat^\perp}a(\theta)a(\theta)^*-a(0)a(0)^*$, combinations of the $N$ atoms at the points of $\Lat^\perp$ with a single negative coefficient. In the setting of Theorem~\ref{thm:twogroups} these $N=2M$ vectors $a(\theta)$ span a space of dimension $N-1$, because their entries depend only on the class of the index and the characters are linearly independent on the $N-1$ classes met by $\Lambda$. This is the configuration of $p+2$ points in a projective space of dimension $p=2M-2$ used by Blekherman, Sinn and Velasco to bound the Hankel index from above \cite[Theorem~4.1]{BSV2017}, and $T_\star$ is the corresponding extreme ray, made explicit.

For $m=(2,2)$ the state $\rho_{T_\star}$ is a four-qubit PPT state of rank six that is entangled exactly across the two bipartitions $A_1B_1|A_2B_2$ and $A_1B_2|A_2B_1$. Gabardo's family \cite[Theorem~6.1]{Gabardo1998} has rank $m_1m_2+m_1+m_2-2$, which coincides with $2(m_1+m_2)-2$ only for $m_1=m_2=2$; Theorem~\ref{thm:rank} shows that $T_\star$ has the smallest rank among the non-decomposable PSD Toeplitz matrices of this configuration.

\subsection{Classification of configurations}

\begin{theorem}\label{thm:classif}
\begin{enumerate}
\item[(a)] Every PSD Toeplitz matrix on $\Lambda_m$ is decomposable, equivalently every Toeplitz state of the configuration $m$ is fully separable, if and only if $k=1$, or $k=2$ and $\min(m_1,m_2)=1$.
\item[(b)] In all other configurations the cone $\Dec_m$ is not a spectrahedral shadow; that is, full separability of Toeplitz states cannot be expressed by any finite semidefinite programme with auxiliary variables.
\end{enumerate}
\end{theorem}

Part (a) is not new: for $k=2$ it follows from the work of Rudin \cite{Rudin1963} and of Bakonyi and Naevdal \cite{BakonyiNaevdal2000} on the extension property of finite subsets of $\Z^2$, see also \cite{BakonyiWoerdeman2011}. We include the short proof because it follows directly from Theorem~\ref{thm:cuts}.

\begin{proof}
(a) Suppose $k=1$ and $m_1\ge2$, and let $S$ split the group. Theorem~\ref{thm:cuts}(a), whose hypothesis holds trivially for $k=1$, gives separability across $S|S^c$, and Theorem~\ref{thm:cuts}(c) then gives $T\in\Dec$. For $k=1$, $m_1=1$ a PSD $2\times2$ Toeplitz matrix is decomposable, as noted after \eqref{eq:Phi}. Suppose $k=2$ and $m_2=1$. If $m_1\ge2$, let $S$ consist of one qubit of group~1 and the qubit of group~2; if $m_1=1$, let $S$ consist of the qubit of group~1. In both cases Theorem~\ref{thm:cuts}(a) applies with $j_0=1$ and Theorem~\ref{thm:cuts}(c) applies, so $T\in\Dec$. For $k=1$ the argument rests on Lemma~\ref{lem:block} with blocks of size one, which is the Carath\'eodory--Toeplitz theorem itself, so it gives no new proof of that theorem; for $k=2$, $m_2=1$ it gives a quantum-information proof of Gabardo's result for the grids $\{0,\dots,m_1\}\times\{0,1\}$ \cite[Theorem~5.10]{Gabardo1998}.

Conversely, for $k=2$ and $m_1,m_2\ge2$, Theorem~\ref{thm:twogroups} gives a counterexample. For $k\ge3$ the groups can be divided into two classes with at least two qubits each, unless $m=(1,1,1)$; in the first case Lemma~\ref{lem:merge} transports the matrix $T_\star$ of the two merged groups, and for $m=(1,1,1)$ the three-qubit UPB states of Section~\ref{sec:upb} or Gabardo's example \cite[Theorem~6.3]{Gabardo1998} give PSD Toeplitz matrices of rank four that are not decomposable.

(b) The dual of $\Dec_m$ is the cone of trigonometric polynomials $q(\theta)=\sum_{d\in D}q_de^{i\langle d,\theta\rangle}$ that are nonnegative on $\T^k$, and the dual of $\Toep^+_m$ is the closed cone of Hermitian sums of squares $\sum_i|p_i(\theta)|^2$ with $p_i$ supported on $\Lambda$. By (a) these cones differ, so there is a nonnegative $q$ that is not such a sum of squares. Under the Cayley substitution $e^{i\theta_j}=(1+ix_j)/(1-ix_j)$, $\tilde q(x)=\prod_j(1+x_j^2)^{m_j}q(\theta(x))$ is a real polynomial, nonnegative on $\R^k$, with support in the finite downward closed set $A=\prod_j\{0,\dots,2m_j\}$. If $\tilde q=\sum f_i^2$, the Newton polytope of each $f_i$ is contained in $\prod_j[0,m_j]$, and substituting back gives a Hermitian sum of squares representation of $q$. So $\tilde q$ is not a sum of squares, and Fawzi's theorem \cite[Theorem~3]{Fawzi2021}, whose hypotheses are a finite downward closed support set and a polynomial with this support that is nonnegative on $\R^k$ but not a sum of squares, shows that $C_A=\overline{\conv}\{(x^u)_{u\in A}:x\in\R^k\}$ is not a spectrahedral shadow; the result builds on Scheiderer's work \cite{Scheiderer2018}. The same substitution defines an invertible linear map $\mathcal C$ with $\mathcal C((x^u)_u)=\prod_j(1+x_j^2)^{m_j}\,a(\theta(x))a(\theta(x))^*$. The atoms $a(\theta)a(\theta)^*$ with $\theta_j=\pi$ for some $j$ have no preimage $x\in\R^k$; their preimages under $\mathcal C$ have vanishing coefficient $y_0$ of the constant monomial and are limits of $(x^u)_u/\prod_{j:\theta_j=\pi}x_j^{2m_j}$ as these $x_j$ tend to infinity. Let $R$ be the convex cone that they generate. Since $\Dec_m$ is generated by finitely many atoms at a time, $\mathcal C^{-1}(\Dec_m)\cap\{y_0=1\}=\conv\{(x^u)_u\}+R$. This set is closed, as a section of a closed cone, and it lies between $\conv\{(x^u)_u\}$ and its closure, because for a generator $r$ of $R$ and $t\ge0$ the point $p+tr$ is the limit of the convex combinations $(1-\epsilon)p+\epsilon(x^u)_u$ with $\epsilon=t/\prod_{j:\theta_j=\pi}x_j^{2m_j}$. Hence $C_A=\mathcal C^{-1}(\Dec_m)\cap\{y_0=1\}$. Linear preimages and affine sections of spectrahedral shadows are spectrahedral shadows, so $\Dec_m$ is not one.
\end{proof}

Fawzi \cite{Fawzi2021} showed that the set of all separable states of a bipartite system has no semidefinite representation except in dimensions $2\times2$ and $2\times3$. Theorem~\ref{thm:classif}(b) shows that the same holds within the much smaller Toeplitz class, already for four qubits with $m=(2,2)$ and for three qubits with $m=(1,1,1)$.

\subsection{Rank thresholds}

For a configuration $m$ with $k\ge2$ put
\[
r(m)=\min\Bigl(\{2(m_a+m_b)-2:\ a\ne b,\ m_a,m_b\ge2\}\cup\{2(m_a+m_b+m_c)-2:\ a<b<c\}\Bigr),
\]
with $\min\emptyset=\infty$.

\begin{theorem}[Rank thresholds]\label{thm:rank}
\begin{enumerate}
\item[(a)] If $r(m)<\infty$ there is $T\in\Toep^+_m\setminus\Dec_m$ with $\rank T=r(m)$.
\item[(b)] Every $T\in\Toep^+_m$ with $\rank T<r(m)$ is decomposable, and hence every Toeplitz state of rank less than $r(m)$ is fully separable, in the following cases:
\begin{itemize}
\item $k=2$, $m_1,m_2\ge2$, where $r(m)=2(m_1+m_2)-2$;
\item $m=(1,\dots,1)$ with $k\ge3$, where $r(m)=4$;
\item $k=3$ and $m$ a permutation of $(a,1,1)$ with $a\ge2$, of $(2,2,1)$ or of $(2,2,2)$, where $r(m)=2a+2$, $6$ and $6$;
\item $m$ a permutation of $(3,2,1)$, where $r(m)=8$ (computer-assisted).
\end{itemize}
\item[(c)] For $m=(3,2,2)$, every $T\in\Toep^+_m$ with $\rank T\le4$ is decomposable, while $r(m)=6$ (computer-assisted).
\end{enumerate}
\end{theorem}

\begin{proof}[Proof of (a)]
If the minimum is attained at a pair $a,b$, take $T_\star$ of Theorem~\ref{thm:twogroups} on the levels $a,b$ and tensor it with $a(0)a(0)^*$ on the remaining levels. The result is a PSD Toeplitz matrix of the same rank, and it is not decomposable because its principal submatrix with the remaining indices equal to zero is $T_\star$. If the minimum is attained at a triple, divide it into two classes with at least two qubits each when possible and transport $T_\star$ by Lemma~\ref{lem:merge}, which preserves the rank; if $m_a=m_b=m_c=1$, use a rank-four example on $\{0,1\}^3$ (Section~\ref{sec:upb}). Then tensor with $a(0)a(0)^*$ as before.
\end{proof}

Part (b) is proved in Appendix~\ref{app:rank}. For two groups it is a result of Blekherman, Sinn and Velasco on toric surfaces \cite[Examples~6.16 and~6.18]{BSV2017}: every functional of rank at most $2a+2b-3$ on forms of bidegree $(a,b)$ comes from a measure. The cases with three groups are new. The threshold $r(m)$ is defined combinatorially; by Lemma~\ref{lem:cayley} the smallest rank of a non-decomposable PSD Toeplitz matrix is the Hankel index of \cite{BSV2017} of the embedded variety, and Theorem~\ref{thm:rank} shows that the two agree in the cases listed in (b). The proof identifies $\Toep^+_m$ with the dual cone of sums of squares on the real points of $(\mathbb P^1)^k$ embedded by $\mathcal O(m_1,\dots,m_k)$, and uses the theorem of Blekherman, Sinn and Velasco \cite{BSV2017} that property $N_{2,p}$ forces all extreme rays of rank at most $p$ to have rank one. The required syzygy properties follow from the work of Gallego and Purnaprajna on rational surfaces \cite{GallegoPurnaprajna2001}, of Rubei on Segre embeddings \cite{Rubei2002}, and, for $m=(2,2,2)$, from the Noether--Lefschetz theorem of Moishezon \cite{Moishezon1967}, the computation of the Clifford index of curves on K3 surfaces by Green, Lazarsfeld and Knutsen \cite{GreenLazarsfeld1987,Knutsen2001}, and Green's conjecture for such curves, proved by Aprodu and Farkas \cite{AproduFarkas2011}. For $m=(1,\dots,1)$ part (b) also follows from the separability of multipartite PPT states of rank at most three \cite{ChenDjokovic2013}. In all cases where we can compute it, the threshold equals $r(m)$, and we conjecture that this holds in general. Sufficient rank conditions for Vandermonde decompositions of PSD multilevel Toeplitz matrices were obtained in signal processing \cite{YangXieStoica2016,AnderssonCarlsson2017}; in \cite{YangXieStoica2016} the condition is that the rank be lower than the dimension of each Toeplitz block. Theorem~\ref{thm:rank}(b) gives sharp conditions for the configurations listed there and makes no claim for other configurations.

Table~\ref{tab:rank} summarises the status of the rank thresholds.

\begin{table}[t]
\centering
\small
\begin{tabular}{lll}
\toprule
configuration $m$ & $r(m)$ & status of ``rank $<r(m)$ implies $T\in\Dec_m$''\\
\midrule
$k=2$, $m_1,m_2\ge2$ & $2(m_1+m_2)-2$ & proved\\
$(1,\dots,1)$, $k\ge3$ & $4$ & proved\\
$(a,1,1)$, $a\ge2$ & $2a+2$ & proved\\
$(2,2,1)$ & $6$ & proved\\
$(2,2,2)$ & $6$ & proved\\
$(3,2,1)$ & $8$ & proved, computer-assisted\\
$(3,2,2)$ & $6$ & proved for rank $\le4$, computer-assisted; rank $5$ open\\
other configurations & $r(m)$ & conjectured\\
\bottomrule
\end{tabular}
\caption{Rank thresholds of Theorem~\ref{thm:rank}, up to permutations of $m$. In all cases $r(m)$ is attained by Theorem~\ref{thm:rank}(a).}
\label{tab:rank}
\end{table}

At the threshold, the range decides.

\begin{proposition}\label{prop:rangecrit}
Suppose every $T\in\Toep^+_m$ of rank less than $r_0$ is decomposable, and let $T\in\Toep^+_m$ have rank $r_0$. Then $T\in\Dec_m$ if and only if $\Rg(T)$ contains a vector $a(\theta)$. If $T\notin\Dec_m$, then $T$ spans an extreme ray of $\Toep^+_m$, and $\rho_T$ spans an extreme ray of $\mathcal P_S$ for every bipartition $S|S^c$ that splits all groups.
\end{proposition}

\begin{proof}
If $a(\theta)\in\Rg(T)$, then $T-\varepsilon a(\theta)a(\theta)^*$ is PSD of rank $r_0-1$ for $\varepsilon=(a(\theta)^*T^+a(\theta))^{-1}$, hence decomposable, and so is $T$. The converse is clear. If $T\notin\Dec_m$ and the minimal face of $T$ had dimension at least two, there would be a Hermitian Toeplitz $H\not\parallel T$ with $\Rg(H)\subseteq\Rg(T)$ and $h(0)=0$; moving from $T$ along $\pm H$ until the boundary of the face is reached writes $T$ as a convex combination of two PSD Toeplitz matrices of rank less than $r_0$, which are decomposable, a contradiction. The last statement is Theorem~\ref{thm:faces}.
\end{proof}

The same subtraction argument gives decompositions with the smallest possible number of atoms.

\begin{corollary}\label{cor:vandermonde}
In every configuration covered by Theorem~\ref{thm:rank}(b), every $T\in\Toep^+_m$ of rank $r_0<r(m)$ can be written as $T=\sum_{l=1}^{r_0}p_l\,a(\theta_l)a(\theta_l)^*$ with $p_l>0$, that is, it has a multi-level Vandermonde decomposition with exactly $r_0$ atoms.
\end{corollary}

\begin{proof}
By Theorem~\ref{thm:rank}(b), $T\in\Dec_m$, so the atoms of any decomposition of $T$ lie in $\Rg(T)$. For such an atom $a(\theta)$ and $\varepsilon=(a(\theta)^*T^+a(\theta))^{-1}$, the matrix $T-\varepsilon a(\theta)a(\theta)^*$ is a PSD Toeplitz matrix of rank $r_0-1$, and induction on $r_0$ gives the claim.
\end{proof}

For two groups the corollary is a consequence of \cite[Example~6.18]{BSV2017}. For $m=(2,2)$, Corollary~\ref{cor:vandermonde} covers all $9\times9$ two-level Toeplitz matrices of rank at most $5$, whereas the condition of \cite{YangXieStoica2016}, rank lower than the dimension of each Toeplitz block, covers ranks up to $2$.

\section{Discussion}\label{sec:discussion}

We have studied three questions about bound entanglement in the class of Toeplitz states, where positivity under partial transposition is automatic and full separability is the truncated trigonometric moment problem on the torus. \emph{Which states:} for three qubits, an algebraic argument independent of the moment problem shows that the class contains, up to invertible local operators, all rank-four PPT entangled states with nonzero Lorentz invariant, among them the UPB states and the Bell-nonlocal state of V\'ertesi and Brunner, and the states with vanishing invariant whose quadrilateral is a rhombus. \emph{How robust:} for the UPB states the moment problem gives exact white-noise thresholds, decompositions into eight product states at the threshold, and a witness in closed form that detects every entangled state of the family with eight local Pauli settings. \emph{How small:} the rank below which every Toeplitz state is fully separable is governed by the Hankel index of a product of projective lines; for two groups the extremal states of Theorem~\ref{thm:twogroups} make the sharp bound of Blekherman, Sinn and Velasco explicit, and for three groups Theorem~\ref{thm:rank} determines it in several configurations through syzygies of Segre--Veronese varieties. Along the way, the bipartitions that can carry entanglement within the class are determined by the arrangement of the groups, and across bipartitions that split all groups the minimal faces of the PPT cone at Toeplitz states are minimal faces of the Toeplitz cone.

Several questions remain open.

\begin{enumerate}
\item \emph{The most robust bound entangled states of rank four.} Which three-qubit PPT entangled state of rank four tolerates the most white noise? Table~\ref{tab:thresholds} shows that the answer is not a UPB state; the largest value known to us is $1-1/\sqrt2$, attained by a GHZ-diagonal state of type~II, which is covered by the criterion of G\"uhne and of Han and Kye \cite{Guhne2011,HanKye2017GHZ}, and among type~I states a local search reaches $0.2635$. Whether a state of type~I can exceed $1-1/\sqrt2$ is open.
\item \emph{Other noise models.} Local depolarising noise maps Toeplitz states to Toeplitz states and commutes with the symmetries used in Section~\ref{sec:noise}, so the reduction of Section~\ref{sec:noise} applies to it as well. Exact thresholds for UPB states under local depolarising noise would complement Theorem~\ref{thm:upb}.
\item \emph{Exact thresholds beyond symmetry.} The analytic proof of Theorem~\ref{thm:upb} uses the Klein symmetry of UPB states. For general type~I states, such as the V\'ertesi--Brunner state with $\lambda^*\approx0.0917$, the convex hull of the torus in the full $26$-dimensional moment space would be needed.
\item \emph{Rank thresholds.} We conjecture that the threshold rank equals $r(m)$ for every configuration. The lower bound obtained from property $N_p$ need not be sharp in general: Blekherman, Chen and Jung \cite{BlekhermanChenJung2024} give examples among projections of rational normal curves, which are not arithmetically Cohen--Macaulay. Blekherman, Sinn and Velasco prove that the Hankel index equals the syzygy bound in several situations \cite[Theorem~4.8]{BSV2017}. The Segre--Veronese varieties considered here are arithmetically Cohen--Macaulay, so the examples of \cite{BlekhermanChenJung2024} do not rule out that syzygies suffice for them. Among three groups, the first configurations not covered by Theorem~\ref{thm:rank} are $m=(3,2,2)$, where it remains to exclude non-decomposable matrices of rank five, and $m=(a,b,1)$ with $a\ge b\ge2$ other than $(2,2,1)$ and $(3,2,1)$. By Appendix~\ref{app:rank}, the conjecture would follow from the vanishing of the Koszul cohomology groups $K_{p,2}$ of the corresponding Segre--Veronese varieties for $p<r(m)$.
\item \emph{Bipartitions with intact groups.} When a bipartition splits some groups and leaves a group with $m_j\ge2$ intact, and Theorem~\ref{thm:cuts}(a) does not apply, separability across it corresponds to membership in a cone generated by $\bigotimes_{j\in J}\proj{\theta_j}^{\otimes m_j}\otimes\varphi\otimes\chi$ with arbitrary symmetric $\varphi$ and $\chi$. This cone is strictly larger than $\Dec_m$. For example, for $m=(2,2,2)$ let $T$ be $T_\star$ on levels $1$ and $3$ tensored with $a(0)a(0)^*$ on level $2$, and consider a bipartition that splits groups $1$ and $2$ and leaves group $3$ on one side. Then $\rho_T$ is the product of $\rho_{T_\star}$ on groups $1,3$, which is separable across this bipartition by Theorem~\ref{thm:cuts}(a), and a product state on group $2$; but $T\notin\Dec_m$. A description of this cone in terms of matrix-valued moment problems would complete Theorem~\ref{thm:cuts}.
\item \emph{Entanglement across all non-trivial bipartitions.} Example~\ref{ex:tetracode} gives a four-qubit Toeplitz state entangled across every bipartition that can carry entanglement. We do not know whether such states exist for every number of groups.
\item \emph{Qudits.} Mixtures of products of equatorial qudit states $d^{-1/2}\sum_le^{i\theta_l}\ket l$ lead to moment problems on higher-dimensional tori. Whether the dictionary of Section~\ref{sec:dict} has an analogue for such states, and how the classification of varieties of minimal degree used by Blekherman, Smith and Velasco \cite{BSmV2016} enters, is open.
\end{enumerate}

Code for all numerical and computer-assisted results is available as described in Appendix~\ref{app:comp}.

\section*{Author contributions}
L.~Q.\ and T.~Z.\ jointly carried out the research and wrote the manuscript. The large language model Claude (Anthropic) was used to assist with literature searches, the development and checking of proofs, the numerical and symbolic computations, and the drafting of the text; all results were verified by the authors.

\bibliographystyle{quantum}
\bibliography{refs}

\appendix
\section{Proofs for Sections~\ref{sec:three} and~\ref{sec:noise}}\label{app:proofs}

\begin{proof}[Proof of Lemma~\ref{lem:det}]
Both sides are polynomials in the sixteen real coordinates of $M$, and the identity is verified by direct symbolic expansion (script \texttt{det\_identity.py}, which uses the partial transpose on the first qubit; the two partial transposes differ by a full transposition and have the same determinant). A conceptual proof is as follows. For $A,B\in\SL(2,\C)$ and $M'=(A\otimes B)M(A\otimes B)^{*}$ one has $\det M'=\det M$, $M'^{\Gamma}=(\bar A\otimes B)M^{\Gamma}(\bar A\otimes B)^{*}$, and $M'\widetilde{M'}=(A\otimes B)M\widetilde M(A\otimes B)^{-1}$, so both sides are invariant. If $M=\sum_i\lambda_iB_i$ is diagonal in the Bell basis then $\widetilde M=M$ and $M^{\Gamma}$ has eigenvalues $\sigma/2-\lambda_i$ with $\sigma=\sum_i\lambda_i$, and \eqref{eq:det} becomes the identity $\prod_i(\sigma-2\lambda_i)=8e_4-2p_4+p_2^2$ between symmetric functions of the $\lambda_i$. Since a generic positive definite $M$ can be brought to Bell-diagonal form by such transformations, by the normal form of Verstraete, Dehaene and De Moor \cite{VerstraeteDehaeneDeMoor2001}, the identity holds on an open set and hence everywhere.
\end{proof}

\begin{proof}[Proof of Lemma~\ref{lem:key}]
The operator $\sigma=\rho^{\Gamma_2}$ is a PPT entangled state of rank four, so by Proposition~\ref{prop:CD} $\sigma=\sum_{k}X_k'\otimes W_k$ with $W_k=\proj{w_k}$ and $w_1,\dots,w_4$ linearly independent. On the other hand $\sigma=\sum_iX_i\otimes U_i^{\Gamma}$. Let $(H_i)$ be the basis of Hermitian matrices dual to $(X_i)$. Then
\[
U_i^{\Gamma}=\sum_kC_{ik}W_k,\qquad C_{ik}=\tr(H_iX_k')\in\R,
\]
and $C$ is invertible because the $U_i^{\Gamma}$ are linearly independent. We may rescale $u_i\mapsto c_iu_i$, $x_i\mapsto x_i/|c_i|$ so that $G_{ii}=1$; this multiplies the rows of $C$ by positive numbers. Write $\tilde U=[u_1,\dots,u_4]$ and $\tilde W=[w_1,\dots,w_4]$.

\emph{Step 1: a quartic identity.} For $a\in\R^4$ put $M(a)=\sum_ia_iU_i=\tilde U A\tilde U^{*}$ with $A=\diag(a)$. Then $M(a)^{\Gamma}=\tilde W\diag(a^{\tp}C)\tilde W^{*}$, $M\widetilde M$ is similar to $A\,\overline G\,A\,G$, and $\det M(a)=|\det\tilde U|^2a_1a_2a_3a_4=|\det G|\,a_1a_2a_3a_4$ because $\det G=(\det\tilde U)^2\det F$ and $\det F=1$. Lemma~\ref{lem:det} gives
\begin{equation}\label{eq:quartic}
\Xi(a):=8|\det G|\,a_1a_2a_3a_4-2\tr\bigl[(A\overline GAG)^2\bigr]+\bigl(\tr A\overline GAG\bigr)^2=c_0\prod_{k=1}^{4}l_k(a),
\end{equation}
with $c_0=16|\det\tilde W|^2>0$ and $l_k(a)=\sum_iC_{ik}a_i$.

\emph{Step 2: signs.} $U_i^{\Gamma}=\tilde W\diag(C_{i1},\dots,C_{i4})\tilde W^{*}$ has the inertia of row $i$ of $C$. Since $u_i$ is entangled, $U_i^{\Gamma}$ has three positive eigenvalues and one negative. Hence each row of $C$ has exactly one negative entry, in a column denoted $n(i)$, and three positive entries.

\emph{Step 3: positivity.} For $z\in\C^2$ the operator $(\bra z\otimes I)\sigma(\ket z\otimes I)=M(a)^{\Gamma}$ with $a_i=|\langle z,x_i\rangle|^2$ is positive semidefinite, so $l_k(a)\ge0$ for all $k$. For generic $z$ all $a_i$ are positive. Hence no column of $C$ has only negative entries.

\emph{Step 4: pairs.} Fix $i\ne j$ and set all variables except $a_i=\alpha$, $a_j=\beta$ to zero. With $r=G_{ij}$, $u=|r|^2$, $v=|1-r^2|$ and $t=\alpha^2+2u\alpha\beta+\beta^2$, a direct computation from \eqref{eq:quartic} gives
\[
\Xi=-(t-2v\alpha\beta)(t+2v\alpha\beta)=c_0\prod_k(C_{ik}\alpha+C_{jk}\beta).
\]
The right-hand side is a nonzero product of real linear forms, so the quadratic form $\phi=\alpha^2+2(u-v)\alpha\beta+\beta^2$ has real roots. Since $|u-v|\le1$ by the triangle inequality, $(u-v)^2=1$. There are two cases.
\begin{itemize}
\item[(i)] $v=u+1$, that is $r\in i\R$. Then $\phi=(\alpha-\beta)^2$ and $\psi=t+2v\alpha\beta=\alpha^2+2(2u+1)\alpha\beta+\beta^2$, so $\Xi$ has the double root $\alpha/\beta=1$ and two negative roots.
\item[(ii)] $v=u-1$, that is $r\in\R$ and $|r|\ge1$. Then $\phi=(\alpha+\beta)^2$ and $\psi=\alpha^2+2(2u-1)\alpha\beta+\beta^2$, and all roots of $\Xi$ are negative.
\end{itemize}
The root of the linear form of column $k$ is $\alpha/\beta=-C_{jk}/C_{ik}$. If $n(i)\ne n(j)$, the columns $n(i)$ and $n(j)$ give positive roots, so case (i) holds and both roots equal $1$:
\begin{equation}\label{eq:F1}
C_{j,n(i)}=-C_{i,n(i)},\qquad C_{i,n(j)}=-C_{j,n(j)} .
\end{equation}
If $n(i)=n(j)$ all roots are negative and case (ii) holds.

\emph{Step 5: the fibres of $n$.} By Step 3 the map $n$ is not constant. Suppose $n$ takes exactly two values $n_1,n_2$. By \eqref{eq:F1}, column $n_1$ equals $-c$ on $n^{-1}(n_1)$ and $c$ on $n^{-1}(n_2)$, and column $n_2$ equals $c'$ on $n^{-1}(n_1)$ and $-c'$ on $n^{-1}(n_2)$, with $c,c'>0$. The two columns are proportional, contradicting the invertibility of $C$.

Suppose $n$ takes exactly three values, with $n(1)=n(2)=n_1$, $n(3)=n_2$, $n(4)=n_3$. By \eqref{eq:F1} the rows $1$ and $2$ of $C$ agree in the columns $n_1,n_2,n_3$. For the pair $(1,2)$ three of the four linear forms are therefore multiples of $\alpha+\beta$. We are in case (ii), so $-1$ is a root of $\psi$, which forces $u=1$ and $\psi=(\alpha+\beta)^2$. Then the fourth linear form is also a multiple of $\alpha+\beta$, so rows $1$ and $2$ of $C$ coincide, a contradiction.

Hence $n$ is a bijection. By \eqref{eq:F1} every column $n(k)$ takes the value $-c_k$ in row $k$ and $c_k$ in the other rows. For a pair $(i,j)$ the two columns $n(k)$, $k\ne i,j$, give the linear forms $c_k(\alpha+\beta)$, so $-1$ is a double root of $\Xi$. We are in case (i), hence $-1$ is a root of $\psi=\alpha^2+2(2u+1)\alpha\beta+\beta^2$, which forces $u=0$. Thus $G_{ij}=0$ for all $i\ne j$.
\end{proof}

\begin{proof}[Proof of Lemma~\ref{lem:V}]
With $\xi_1=\cos^2x$, $\xi_2=\cos^2y$, $\xi_3=\cos^2z$ one has $\mu_1^2=(1-\xi_1)(1-\xi_2)$, $\mu_2^2=\xi_1(1-\xi_3)$, $\mu_3^2=\xi_2\xi_3$, $\mu_4^2=\xi_1(1-\xi_2)\xi_3$, $\mu_5^2=(1-\xi_1)\xi_2(1-\xi_3)$, whose sum is $1$, and $\mu_4\mu_5=\mu_1\mu_2\mu_3$ is immediate from \eqref{eq:M}. Conversely let $m\in\mathcal M$ and suppose first that $\mu_1^2,\mu_2^2,\mu_3^2<1$. Put
\[
\xi_1=\frac{\mu_2^2+\mu_4^2}{1-\mu_3^2},\qquad\xi_2=\frac{\mu_3^2+\mu_5^2}{1-\mu_2^2},\qquad\xi_3=\frac{\mu_3^2+\mu_4^2}{1-\mu_1^2},
\]
which lie in $[0,1]$. Using $|m|=1$ and $\mu_4^2\mu_5^2=\mu_1^2\mu_2^2\mu_3^2$ one verifies the five equations for the squares; for instance $(1-\xi_1)(1-\xi_2)=(\mu_1^2+\mu_5^2)(\mu_1^2+\mu_4^2)/((1-\mu_3^2)(1-\mu_2^2))=\mu_1^2$. It remains to choose the signs of $\sin x,\cos x,\sin y,\cos y,\sin z,\cos z$, which can be prescribed independently wherever these numbers are nonzero. If $\mu_4\ne0$, choose the signs of $\cos x$, $\sin y$ and $\cos z$ so that $\mu_4$ has the right sign, and then the signs of $\sin x$, $\sin z$ and $\cos y$ from $\mu_1$, $\mu_2$ and $\mu_3$; the sign of $\mu_5$ is then correct because $\mu_4\mu_5=\mu_1\mu_2\mu_3$. If $\mu_4=0$, then $\cos x$, $\sin y$ or $\cos z$ vanishes. In these three cases the sign of $\sin z$, of $\sin x$ or of $\cos y$, respectively, occurs among the coordinates only in $\mu_5$ and in a coordinate that vanishes, so after matching $\mu_1,\mu_2,\mu_3$ as before it can be used to match the sign of $\mu_5$. If some $m_j^2=1$ with $j\le3$, then $m=\pm e_j$, which is in the image.
\end{proof}

\begin{proof}[Proof of Lemma~\ref{lem:concave}]
The positive definite matrices with unit diagonal form the relative interior of this convex set and are dense in it. Since $\sqrt{\det}$ is continuous, concavity on this interior extends to the whole set by taking limits along segments, so it suffices to consider positive definite $\Omega$. Let $H\ne0$ be symmetric with zero diagonal and $H'=\Omega^{-1/2}H\Omega^{-1/2}$. The second derivative of $\sqrt{\det}$ at $\Omega$ in the direction $H$ is $\sqrt{\det\Omega}\,\bigl[\tfrac14(\tr H')^2-\tfrac12\tr H'^2\bigr]$. Since $\tr H=0$, $H$ is indefinite, and so is $H'$ by Sylvester's law. Replacing $H'$ by $-H'$ we may assume that its eigenvalues are $k_1,k_2\ge0$ and $-k_3\le0$. Then $2\tr H'^2-(\tr H')^2=(k_1-k_2)^2+k_3^2+2k_3(k_1+k_2)\ge0$.
\end{proof}

\section{Proof of the rank thresholds}\label{app:rank}

\subsection{Toeplitz matrices as moment functionals on \texorpdfstring{$(\mathbb P^1)^k$}{(P1)k}}

Let $X=(\mathbb P^1)^k$ over $\R$, embedded by $L=\mathcal O(m_1,\dots,m_k)$. Its homogeneous coordinate ring has $R_1$ equal to the real forms of multidegree $m$ and, since the embedding is projectively normal, $R_2$ equal to the real forms of multidegree $2m$. Following \cite{BSV2017}, see also \cite{Blekherman2012}, let $\Sigma_X\subset R_2$ be the cone of sums of squares of elements of $R_1$ and $\Sigma_X^*$ its dual, the Hankel spectrahedron. The rank of $\ell\in\Sigma_X^*$ is the rank of the quadratic form $f\mapsto\ell(f^2)$ on $R_1$.

For each factor with coordinates $(x_0,x_1)$ put $\zeta=x_0+ix_1$. The forms $\prod_j\zeta_j^{m_j+d_j}\bar\zeta_j^{m_j-d_j}$, $d\in D$, form a basis of the complexification of $R_2$, and a real form has coefficients $f_d$ with $f_{-d}=\overline{f_d}$. For a Hermitian Toeplitz matrix $T$ define $\ell_T\in R_2^*$ by
\[
\ell_T\Bigl(\prod_j\zeta_j^{m_j+d_j}\bar\zeta_j^{m_j-d_j}\Bigr)=t(-d).
\]

\begin{lemma}\label{lem:cayley}
The map $T\mapsto\ell_T$ is a linear bijection from Hermitian Toeplitz matrices on $\Lambda_m$ onto $R_2^*$. It satisfies $\ell_T\in\Sigma_X^*$ if and only if $T\ge0$, and $\rank\ell_T=\rank T$. The evaluation at a real point of $X$ with $\zeta_j=|\zeta_j|e^{i\theta_j/2}$ is $\ell_T$ for $T=\prod_j|\zeta_j|^{2m_j}\,a(-\theta)a(-\theta)^*$. In particular $\Toep^+_m\cong\Sigma_X^*$, and $\Dec_m$ corresponds to the cone generated by evaluations at real points.
\end{lemma}

\begin{proof}
For a complex form $F=\sum_{\alpha\in\Lambda}c_\alpha\prod_j\zeta_j^{\alpha_j}\bar\zeta_j^{m_j-\alpha_j}$ of multidegree $m$ one has $|F|^2=\sum_{\alpha,\beta}c_\alpha\bar c_\beta\prod_j\zeta_j^{m_j+\alpha_j-\beta_j}\bar\zeta_j^{m_j-\alpha_j+\beta_j}$ on real points, so $\ell_T(|F|^2)=c^*Tc$. Writing $F=f+ig$ with real $f,g$ gives $\ell_T(f^2)+\ell_T(g^2)=c^*Tc$, so the Hermitian form $c\mapsto c^*Tc$ is the Hermitian extension of the real quadratic form $f\mapsto\ell_T(f^2)$; both statements about positivity and rank follow. The dimensions agree, and the formula for evaluations is immediate.
\end{proof}

\begin{lemma}\label{lem:extreme}
Every PSD Toeplitz matrix of rank one is a multiple of some $a(\theta)a(\theta)^*$. If every extreme ray of $\Toep^+_m$ of rank at most $p$ has rank one, then every $T\in\Toep^+_m$ with $\rank T\le p$ is decomposable.
\end{lemma}

\begin{proof}
If $uu^*$ is Toeplitz, all $|u_\alpha|$ are equal and $u_{\alpha+e_j}\bar u_\alpha$ is independent of $\alpha$, so $u$ is a multiple of $a(\theta)$. The cone $\Toep^+_m$ is closed and pointed, so $T$ is a sum of elements of extreme rays, whose ranges lie in $\Rg(T)$ and which therefore have rank at most $p$.
\end{proof}

By Lemma~\ref{lem:cayley}, the smallest rank larger than one of an extreme ray of $\Toep^+_m$ is the Hankel index $\eta(X)$ of \cite{BSV2017}. Blekherman, Sinn and Velasco proved that if $X$ satisfies property $N_{2,p}$, then $\eta(X)\ge p+1$ \cite[Corollary~2.7]{BSV2017}. For the projectively normal varieties considered here, $N_{2,p}$ is Green's property $N_p$ \cite{Green1984,EGHP2005}, which holds if and only if the Koszul cohomology groups satisfy $K_{i,q}(X,L)=0$ for $1\le i\le p$ and $q\ge2$. Combining with Lemma~\ref{lem:extreme}:

\begin{proposition}\label{prop:Np}
If $(\mathbb P^1)^k$ embedded by $\mathcal O(m)$ satisfies $N_p$, then every $T\in\Toep^+_m$ with $\rank T\le p$ is decomposable.
\end{proposition}

\subsection{The syzygy input}

\emph{Two groups.} This case is contained in the work of Blekherman, Sinn and Velasco on toric surfaces \cite[Example~6.16]{BSV2017}, since $X$ is the toric surface of the rectangle $[0,m_1]\times[0,m_2]$, whose boundary contains $2(m_1+m_2)$ lattice points. We include the short argument for completeness. $X=\mathbb P^1\times\mathbb P^1$ is a smooth rational surface with $-K_X=\mathcal O(2,2)$, and $L=\mathcal O(m_1,m_2)$ is ample and base point free. Gallego and Purnaprajna proved that $N_p$ holds when $-K_X\cdot L\ge p+3$, for a rational surface $X$ over $\C$ and $L$ ample and base point free \cite[Theorem~1.3]{GallegoPurnaprajna2001}; this direction of their theorem needs no further hypothesis on $X$, and the assumption that $X$ be anticanonical enters only in the converse. Graded Betti numbers do not change under extension of the base field from $\R$ to $\C$, so the result applies to the real embeddings considered here. Here $-K_X\cdot L=2(m_1+m_2)$, so $N_p$ holds for $p\le2(m_1+m_2)-3$, which gives the first case of Theorem~\ref{thm:rank}(b).

\emph{Segre embeddings.} For $L=\mathcal O(1,\dots,1)$ on $(\mathbb P^1)^k$ with $k\ge3$, Rubei proved that $N_p$ holds exactly for $p\le3$ \cite{Rubei2002}. This gives the second case.

\emph{Three factors.} Let $X=(\mathbb P^1)^3$ and $L=\mathcal O(m_1,m_2,m_3)$. By the K\"unneth formula $X$ is arithmetically Cohen--Macaulay in this embedding, so its graded Betti numbers coincide with those of a general hyperplane section $Y\in|L|$, and $K_Y=\mathcal O(m_1-2,m_2-2,m_3-2)|_Y$ by adjunction.

For $m=(a,1,1)$ and $m=(2,2,1)$ the defining form of $Y$ is linear in the coordinates of the last factor, so $Y$ is birational to $\mathbb P^1\times\mathbb P^1$ and is a smooth rational surface. Intersection numbers on $X$ give $-K_Y\cdot L|_Y=(2-m_1,2-m_2,2-m_3)\cdot L^2$, which equals $2a+4$ and $8$ respectively. The restriction $L|_Y$ is very ample, and the sufficient condition of \cite[Theorem~1.3]{GallegoPurnaprajna2001}, which requires only a rational surface and an ample base point free line bundle, shows that $N_p$ holds for $p\le2a+1$ and for $p\le5$; no anticanonical divisor on $Y$ is needed.

For $m=(2,2,2)$, adjunction gives $K_Y=0$, so $Y$ is a K3 surface, and a general hyperplane section $C$ of $Y$ is a canonical curve of genus $25$ ($2g-2=L^3=48$), embedded by the complete canonical series because $X$ is arithmetically Cohen--Macaulay; hence $C$ has the same graded Betti numbers as $X$. For a canonical curve the Betti table is symmetric, $\beta_{i,j}=\beta_{g-2-i,g+1-j}$, so $K_{p,3}=0$ for $p\ne g-2$, and $K_{p,q}=0$ for $q\ge4$ because the homogeneous coordinate ring of a canonical curve has regularity three. It remains to show $K_{p,2}=0$ for $p\le5$, and since the Betti numbers do not depend on the choice of $Y$ and $C$, we may take $Y$ very general. Because $L=\mathcal O(2,2,2)$ is very ample and $h^{2,0}(Y)=1>0=h^{2,0}(X)$, Moishezon's Noether--Lefschetz theorem \cite[Sec.~7.5]{Moishezon1967} shows that $\mathrm{Pic}\,Y$ and $\mathrm{Pic}\,X$ have the same rank for very general $Y$. The image of $\mathrm{Pic}\,X$ is primitive: by the theorem of Andreotti and Frankel \cite{AndreottiFrankel1959}, the affine variety $X\setminus Y$ has the homotopy type of a CW complex of real dimension at most three, so by Lefschetz duality $H^i(X,Y;\Z)\cong H_{6-i}(X\setminus Y;\Z)$ vanishes for $i=2$ and is torsion-free for $i=3$, and the restriction $H^2(X,\Z)\to H^2(Y,\Z)$ is injective with torsion-free cokernel. Hence $\mathrm{Pic}\,X\to\mathrm{Pic}\,Y$ is an isomorphism for very general $Y$; the same statement for generic surfaces of tridegree $(2,2,2)$ is Lemma~3.8 of Cantat and Oguiso \cite{CantatOguiso2011}, who derive it from the Noether--Lefschetz theorem. Thus $\mathrm{Pic}\,Y$ is generated by the restrictions $H_1,H_2,H_3$ of the three rulings, with $H_i^2=0$ and $H_i\cdot H_j=2$ for $i\ne j$, and $L|_Y=2(H_1+H_2+H_3)$. By Green and Lazarsfeld \cite{GreenLazarsfeld1987}, the Clifford index of a smooth curve $C\in|L|_Y|$ satisfies $\mathrm{Cliff}(C)\le\lfloor(g-1)/2\rfloor$, and if the inequality is strict there is an effective divisor $E$ on $Y$ with $h^0(E)\ge2$, $h^0(L-E)\ge2$ and $\mathrm{Cliff}(C)=E\cdot(L-E)-2$ \cite[(2.3)]{Martens1989}, \cite[Lemma~8.3]{Knutsen2001}. Conversely, every such $E$ gives $\mathrm{Cliff}(C)\le\mathrm{Cliff}(\mathcal O_C(E))\le E\cdot(L-E)-2$. Hence $\mathrm{Cliff}(C)=\min\{\lfloor(g-1)/2\rfloor,\mu\}$, where $\mu$ is the minimum of $E\cdot(L-E)-2$ over these divisors. For $E=aH_1+bH_2+cH_3$ with integers $a,b,c$, which may be negative, put $\sigma_1=a+b+c$ and $\sigma_2=ab+bc+ca$; then $E\cdot L=8\sigma_1$, $E^2=4\sigma_2$ and $E\cdot(L-E)-2=8\sigma_1-4\sigma_2-2$. In particular $E^2\equiv0\pmod4$, so $Y$ contains no smooth rational curves. If $h^0(E)\ge2$ and $h^0(L-E)\ge2$, then $E$ and $L-E$ are nonzero and effective, and since $L$ is ample, $0<8\sigma_1<48$ and $1\le \sigma_1\le5$. Since $\sigma_1^2-3\sigma_2=\tfrac12\bigl((a-b)^2+(b-c)^2+(c-a)^2\bigr)\ge0$ for all real $a,b,c$, we get $8\sigma_1-4\sigma_2-2\ge8\sigma_1-\tfrac43\sigma_1^2-2>4$, and $8\sigma_1-4\sigma_2-2\equiv2\pmod4$, so $E\cdot(L-E)-2\ge6$, with equality for $E=H_1$. Hence $\mathrm{Cliff}(C)=\min\{12,6\}=6$. Green's conjecture, $K_{p,2}(C,K_C)=0$ for $p<\mathrm{Cliff}(C)$, holds for every smooth curve on a K3 surface \cite{AproduFarkas2011}. Therefore $K_{p,2}=0$ for $p\le5$, and $N_p$ holds for $p\le5$. The computation in Table~\ref{tab:koszul} confirms this vanishing independently.

In each case the bound $r(m)$ of Theorem~\ref{thm:rank}(a) matches. This proves the third item of Theorem~\ref{thm:rank}(b); the thresholds are sharp by Theorem~\ref{thm:rank}(a).

\emph{Computer-assisted cases.} The group $\mathrm{SL}_2(\C)^3$ acts on $K_{p,q}(X,L)$, which is concentrated in multidegree $(p+q)m$. Every irreducible representation occurring there contains a nonzero vector of the central weight, that is of multidegree $\lfloor(p+q)m/2\rfloor$ in the coordinates $x_1^{(j)}$. It therefore suffices to show that the central weight space of $K_{p,q}$ vanishes. We computed this weight space as the homology of the Koszul complex $\Lambda^{p+1}R_1\otimes R_{q-1}\to\Lambda^pR_1\otimes R_q\to\Lambda^{p-1}R_1\otimes R_{q+1}$ by exact sparse elimination over $\mathbb F_P$, $P=2^{31}-1$. Since ranks over $\mathbb F_P$ do not exceed ranks over $\mathbb Q$, vanishing over $\mathbb F_P$ implies vanishing over $\mathbb Q$. Table~\ref{tab:koszul} lists every case computed, with the central multidegree, the dimensions of the three terms, the ranks of the two maps and the running time.

For $m=(2,2,2)$ the computation gives $K_{p,2}=0$ for $1\le p\le5$, in agreement with the argument above. For $m=(3,2,1)$ it gives $K_{p,2}=0$ for $1\le p\le7$. Since $H^3(X,L^t)\ne0$ only for $t\le-2$, the ring is Cohen--Macaulay with $a$-invariant $-2$ and regularity two, so $K_{p,q}=0$ for $q\ge3$, and $N_7$ holds. For $m=(3,2,2)$ the regularity is three, and the computation gives $K_{p,2}=K_{p,3}=0$ for $1\le p\le4$, so $N_4$ holds. Together with Proposition~\ref{prop:Np} this proves the last item of Theorem~\ref{thm:rank}(b) and Theorem~\ref{thm:rank}(c).

\begin{table}[t]
\centering
\small
\begin{tabular}{cccccc}
\toprule
$m$ & $(p,q)$ & multidegree & dimensions & ranks & time (s)\\
\midrule
$(2,2,2)$ & $(1,2)$ & $(3,3,3)$ & $171,\ 27,\ 1$ & $26,\ 1$ & $0.0$\\
$(2,2,2)$ & $(2,2)$ & $(4,4,4)$ & $1081,\ 351,\ 27$ & $325,\ 26$ & $0.0$\\
$(2,2,2)$ & $(3,2)$ & $(5,5,5)$ & $4990,\ 2433,\ 351$ & $2108,\ 325$ & $0.1$\\
$(2,2,2)$ & $(4,2)$ & $(6,6,6)$ & $18802,\ 12682,\ 2925$ & $10082,\ 2600$ & $2.5$\\
$(2,2,2)$ & $(5,2)$ & $(7,7,7)$ & $58986,\ 51706,\ 16794$ & $37512,\ 14194$ & $88.7$\\
$(3,2,1)$ & $(1,2)$ & $(4,3,1)$ & $125,\ 24,\ 1$ & $23,\ 1$ & $0.0$\\
$(3,2,1)$ & $(2,2)$ & $(6,4,2)$ & $796,\ 276,\ 24$ & $253,\ 23$ & $0.0$\\
$(3,2,1)$ & $(3,2)$ & $(7,5,2)$ & $2901,\ 1585,\ 276$ & $1332,\ 253$ & $0.1$\\
$(3,2,1)$ & $(4,2)$ & $(9,6,3)$ & $10752,\ 7836,\ 2024$ & $6065,\ 1771$ & $0.7$\\
$(3,2,1)$ & $(5,2)$ & $(10,7,3)$ & $26858,\ 26053,\ 9846$ & $17978,\ 8075$ & $12.0$\\
$(3,2,1)$ & $(6,2)$ & $(12,8,4)$ & $67584,\ 80188,\ 38752$ & $50291,\ 29897$ & $262.8$\\
$(3,2,1)$ & $(7,2)$ & $(13,9,4)$ & $121466,\ 179045,\ 112852$ & $98836,\ 80209$ & $1834.5$\\
$(3,2,2)$ & $(1,2)$ & $(4,3,3)$ & $293,\ 36,\ 1$ & $35,\ 1$ & $0.0$\\
$(3,2,2)$ & $(1,3)$ & $(6,4,4)$ & $630,\ 36,\ 1$ & $35,\ 1$ & $0.0$\\
$(3,2,2)$ & $(2,2)$ & $(6,4,4)$ & $2548,\ 630,\ 36$ & $595,\ 35$ & $0.0$\\
$(3,2,2)$ & $(2,3)$ & $(7,5,5)$ & $5828,\ 630,\ 36$ & $595,\ 35$ & $0.0$\\
$(3,2,2)$ & $(3,2)$ & $(7,5,5)$ & $15511,\ 5828,\ 630$ & $5233,\ 595$ & $0.6$\\
$(3,2,2)$ & $(3,3)$ & $(9,6,6)$ & $42101,\ 7140,\ 630$ & $6545,\ 595$ & $1.1$\\
$(3,2,2)$ & $(4,2)$ & $(9,6,6)$ & $81752,\ 42101,\ 7140$ & $35556,\ 6545$ & $99.3$\\
$(3,2,2)$ & $(4,3)$ & $(10,7,7)$ & $233226,\ 55933,\ 7140$ & $49388,\ 6545$ & $110.3$\\
\bottomrule
\end{tabular}
\caption{Central weight spaces of the Koszul complexes $\Lambda^{p+1}R_1\otimes R_{q-1}\to\Lambda^pR_1\otimes R_q\to\Lambda^{p-1}R_1\otimes R_{q+1}$ used in Appendix~\ref{app:rank}, computed over $\mathbb F_P$, $P=2^{31}-1$. In every case the rank of the incoming map plus the rank of the outgoing map equals the dimension of the middle term, so the central weight space of $K_{p,q}$ vanishes. Wall-clock times on one core of an Intel Xeon processor at $2.1$\,GHz.}
\label{tab:koszul}
\end{table}

For the toric rings considered here, the multigraded Betti numbers can also be computed as reduced homology of squarefree divisor complexes. Computing these over $\mathbb F_2$, which again bounds the rational Betti numbers from above, we found $K_{p,2}=0$ for all $p<r(m)$ in the configurations $(2,2)$, $(2,3)$, $(1,1,1)$, $(2,1,1)$, $(3,1,1)$, $(2,2,1)$, $(2,2,2)$, $(1,1,1,1)$ and $(2,1,1,1)$, and $K_{r(m),2}\ne0$ over $\mathbb F_2$ in each of them. Betti numbers over $\mathbb F_2$ can exceed those over $\mathbb Q$, so the nonvanishing over $\mathbb F_2$ does not imply nonvanishing over $\mathbb Q$; these computations are not used in the proofs, and the sharpness of the thresholds follows from Theorem~\ref{thm:rank}(a).

\section{Further results}\label{app:further}

This appendix collects short examples for the three questions that are not needed for Theorems~A--C: robustness to collective dephasing and a family of $k$-qubit states whose white-noise threshold tends to one, for the second question; a four-qubit state entangled across all balanced bipartitions and the Bell-nonlocal state of V\'ertesi and Brunner, for the first.

\subsection{Collective dephasing and extendibility}\label{app:dephasing}

For $N\in\Z^k$ with $N_j\ge m_j$, a PSD Toeplitz extension of $T$ to $\Lambda_N$ is a PSD Toeplitz matrix $\widetilde T$ on $\Lambda_N$ whose symbol agrees with $t$ on $D_m$. A decomposable $T$ has extensions to all $\Lambda_N$. For a unit vector $w\in\C^{\Lambda_N}$ (extended by zero to $\Z^k$) put
\[
r_w(d)=\sum_{\gamma}\overline{w_\gamma}\,w_{\gamma-d},\qquad dP_w(\varphi)=\Bigl|\sum_\gamma w_\gamma e^{i\langle\gamma,\varphi\rangle}\Bigr|^2\frac{d\varphi}{(2\pi)^k}.
\]
$P_w$ is a probability measure on $\T^k$ with $\int e^{i\langle d,\varphi\rangle}dP_w=r_w(d)$. Let $\hat n_j$ be the number of excitations in group $j$ and $U_\varphi=\bigotimes_je^{i\varphi_j\hat n_j}$.

\begin{theorem}[Collective dephasing]\label{thm:dephasing}
Let $T\in\Toep^+_m$ have a PSD Toeplitz extension $\widetilde T$ to $\Lambda_N$, and let $w\in\C^{\Lambda_N}$ be a unit vector.
\begin{enumerate}
\item[(a)] The matrix $T\odot r_w$ with entries $t(\alpha-\beta)r_w(\alpha-\beta)$ is decomposable:
\[
T\odot r_w=\int_{\T^k}a(\theta)a(\theta)^*F(\theta)\frac{d\theta}{(2\pi)^k},\qquad F(\theta)=v(\theta)^*\widetilde Tv(\theta)\ge0,\quad v(\theta)_\gamma=w_\gamma e^{i\langle\gamma,\theta\rangle}.
\]
Replacing the integral by the uniform grid with $N_j+m_j+1$ points in coordinate $j$ gives an explicit decomposition into at most $\prod_j(N_j+m_j+1)$ atoms.
\item[(b)] $\rho_{T\odot r_w}=\int U_\varphi\rho_TU_\varphi^*\,dP_w(\varphi)$. Thus the state obtained from $\rho_T$ by collective dephasing with phase distribution $P_w$ is fully separable.
\item[(c)] For the product of sine windows $w_\gamma=\prod_j\sqrt{2/(N_j+2)}\,\sin\bigl(\pi(\gamma_j+1)/(N_j+2)\bigr)$ one has $r_w(d)=\prod_jr_{N_j+1}(d_j)$ with
\[
r_L(b)=\frac{(L-|b|)\cos(b\phi_L)+\sin((|b|+1)\phi_L)/\sin\phi_L}{L+1},\qquad\phi_L=\frac{\pi}{L+1},
\]
and $0\le1-r_L(b)\le b^2(1-\cos\phi_L)\le\pi^2b^2/(2(L+1)^2)$.
\end{enumerate}
Consequently $T\in\Dec_m$ if and only if $T$ has PSD Toeplitz extensions to $\Lambda_N$ for all $N$. Moreover, if the matrix with entries $t(d)/r_w(d)$, $d\in D_m$, has a PSD Toeplitz extension to $\Lambda_N$, then $T\in\Dec_m$.
\end{theorem}

\begin{proof}
(a) The coefficient of $e^{-i\langle d,\theta\rangle}$ in $F$ is $r_w(d)\tilde t(d)$, which gives the integral formula, and the integrand has degree at most $N_j+m_j$ in $\theta_j$, so the uniform grid integrates it exactly. (b) $\bra xU_\varphi\rho_TU_\varphi^*\ket{x'}=e^{i\langle\varphi,w(x)-w(x')\rangle}t(w(x)-w(x'))$. (c) The closed form is a direct computation, and $1-r_L(b)=\int(1-\cos b\varphi)\,dP_w\le b^2(1-r_L(1))$ by $|\sin(bx/2)|\le b|\sin(x/2)|$. The consequences follow from (a), since $r_w\to1$ pointwise for sine windows of growing size and $\Dec_m$ is closed.
\end{proof}

The equivalence of decomposability with PSD Toeplitz extensions to all grids is classical \cite{BakonyiWoerdeman2011,GeronimoWoerdeman2004}; Theorem~\ref{thm:dephasing} interprets it as collective dephasing, in analogy with the extendibility hierarchy of \cite{DPS2004}. For $T_\star$ of Theorem~\ref{thm:twogroups} with $m=(2,2)$ and $t(0)=1$, the threshold for mixing with the state $\sigma_0$ of uniformly distributed phases is $0.1068715$, from the necessary condition on $\Lambda_{(3,3)}$ and from column generation; the white-noise threshold of $\rho_{T_\star}$, computed in the finest partition (Lemma~\ref{lem:merge}), is $\lambda^*\approx0.1420$.

\subsection{A family with thresholds tending to one}\label{app:mabk}

Let $m=(1,\dots,1)$ with $k\ge2$ qubits. Define sign vectors $a_n,b_n\in\{\pm1\}^{\{\pm1\}^n}$ recursively by $a_1=(1,1)$, $b_1=(1,-1)$ (indexed by $v_1=+1,-1$) and
\[
a_{n+1}(v,v_{n+1})=\begin{cases}a_n(v),&v_{n+1}=+1,\\ b_n(v),&v_{n+1}=-1,\end{cases}\qquad b_{n+1}(v,v_{n+1})=\begin{cases}a_n(v),&v_{n+1}=+1,\\ -b_n(v),&v_{n+1}=-1.\end{cases}
\]
For $c\in\R$ let $T^{(k)}(c)$ be the Toeplitz matrix on $\{0,1\}^k$ with $t(0)=1$, $t(\pm(1,v))=c\,a_{k-1}(v)$ for $v\in\{\pm1\}^{k-1}$, and $t(d)=0$ otherwise.

\begin{proposition}\label{prop:mabk}
$T^{(k)}(c)\ge0$ if and only if $|c|\le1$, $\rank T^{(k)}(1)=2^{k-1}$, and $T^{(k)}(c)\in\Dec$ if and only if $|c|\le2^{1-k/2}$. Hence the $k$-qubit PPT state $\rho_{T^{(k)}(1)}/2^k$, which is separable across every bipartition of type $1|(k-1)$, becomes fully separable under white noise exactly at $\lambda^*=1-2^{1-k/2}$.
\end{proposition}

\begin{proof}
Nonzero off-diagonal entries pair each $\alpha\in\{0,1\}^k$ with its complement, so $T^{(k)}(c)$ is a direct sum of $2\times2$ blocks $\begin{psmallmatrix}1&\pm c\\\pm c&1\end{psmallmatrix}$, which gives the first two statements. Let $H\subset\{0,\pi\}^k$ be the subgroup of vectors with an even number of entries $\pi$. Translation by $\delta\in H$ multiplies the moment of order $d$ by $e^{i\langle d,\delta\rangle}$, which equals $1$ when all entries of $d$ are nonzero, and averaging over $H$ annihilates every moment whose support is nonempty and not everything. Hence $T^{(k)}(c)\in\Dec$ if and only if $c\,a_{k-1}$ lies in the convex hull $K_{k-1}$ of the vectors $x(\theta)=(e^{i(\theta_0+\langle v,\theta'\rangle)})_v$, $\theta=(\theta_0,\theta')\in\T^k$.

Upper bound. Let $A_n(\theta)=\sum_va_n(v)e^{i\langle v,\theta'\rangle}$ and define $B_n$ with $b_n$. The recursion gives $A_{n+1}=e^{i\theta_{n+1}}A_n+e^{-i\theta_{n+1}}B_n$ and $B_{n+1}=e^{i\theta_{n+1}}A_n-e^{-i\theta_{n+1}}B_n$, so $|A_{n+1}|^2+|B_{n+1}|^2=2(|A_n|^2+|B_n|^2)=2^{n+2}$. If $c\,a_{k-1}\in K_{k-1}$, pairing with $a_{k-1}$ gives $c\,2^{k-1}\le\max_\theta|A_{k-1}(\theta)|\le2^{k/2}$.

Lower bound. Let $c_n=\max\{c:c\,a_n\in K_n\}$, so $c_1=1$. By induction $b_n(v)=v_n\,a_n(v)$. The set $K_n$ is invariant under the gauge transformations $x_v\mapsto e^{i(\delta_0+\langle v,\delta'\rangle)}x_v$, and $a_n\pm ib_n=\sqrt2\,g_\pm(a_n)$ for the gauge $g_\pm$ that multiplies the entries with $v_n=+1$ by $e^{\pm i\pi/4}$ and those with $v_n=-1$ by $e^{\mp i\pi/4}$. With $u=e^{i\pi/4}$ and $c'=c_n/\sqrt2$, the vectors $x_+=c'\bar u(a_n+ib_n)$ and $x_-=c'u(a_n-ib_n)$ lie in $K_n$. Choosing $\theta_{n+1}=\pi/4$ with the measure for $x_+$ and $\theta_{n+1}=-\pi/4$ with the measure for $x_-$, each with probability $1/2$, produces the moments $\tfrac12(ux_++\bar ux_-)=c'a_n$ for $v_{n+1}=+1$ and $\tfrac12(\bar ux_++ux_-)=c'b_n$ for $v_{n+1}=-1$, that is $c'a_{n+1}$. Hence $c_{n+1}\ge c_n/\sqrt2$, and $c_{k-1}\ge2^{1-k/2}$. Finally $T^{(k)}(1)_\lambda=T^{(k)}(1-\lambda)$.
\end{proof}

The states of Proposition~\ref{prop:mabk} are diagonal in a basis of GHZ-type states, and the polynomial $2^{k/2}-\mathrm{Re}\,A_{k-1}$ is a nonnegative trigonometric polynomial whose witness is a Mermin--Ardehali--Belinskii--Klyshko-type operator with equatorial settings. Bound entangled GHZ-diagonal states of many qubits were first studied by D\"ur \cite{Dur2001}, entanglement detection in GHZ-diagonal states by Kay \cite{Kay2011}, and separability criteria for three-qubit X-shaped states, which include the GHZ-diagonal states, in \cite{HanKye2017GHZ,HanKye2017,ChenHanKye2017}. The states of Proposition~\ref{prop:mabk} have nonzero entries only on the diagonal and the anti-diagonal, and by \cite[Corollary~4.6]{HaHanKye2018} their full separability is equivalent to an inequality between the diagonal entry and a dual norm of the anti-diagonal entries, which is difficult to evaluate in general. For $k=3$ the threshold $1-1/\sqrt2$ also follows from the criterion of G\"uhne \cite{Guhne2011}, which Han and Kye proved to be necessary and sufficient for three-qubit GHZ-diagonal states \cite{HanKye2017GHZ}. For $k\ge4$, Proposition~\ref{prop:mabk} evaluates the criterion of \cite{HaHanKye2018} for this family and gives the closed form $1-2^{1-k/2}$ of the threshold, which tends to one as $k$ grows.

\subsection{A four-qubit state entangled across all balanced bipartitions}\label{app:tetracode}

For four qubits forming four groups, Theorem~\ref{thm:cuts}(a) shows that every Toeplitz state is separable across the four bipartitions of type $1|3$. A single state can be entangled across all three balanced bipartitions. Let
\[
\Lat=\{d\in\Z^4:\ d_3\equiv d_1+2d_2,\ d_4\equiv d_1+d_2\ (\mathrm{mod}\ 3)\},
\]
of index $N=9$, whose reduction modulo $3$ is the ternary tetracode.

\begin{example}\label{ex:tetracode}
For this lattice, $\Lambda=\{0,1\}^4$ meets $8$ classes and (L1), (L2) hold. The matrix $T_8$ of Theorem~\ref{thm:lattice}, with $t(d)=7$ for $d\in\Lat$ and $t(d)=-1$ otherwise, is PSD of rank $7$, with the single nonzero eigenvalue $16$, and not decomposable; its Toeplitz face has dimension one, by an exact computation over $\mathbb Q$ (Appendix~\ref{app:comp}), so it also spans an extreme ray. For each of the three balanced bipartitions the range of $\rho_{T_8}$ contains no vector that is a product across the bipartition. Hence $\rho_{T_8}$ is a four-qubit state that is PPT with respect to every bipartition, separable across every bipartition of type $1|3$, and entangled across every bipartition of type $2|2$.
\end{example}

The absence of product vectors was verified by exact elimination over all support patterns (Lemma~\ref{lem:prodvec} and Appendix~\ref{app:comp}).

\subsection{Bell nonlocality}\label{sec:bell}

Peres conjectured that PPT states admit local hidden variable models \cite{Peres1999}; D\"ur showed that multipartite bound entangled states can violate Bell inequalities \cite{Dur2001}, and V\'ertesi and Brunner disproved the conjecture with a three-qubit state $\rho_{\rm VB}$ that is PPT with respect to every bipartition and violates inequality number~5 of \'Sliwa \cite{VertesiBrunner2012,Sliwa2003}. In the parametrisation of \cite{VertesiBrunner2012}, with $(\alpha,\beta,\gamma)=(\pi/12,\pi/4,5\pi/12)$, invariance under partial transposition and normalisation determine $\omega=0.568177$, $p_1=0.063604$, $p_2=p_3=0.273734$ and $p_4=0.388929$, in agreement with the values reported there.

\begin{remark}\label{rem:vb}
Since $\rho_{\rm VB}$ is real and invariant under all partial transposes, it belongs to $\Hank_3$, and by Lemma~\ref{lem:pauli3} it is a Toeplitz state after a local unitary. It has rank four and $Q=3.71\times10^{-2}$ at unit trace, so it is of type~I. No diagonal entry of $\widehat G^{-1}$ vanishes, so by Theorem~\ref{thm:moduli}(d) $\rho_{\rm VB}$ does not arise from a UPB by invertible local operators, as shown by Singh and Banik \cite{SinghBanik2026} by the same count of product vectors in the kernel.
\end{remark}

With the settings of \cite{VertesiBrunner2012} we obtain the Bell value $S=3.006852$ against the local bound $3$, and the white-noise threshold of $\rho_{\rm VB}$ is $\lambda^*=0.091736$ (Table~\ref{tab:thresholds}). A heuristic search over Toeplitz states and over PPT states near the rank-four family, with dichotomic observables $n\cdot\vec\sigma$, $|n|=1$, found violations only of inequality number~5 among the $46$ classes of \'Sliwa, with largest value $S=3.0187072$, in agreement with \cite{VertesiBrunner2012}; settings with deterministic outcomes for one observable of a party were not searched, and these values are lower bounds (Appendix~\ref{app:comp}).

\section{Computations}\label{app:comp}
\sloppy

All computations can be reproduced with the code in the supplementary material, which is also provided as ancillary files of the arXiv version (directory \texttt{anc}). It uses Python~3 with NumPy, SciPy, SymPy, mpmath and CVXPY, the program \texttt{scdd} of cddlib for the facet enumeration, and a C++ compiler for the Koszul program. Each script is run from the directory that contains it; we list the commands together with the statements they support.

\paragraph{Moment problem oracle.} The module \texttt{oracle.py} computes membership $T_\lambda\in\Dec_m$ and the noise threshold $\lambda^*(T)$ by column generation: a linear programme over a finite set of atoms $a(\theta)a(\theta)^*$ determines the smallest $\lambda$, its dual gives a trigonometric polynomial $q$, and new atoms are added at the minima of $q$ found by local optimisation from many starting points, until $\min q\ge0$ up to $10^{-10}$. The command \texttt{python3 oracle.py} gives $\lambda^*=0.1715728753$ for the Shifts state, which agrees with $3-2\sqrt2$ to within $6\times10^{-11}$, and $\lambda^*=0.1932774$ for Gabardo's state; \texttt{python3 vb\_threshold.py} gives $\lambda^*=0.091736$ for the V\'ertesi--Brunner state, and \texttt{python3 four\_qubit\_thresholds.py} gives the white-noise thresholds $0.1419685$ of $\rho_{T_\star}$ and $0.1366822$ of the tetracode state. \texttt{python3 robust\_search.py 12 45} runs the local search of Section~\ref{sec:noise} from Gabardo's state, with seed $12$ of the random number generator and at most $45$ accepted or rejected steps; the search is deterministic for given arguments and prints the threshold of every trial state. The best state found in our run, which was stopped before completion, is stored in \texttt{best\_type1.json}; this file is the reference object for the value $0.2635$, and \texttt{python3 robust\_search.py verify best\_type1.json} checks that it has rank four, is PPT, has $Q\approx0.0305$ and no product vector in its range (smallest distance $0.084$), and recomputes $\lambda^*=0.263459$ with a dual certificate that is nonnegative on a grid of $181^3$ points.

\texttt{python3 certify\_thresholds.py} recomputes the numerical thresholds of Table~\ref{tab:thresholds} with the dual trigonometric polynomial $q$, evaluates $q$ on a grid of $181^3$ points for three qubits and $61^4$ points for four qubits, refines the $60$ smallest grid values by local minimisation, and prints the interval $[\lambda_{\rm LP}+\min(0,\min q),\lambda_{\rm LP}]$ given by weak duality. The minima of $q$ are $-2.1\times10^{-10}$ for Gabardo's state, $-1.1\times10^{-10}$ for the V\'ertesi--Brunner state, $-1.8\times10^{-8}$ for the state of \texttt{best\_type1.json}, $-1.1\times10^{-7}$ for $\rho_{T_\star}$ and $-7.3\times10^{-8}$ for the tetracode state, and the resulting intervals are $[0.1932774,0.1932774]$, $[0.0917361,0.0917361]$, $[0.2634590,0.2634590]$, $[0.1419684,0.1419685]$ and $[0.1366822,0.1366822]$ to seven digits. \texttt{python3 certify\_thresholds.py rigorous} turns this into a computer-assisted proof for the three-qubit states. On each cell of a grid of $121^3$ cells it bounds $q$ from below by $q(c)-\|\nabla q(c)\|_1r-\tfrac12Mr^2$, where $c$ is the centre, $r$ the half-width in the maximum norm and $M=\sum_d|\hat q(d)|\,|d|_1^2$, minus a margin of $10^{-12}(1+|y_0|+\sum_d|\hat q(d)|)$ for rounding errors; cells with bound below $-10^{-8}$ are divided into $27$ subcells, up to nine times. The resulting lower bounds for $\min q$ are $-1.0\times10^{-8}$, $-1.0\times10^{-8}$ and $-1.2\times10^{-8}$. For the upper bound, the moments of the final primal decomposition differ from those of $T_{\lambda_{\rm LP}}$ by at most $7\times10^{-15}$ in total; adding the trigonometric polynomial with these differences as coefficients to the uniform measure gives a decomposition of $T_\lambda$ for $\lambda$ larger than $\lambda_{\rm LP}$ by at most the same amount. The certified intervals agree with the values above to seven digits.

\paragraph{UPB states.} \texttt{python3 shifts\_recipe.py} checks the measurement recipe of Section~\ref{sec:recipe}: the Shifts frame, the Pauli form of $W_{\rm S}'$ and its expectation values, the equality of the Pauli supports of $W_{\rm S}'$ and $\Pi$, the eight-state decomposition of Table~\ref{tab:shifts} (error below $10^{-16}$), the vanishing of $W_{\rm S}'$ on these product states and their linear independence, and the statistics of the ratio of thresholds quoted after Proposition~\ref{prop:unique}, including a bootstrap interval for the median from $10000$ resamples. \texttt{python3 upb\_symmetric.py} computes the dashed curve in Figure~\ref{fig:upb}. \texttt{python3 upb\_identities.py} checks the formulas for the orbit sums $O_d$ in Section~\ref{sec:noise} at random points in $50$-digit arithmetic (largest deviation below $10^{-49}$), verifies symbolically the identities $|\mathbf m_0|=1$, $A_0B_0C_0=-s/8$, $e_1e_2=s/8$, the formula for $\ell\cdot \mathbf m_0$ and the final equality in the proof of Theorem~\ref{thm:upb}, and checks the decomposition at the threshold (largest deviation below $10^{-50}$). \texttt{python3 upb\_certificate.py} reproduces $T_{\lambda^*}$ from the eight-atom decomposition with error below $2\times10^{-15}$ and gives $\max\ell\cdot m=1+ABC$ to within $3\times10^{-16}$ on six parameter triples. \texttt{python3 upb\_optimal\_witness.py} computes the witness of Corollary~\ref{cor:upb}(d) from $q$ by Theorem~\ref{thm:dict}(c) and transforms it to the frame of Bravyi's form. For $\theta_A=\theta_B=\theta_C=\pi/4$ it coincides with $W_{\rm S}$ up to a positive factor, the minimum of $\bra\psi W_{\rm S}\ket\psi$ over product vectors from $300$ starting points is $0$ to within $10^{-14}$, and $\tr\bigl(W_{\rm S}[(1-\lambda)\rho_{\rm UPB}+\lambda I/8]\bigr)$ changes sign at $\lambda=3-2\sqrt2$. For five random angle triples the witness contains no strings with a factor $Y$ and no one-body strings, it contains $12$ two-body and $8$ three-body strings, its minimum over product vectors is $0$ to within $3\times10^{-11}$, and the trace changes sign at $\lambda^*$; it agrees with the closed form of Corollary~\ref{cor:upb}(d) to within $3\times10^{-16}$, its coefficient of $XXX$ equals $\tau^*(e_1^2-e_2^2)/8$, and the coefficient of $XXX$ in $\Pi$ vanishes. \texttt{python3 upb\_uniqueness.py} constructs the eight points of the representing measure from the explicit formulas in the proof of Theorem~\ref{thm:upb} for the Shifts UPB and nine further angle triples, checks that $q$ vanishes there (largest value $4\times10^{-16}$) and that local minimisation of $q$ from $400$ starting points finds exactly these eight zeros, and shows that the $32$ linear conditions of Proposition~\ref{prop:unique} on the $27$ coefficients have a one-dimensional solution space in each case, with smallest nonzero singular value at least $0.098$. \texttt{python3 upb\_shifts\_exact.py} carries out the exact computation in the proof of Corollary~\ref{cor:unique}: for the Shifts UPB it identifies the eight points with coordinates $(\pm\beta,\pm\alpha)$ and $(\pm\alpha,\pm\beta)$, verifies symbolically that $q$ and its gradient vanish and that the Hessian of $q$ is positive definite there, and computes the rank $26$ of the $32\times27$ matrix of conditions over the field $\mathbb Q(\alpha)$. \texttt{python3 upb\_projector\_witness.py} computes the detection thresholds of the projector witnesses in Figure~\ref{fig:upb} by minimising $\bra\psi\Pi\ket\psi$ over product vectors from $60$ starting points per UPB and stores the angles, $s$, $\lambda^*$ and the detection thresholds in \texttt{upb\_points.json}, and \texttt{python3 upb\_figure.py} draws the figure.

\paragraph{Three qubits.} \texttt{python3 det\_identity.py} verifies Lemma~\ref{lem:det}, with the partial transpose on the first qubit, by symbolic expansion in the sixteen real coordinates of $M$. \texttt{python3 exact\_checks.py} verifies in exact arithmetic the statements of Remark~\ref{rem:gabardo} and Example~\ref{ex:tetracode}: for Gabardo's matrix, rank four, $Q=15/512$ at unit trace, the binary quartic whose roots are the points $[x_i]$ has invariant $J=0$, its roots are $\pm1$ and $e^{\pm2\pi i/3}$ with cross-ratio $1/2$, and the diagonal entries of $G^{-1}$ are nonzero; for the tetracode matrix, rank seven, $T_8^2=16\,T_8$, and the Toeplitz face has dimension one. As an independent test of Theorem~\ref{thm:main3}, \texttt{python3 rank4\_filter\_test.py 1 200} and \texttt{python3 rank4\_filter\_test.py 2 200} generate rank-four states that are PPT with respect to all three bipartitions by solving $\rho=RR^*$, $\rho^{\Gamma_j}=S_jS_j^*$ with $R,S_j\in\C^{8\times4}$ by the Levenberg--Marquardt method from random starting points, without using the structure established in Section~\ref{sec:three}. Of the $400$ states obtained, $362$ have a product vector in the range, up to $10^{-8}$, or a partial transpose of rank different from four. The remaining $38$ are entangled and of type~I, with $Q$ between $1.1\times10^{-3}$ and $0.202$ at unit trace. The sampling does not exclude states of type~II, but none occurred among the $400$ samples; for type~II the explicit form of Cheng and Chen \cite{ChengChen2026} is available, and the test concerns type~I only. For each of these $38$ states and each qubit $j$, the linear map $H\mapsto\tr_j[(H\otimes I)\rho]$ on Hermitian $2\times2$ matrices is singular, with ratio of the smallest to the largest singular value below $9\times10^{-16}$, its kernel vector is an indefinite matrix, and after the resulting filter in $\SL(2,\C)^3$ the largest coefficient of a Pauli string containing $Z$ is below $3\times10^{-15}$, as predicted by the proof of Theorem~\ref{thm:main3}. \texttt{python3 moduli\_test.py} constructs $200$ random states of the form in Theorem~\ref{thm:normal}(a), recovers their decompositions from the states alone, and checks that the triple $(\mathrm{cr},\widehat G,o)$ is recovered, that it is unchanged up to the finite group of Section~\ref{sec:three} under random real local operators, and that the exchange of qubits $2$ and $3$ changes $o$; it also computes the $j$-invariants $j_2,j_3$ of Remark~\ref{rem:jinv} from the decompositions across $2\,|\,13$ and $3\,|\,12$. All checks succeed for all $200$ states, $j_2\ne j_3$ in all cases with smallest relative difference $3.9\times10^{-4}$, and the exchange interchanges $j_2$ and $j_3$. \texttt{python3 vb\_state.py} solves for the parameters of the V\'ertesi--Brunner state and checks its rank and invariance under partial transposition, and \texttt{python3 vb\_bell\_value.py} evaluates \'Sliwa's inequality number~5 at the settings of \cite{VertesiBrunner2012}, which gives $3.006852$.

\paragraph{Extension hierarchy and the family of Proposition~\ref{prop:mabk}.} \texttt{python3 extension\_hierarchy.py} computes, by semidefinite programmes over Toeplitz matrices on $\Lambda_N$, the necessary bound $0.1068715$ quoted after Theorem~\ref{thm:dephasing} and the sufficient bounds $0.801$, $0.661$, $0.543$, $0.448$ and $0.374$ from sine windows on $\Lambda_{(L-1,L-1)}$ for $L=4,\dots,8$; it stops at $L=8$, since the next grid needs more than $7$\,GB of memory, and \texttt{python3 tstar\_threshold.py} computes $\lambda^*(T_\star)=0.1068715$ by column generation. \texttt{python3 recursive\_family.py} computes the largest $c$ with $T^{(k)}(c)\in\Dec$ for $k=2,\dots,5$ by column generation, in agreement with $2^{1-k/2}$ to within $2\times10^{-8}$, and checks the closed form of the sine windows in Theorem~\ref{thm:dephasing}(c).

\paragraph{Bell inequalities.} In the directory \texttt{bell}, \texttt{python3 mkpoly.py} writes the $64$ deterministic points, \texttt{scdd loc.ext} enumerates the facets of the local polytope, \texttt{python3 classify.py} verifies each of the $53856$ facets in integer arithmetic to be valid and tight on $26$ affinely independent vertices and groups them into orbits of the symmetry group of order $3072$, and \texttt{python3 mapsliwa.py} matches \'Sliwa's list to the $46$ orbits by computing the orbit of each listed inequality. \texttt{python3 opt2.py T 30 1} and \texttt{python3 opt2.py P 20 1} run the optimisation of Appendix~\ref{sec:bell} over Toeplitz states and over general PPT states; it alternates an $8\times8$ semidefinite programme for the state with closed-form updates of the observables, since for fixed other observables the Bell value is affine in the Bloch vector of each observable.

\paragraph{Example~\ref{ex:tetracode}.} \texttt{python3 tetracode.py} checks the conditions (L1), (L2) and the count of classes by enumeration. For each balanced bipartition, a product vector $\varphi\otimes\chi$ in the range must be orthogonal to the binomial kernel vectors and to the class-sum vectors. For each pair of supports of $\varphi$ and $\chi$ that is compatible with the binomial conditions, these conditions define a subgroup of an algebraic torus, which the script computes in Smith normal form; the remaining condition is then a Laurent polynomial on this subgroup, and the system has a solution if and only if this polynomial has at least two terms or is identically zero. For the tetracode lattice no support pattern admits a solution for any of the three balanced bipartitions.

\paragraph{Koszul cohomology.} The program \texttt{koszul/koszul.cpp}, compiled with \texttt{g++ -O2 -std=c++17} and run as \texttt{./koszul m1 m2 m3 p q}, computes the central weight space of $K_{p,q}$ in Appendix~\ref{app:rank} by sparse elimination over $\mathbb F_P$, $P=2^{31}-1$; it prints the central multidegree, the dimensions of the three terms, the ranks of the two maps and the dimension of the homology. All cases computed are listed in Table~\ref{tab:koszul} with their wall-clock running times on one core of an Intel Xeon processor at $2.1$\,GHz; the longest, $m=(3,2,1)$ with $p=7$, takes about half an hour. The cases with $m=(2,2,2)$ confirm a vanishing that is also proved without computer in Appendix~\ref{app:rank}. \texttt{python3 koszul/betti\_f2.py} computes multigraded Betti numbers over $\mathbb F_2$ from the squarefree divisor complexes $\Delta_b$ for all multidegrees $b$ up to symmetry, for the configurations given as arguments.

\end{document}